\documentclass{article}

\usepackage{inputenc}
\usepackage{hyperref}
\usepackage{amsmath}
\usepackage{cleveref}

\title{Agentic Very Much! \\Adoption of Coding Agent in New GitHub Projects}
\author{Romain Robbes, Th\'eo Matricon, Thomas Degueule, Andre Hora, Stefano Zacchiroli}
\date{\today}

\usepackage{graphicx}
\usepackage{pdfpages}
\usepackage{hyperref}
\usepackage{booktabs}
\usepackage{geometry}
\usepackage{xspace}
\usepackage{array}
\usepackage{url}
\usepackage[table,xcdraw]{xcolor}
\usepackage{paralist}
\usepackage[size=tiny]{todonotes}
\usepackage{multirow}
\usepackage{float}
\usepackage[stable]{footmisc}

\usepackage{pifont}          
\usepackage{amsmath}
\usepackage{amssymb}
\usepackage[scaled=.8]{beramono}
\usepackage{cleveref}
\usepackage{ifthen}
\usepackage[normalem]{ulem}
\usepackage{environ}

\newcommand{\bothcells}[1]{\multicolumn{2}{c}{\text{#1}}}
\usepackage{tikz}
\usepackage{pgfplots}
\usepackage{pgfplotstable}
\usepackage{xparse}

\pgfplotsset{compat=1.18}

\pgfplotsset{
    sparkline/.style={
        width=4em,
        height=1.2ex,
        axis lines=none,
        ticks=none,
        xlabel={},
        ylabel={},
        every axis plot/.append style={line width=0.4pt},
        clip=false,
        baseline,
        scale only axis,
        xtick=\empty,
        ytick=\empty,
        axis line style={opacity=0},
        enlarge x limits=0.05,
        enlarge y limits=0.1,
    },
    sparkline bar/.style={
        sparkline,
        ybar,
        bar width=0.3em,
        bar shift=0pt,
        enlarge x limits=0.1,
    },
    sparkline line/.style={
        sparkline,
        every axis plot/.append style={mark=none, solid},
    },
       sparkline box/.style={
    width=3em,              
    height=0.8ex,           
    axis lines=none,
    ticks=none,
    xlabel={},
    ylabel={},
    baseline,
    scale only axis,
    xtick=\empty,
    ytick=\empty,
    enlarge x limits=0.05,  
    ymin=-0.5, ymax=0.5,    
}
}

\NewDocumentCommand{\sparklineBarsSimple}{m}{%
    \begin{tikzpicture}[baseline]
        \begin{axis}[sparkline bar]
            \foreach \value [count=\i from 0] in {#1} {
                \pgfmathparse{{\mycolors}[\i]}
                \let\currentcolor\pgfmathresult
                \addplot+[fill=\currentcolor, draw=\currentcolor] coordinates {(\i,\value)};
            }
        \end{axis}
    \end{tikzpicture}%
}

\def\mycolors{{"blue", "red", "green", "orange", "purple", "cyan", "magenta"}}

\NewDocumentCommand{\sparklineBarsCSVbug}{m}{%
    \begin{tikzpicture}[baseline]
        \begin{axis}[sparkline bar]
            \pgfplotstableread[col sep=comma]{#1}\datatable
            \pgfplotstablegetrowsof{\datatable}
            \pgfmathtruncatemacro{\numrows}{\pgfplotsretval-1}
            \foreach \i in {0,...,\numrows} {
                \pgfplotstablegetelem{\i}{value}\of{\datatable}
                \let\myvalue\pgfplotsretval
                \pgfplotstablegetelem{\i}{color}\of{\datatable}
                \let\mycolor\pgfplotsretval
                \addplot+[fill=\mycolor, draw=\mycolor] coordinates {(\i,\myvalue)};
            }
        \end{axis}
    \end{tikzpicture}%
}

\NewDocumentCommand{\sparklineBarsCSV}{m}{%
    \begin{tikzpicture}[baseline]
        \begin{axis}[sparkline bar]
            \pgfplotstableread[col sep=comma]{#1}\datatable
            \pgfplotstablegetrowsof{\datatable}
            \pgfmathtruncatemacro{\numrows}{\pgfplotsretval-1}
            \foreach \i in {0,...,\numrows} {
                \pgfplotstablegetelem{\i}{value}\of{\datatable}
                \let\myvalue\pgfplotsretval
                \pgfplotstablegetelem{\i}{color}\of{\datatable}
                \edef\temp{\noexpand\addplot+[fill=\pgfplotsretval, draw=\pgfplotsretval] coordinates {(\i,\myvalue)};}
                \temp
            }
        \end{axis}
    \end{tikzpicture}%
}

\NewDocumentCommand{\sparklineTikzBars}{O{} m}{%
    \begin{tikzpicture}[baseline, x=0.4em, y=0.6ex]
        \foreach \value [count=\i from 0] in {#2} {
            \pgfmathparse{\value/10}
            \let\scaledvalue\pgfmathresult
            \fill[blue] (\i-0.25, 0) rectangle (\i+0.25, \scaledvalue);
        }
    \end{tikzpicture}%
}

\begin{document}
\maketitle

\begin{abstract}
    In previous work, we investigated the adoption of coding agents in GitHub projects, finding that it was very significant. This study follows this line of work, but analyses new projects, that were created after the previous study. In this new sample, we find that the adoption of coding agents is more than twice as high. We also find that the adoption is significantly more intensive, as the proportion of AI-assisted commits is sensibly higher, despite strong signs that we do not detect all of it.
\end{abstract}

\section{Introduction and methodology reminder}

This is a companion paper to our previous study on coding agent adoption in GitHub projects (the ``Agentic Much?'' paper \cite{robbes2026agentic}). It uses the sames heuristics, and share the same perils \cite{agentminingpaper}. We refer to the original paper for details on the methodology, but give a brief summary:

\begin{itemize}
    \item We use the sampling tool of Dabic et al \cite{dabic2021sampling}, to select projects with at least 100 commits and 5,000 lines of code, that are not forks, and are created after the date of the initial study \cite{robbes2026agentic} (29/08/2025). The sampling tool itself only monitors projects with at least 10 GitHub stars. We access the tool on 04/04/2025, and download a list of ~13,000 such projects.
    \item For each project, we retrieve its file list and gitignore file via the GitHub API. We use these to search for file-based heuristics \cite{agentminingpaper}.
    \item We also retrieve its pull requests (starting in 01/01/2025, taking at most 10,000 PRs) to search for PR-based heuristics.
    \item We do a partial clone from 01/01/2025 to 04/04/2025, to extract the commits from the main branch, and look for commit-based heuristics.
\end{itemize}

Based on this data, we compute metrics on the adoption of these projects. In particular, we compute:

\begin{itemize}
    \item The number of files identified by file-based heuristics as containing agent traces.
    \item The number of lines and changes in these files.
    \item The number of PRs merged in the main branch, in order to more precisely map the commits.
    \item the Ratio of AI-assisted commits over all human + AI-assisted commits, which we call from now on the Commit Ratio.
    \item The ratio of commits is also computed for specific programming language.
\end{itemize}

We also use metrics coming from the original dataset, notably, the number of lines of code in repositories, the number of commits, the number of issues, pull requests, and contributors. We also use the GitHub topics of the repositories.

We run the exact same analysis as in the ``Agentic Much'' paper, on a dataset of newer projects. We compare the results with the analysis on the same dataset of the ``Agentic Much''. In the remainder of this paper, we refer to these results as the ``older projects''.

Note that at this time of writing, the results from the ``Agentic Much'' paper itself are from a run of the analysis on 21/02/2026; the results presented here are from a newer run of the analysis, in early April--the ``Agentic Much'' paper will be updated in due time.

\section{RQ1: Estimation of Overall Adoption}

We begin with overall statistics of tool adoption across the surveyed repositories. \Cref{tab:overall_adoption} compares the adoption of new projects with older projects. We measure adoption at the file level (identified by presence of files matching heuristics in the file list of the paper, or in the .gitignore file). We also measure it at the commit level (identified by matching author and co-authors with known agents, or via identifying pull requests authored by known agents). We can see several striking differences.


\newcommand\VeryTotalProjects{12,794\xspace}
\newcommand\VeryToolUseDenominator{12,794\xspace}
\newcommand\VeryToolUseCount{6,169\xspace}
\newcommand\VeryToolUsePercent{48.22\%\xspace}
\newcommand\VeryCommitUseInToolDenominator{6,169\xspace}
\newcommand\VeryCommitUseInToolCount{4,741\xspace}
\newcommand\VeryCommitUseInToolPercent{76.85\%\xspace}
\newcommand\VeryIgnoredToolsOnlyDenominator{12,794\xspace}
\newcommand\VeryIgnoredToolsOnlyCount{1,511\xspace}
\newcommand\VeryIgnoredToolsOnlyPercent{11.81\%\xspace}
\newcommand\VeryIgnoredToolsAllDenominator{12,794\xspace}
\newcommand\VeryIgnoredToolsAllCount{4,149\xspace}
\newcommand\VeryIgnoredToolsAllPercent{32.43\%\xspace}
\newcommand\VeryAllFileBasedUseDenominator{12,794\xspace}
\newcommand\VeryAllFileBasedUseCount{7,680\xspace}
\newcommand\VeryAllFileBasedUsePercent{60.03\%\xspace}
\newcommand\VeryCommitUseAllFileUsersDenominator{7,680\xspace}
\newcommand\VeryCommitUseAllFileUsersCount{5,624\xspace}
\newcommand\VeryCommitUseAllFileUsersPercent{73.23\%\xspace}
\newcommand\VerySampledCommitUseDenominator{5,114\xspace}
\newcommand\VerySampledCommitUseCount{1,510\xspace}
\newcommand\VerySampledCommitUsePercent{29.53\%\xspace}
\newcommand\VeryExtrapolationGroupCount{5,114\xspace}
\newcommand\VeryExtrapolatedCommitUsers{1,510\xspace}
\newcommand\VeryKnownUsersCount{7,680\xspace}
\newcommand\VeryAllTotalUsersEstimate{9,190\xspace}
\newcommand\VeryOverallAdoptionRatePercent{71.83\%\xspace}
\newcommand\VeryHighEstimateCommitUsers{2,062\xspace}
\newcommand\VeryHighEstimateTotalUsersEstimate{9,742\xspace}
\newcommand\VeryHighEstimateOverallAdoptionRatePercent{76.15\%\xspace}
\newcommand\TotalAnalyzedProjects{12,800\xspace}


\newcommand\AllTotalProjects{127,670\xspace}
\newcommand\AllToolUseDenominator{127,670\xspace}
\newcommand\AllToolUseCount{15,685\xspace}
\newcommand\AllToolUsePercent{12.29\%\xspace}
\newcommand\AllCommitUseInToolDenominator{15,685\xspace}
\newcommand\AllCommitUseInToolCount{10,637\xspace}
\newcommand\AllCommitUseInToolPercent{67.82\%\xspace}
\newcommand\AllIgnoredToolsOnlyDenominator{127,670\xspace}
\newcommand\AllIgnoredToolsOnlyCount{3,083\xspace}
\newcommand\AllIgnoredToolsOnlyPercent{2.41\%\xspace}
\newcommand\AllIgnoredToolsAllDenominator{127,670\xspace}
\newcommand\AllIgnoredToolsAllCount{6,409\xspace}
\newcommand\AllIgnoredToolsAllPercent{5.02\%\xspace}
\newcommand\AllAllFileBasedUseDenominator{127,670\xspace}
\newcommand\AllAllFileBasedUseCount{18,768\xspace}
\newcommand\AllAllFileBasedUsePercent{14.70\%\xspace}
\newcommand\AllCommitUseAllFileUsersDenominator{18,768\xspace}
\newcommand\AllCommitUseAllFileUsersCount{12,135\xspace}
\newcommand\AllCommitUseAllFileUsersPercent{64.66\%\xspace}
\newcommand\AllSampledCommitUseDenominator{108,902\xspace}
\newcommand\AllSampledCommitUseCount{15,015\xspace}
\newcommand\AllSampledCommitUsePercent{13.79\%\xspace}
\newcommand\AllExtrapolationGroupCount{108,902\xspace}
\newcommand\AllExtrapolatedCommitUsers{15,015\xspace}
\newcommand\AllKnownUsersCount{18,768\xspace}
\newcommand\AllAllTotalUsersEstimate{33,783\xspace}
\newcommand\AllOverallAdoptionRatePercent{26.46\%\xspace}
\newcommand\AllHighEstimateCommitUsers{23,222\xspace}
\newcommand\AllHighEstimateTotalUsersEstimate{41,990\xspace}
\newcommand\AllHighEstimateOverallAdoptionRatePercent{32.89\%\xspace}
\newcommand\AllTotalAnalyzedProjects{128,225\xspace}

\newcommand{\cmark}{\ding{51}}   
\newcommand{\xmark}{\ding{55}}   

\begin{table}[ht]
\centering
\caption{Overall adoption statistics. For each statistic, the middle rows show the subset of the data the adoption is measured in.}
\label{tab:overall_adoption}
\scriptsize
\begin{tabular}{l|ccc|rrr|rrr}
\toprule
\textbf{Metric} & \textbf{File} & \textbf{Ignored} & \textbf{Commit} & \textbf{Count} & \textbf{Out of} & \textbf{Percent} & \textbf{Count} & \textbf{Out of} & \textbf{Percent}\\
\midrule
\midrule
\multicolumn{4}{l}{\textbf{File-level tool usage}} & \multicolumn{3}{c}{\textbf{New projects}} & \multicolumn{3}{c}{\textbf{Older projects}}\\
File-level use & \cmark &  -- & -- & \VeryToolUseCount & \VeryTotalProjects & \VeryToolUsePercent
& \AllToolUseCount & \AllTotalProjects & \AllToolUsePercent\\
Ignored files (all) & -- & \cmark & -- & \VeryIgnoredToolsAllCount & \VeryTotalProjects & \VeryIgnoredToolsAllPercent
& \AllIgnoredToolsAllCount & \AllTotalProjects & \AllIgnoredToolsAllPercent \\
Ignored files (only) & \xmark & \cmark & -- & \VeryIgnoredToolsOnlyCount & \VeryTotalProjects & \VeryIgnoredToolsOnlyPercent
& \AllIgnoredToolsOnlyCount & \AllTotalProjects & \AllIgnoredToolsOnlyPercent \\
All file-level use & \bothcells{\cmark (either)} & -- & \VeryAllFileBasedUseCount & \VeryTotalProjects & \VeryAllFileBasedUsePercent
& \AllAllFileBasedUseCount & \AllTotalProjects & \AllAllFileBasedUsePercent \\
\midrule
\multicolumn{10}{l}{\textbf{Commit-level tool usage}} \\
Commit use (file level) & \cmark & -- & \cmark & \VeryCommitUseInToolCount & \VeryCommitUseInToolDenominator & \VeryCommitUseInToolPercent
& \AllCommitUseInToolCount & \AllCommitUseInToolDenominator & \AllCommitUseInToolPercent \\
Commit use (all files) & \bothcells{\cmark (either)} & \cmark & \VeryCommitUseAllFileUsersCount & \VeryCommitUseAllFileUsersDenominator & \VeryCommitUseAllFileUsersPercent
& \AllCommitUseAllFileUsersCount & \AllCommitUseAllFileUsersDenominator & \AllCommitUseAllFileUsersPercent \\
Commit use, no files & \xmark & \xmark & \cmark & \VerySampledCommitUseCount & \VerySampledCommitUseDenominator & \VerySampledCommitUsePercent
& \AllSampledCommitUseCount & \AllSampledCommitUseDenominator & \AllSampledCommitUsePercent \\
High Estimate Commit Users & \xmark & \xmark  & \cmark & $\approx$ \VeryHighEstimateCommitUsers & \VeryExtrapolationGroupCount & $\frac{\VerySampledCommitUsePercent}{\VeryCommitUseAllFileUsersPercent}$
& $\approx$ \AllHighEstimateCommitUsers & \AllExtrapolationGroupCount & $\frac{\AllSampledCommitUsePercent}{\AllCommitUseAllFileUsersPercent}$ \\
\midrule
\multicolumn{10}{l}{\textbf{Adoption estimates}} \\
Conservative Adoption Estimate & -- & -- & -- & \VeryAllTotalUsersEstimate & \VeryTotalProjects & \VeryOverallAdoptionRatePercent
& \AllAllTotalUsersEstimate & \AllTotalProjects & \AllOverallAdoptionRatePercent \\
High Estimate Overall Adoption & -- & -- & -- & $\approx$ \VeryHighEstimateTotalUsersEstimate & \VeryTotalProjects & \VeryHighEstimateOverallAdoptionRatePercent
& $\approx$ \AllHighEstimateTotalUsersEstimate & \AllTotalProjects & \AllHighEstimateOverallAdoptionRatePercent \\
\bottomrule
\end{tabular}
\end{table}

\begin{itemize}
    \item The first is that file-level adoption (``All file level use'' row) is \emph{four times larger} in the newest projects, than in the older projects.
    \item Similarly, the amount of adopters detectable by searching for ignored files (``Ignored files (only)'' row) is 4 to 5 times larger in newer projects, which is roughly in line with the increase in file-level adoption.
    \item In contrast, the commit-only level adoption (``Commit use, no files'' row) is a bit more than twice as large for new projects, so it is comparatively smaller than the increase in file-level adoption.
    \item There is also a higher proportion of commit-level adoption among file users (``Commit use, all files'' row), which goes from a bit less than two thirds, to a bit less than three quarters.
    \item The overall adoption (``Conservative Adoption Estimate'') goes from a bit more than one quarter, to more than two thirds, which is a very significant increase.
\end{itemize}

All in all, this points to two principal findings: in newer projects, the adoption is much more extensive than in older projects. There are also some signs that the adoption is a more ``Explicit'', in newer projects, as the proportion of projects that lack some kind of adoption is, in relative terms, smaller in the newer projects. This effect is limited however.

\newpage

\section{RQ2: Adoption and Project Characteristics}

\newcommand{\VeryAllFilescodelinesNoneSparkline}{\sparklineTikzBars{26.27, 20.35, 23.11, 20.03, 20.20, 18.46, 16.40, 15.41, 13.81, 15.97}}
\newcommand{\VeryAllFilescodelinesExperimentalSparkline}{\sparklineTikzBars{6.24, 4.07, 6.10, 5.37, 6.40, 7.56, 9.29, 9.16, 9.88, 7.84}}
\newcommand{\VeryAllFilescodelinesLimitedSparkline}{\sparklineTikzBars{14.80, 16.42, 11.48, 13.21, 12.65, 15.26, 15.97, 15.84, 16.42, 16.11}}
\newcommand{\VeryAllFilescodelinesConsistentSparkline}{\sparklineTikzBars{16.55, 17.30, 15.70, 19.30, 15.70, 15.12, 17.85, 17.73, 18.31, 16.98}}
\newcommand{\VeryAllFilescodelinesPervasiveSparkline}{\sparklineTikzBars{36.14, 41.86, 43.60, 42.09, 45.06, 43.60, 40.49, 41.86, 41.57, 43.11}}
\newcommand{\VeryAllFilescontributorsNoneSparkline}{\sparklineTikzBars{31.60, 21.79, 17.09, 14.22, 11.52, 7.13, 5.46}}
\newcommand{\VeryAllFilescontributorsExperimentalSparkline}{\sparklineTikzBars{4.75, 7.53, 7.44, 7.59, 9.18, 8.56, 8.85}}
\newcommand{\VeryAllFilescontributorsLimitedSparkline}{\sparklineTikzBars{9.68, 12.44, 14.47, 15.51, 15.45, 20.13, 27.88}}
\newcommand{\VeryAllFilescontributorsConsistentSparkline}{\sparklineTikzBars{10.46, 14.59, 15.38, 20.84, 20.85, 22.35, 30.09}}
\newcommand{\VeryAllFilescontributorsPervasiveSparkline}{\sparklineTikzBars{43.51, 43.64, 45.63, 41.84, 43.00, 41.84, 27.73}}
\newcommand{\VeryAllFilesageyearsNoneSparkline}{\sparklineTikzBars{17.55, 18.48, 20.60, 20.55, 19.93, 15.98}}
\newcommand{\VeryAllFilesageyearsExperimentalSparkline}{\sparklineTikzBars{5.20, 6.67, 7.50, 8.85, 8.42, 9.92}}
\newcommand{\VeryAllFilesageyearsLimitedSparkline}{\sparklineTikzBars{12.01, 13.34, 15.47, 15.36, 19.68, 19.83}}
\newcommand{\VeryAllFilesageyearsConsistentSparkline}{\sparklineTikzBars{15.54, 15.78, 17.36, 19.02, 17.82, 22.04}}
\newcommand{\VeryAllFilesageyearsPervasiveSparkline}{\sparklineTikzBars{49.70, 45.72, 39.07, 36.22, 34.16, 32.23}}
\newcommand{\VeryAllFileswithcommitscodelinesNoneSparkline}{\sparklineTikzBars{0.00, 0.00, 0.00, 0.00, 0.00, 0.00, 0.00, 0.00, 0.00, 0.00}}
\newcommand{\VeryAllFileswithcommitscodelinesExperimentalSparkline}{\sparklineTikzBars{8.24, 5.56, 7.18, 7.35, 8.62, 9.50, 10.95, 11.11, 10.77, 9.50}}
\newcommand{\VeryAllFileswithcommitscodelinesLimitedSparkline}{\sparklineTikzBars{20.25, 19.71, 15.08, 16.49, 17.24, 17.56, 21.01, 18.46, 17.41, 19.71}}
\newcommand{\VeryAllFileswithcommitscodelinesConsistentSparkline}{\sparklineTikzBars{21.86, 22.40, 21.72, 23.12, 18.49, 19.18, 21.01, 20.43, 22.26, 20.07}}
\newcommand{\VeryAllFileswithcommitscodelinesPervasiveSparkline}{\sparklineTikzBars{49.64, 52.33, 56.01, 53.05, 55.66, 53.76, 47.04, 50.00, 49.55, 50.72}}
\newcommand{\VeryAllFileswithcommitscontributorsNoneSparkline}{\sparklineTikzBars{0.00, 0.00, 0.00, 0.00, 0.00, 0.00, 0.00, 0.00}}
\newcommand{\VeryAllFileswithcommitscontributorsExperimentalSparkline}{\sparklineTikzBars{6.95, 9.63, 8.97, 8.85, 10.53, 10.59, 8.89, 9.11}}
\newcommand{\VeryAllFileswithcommitscontributorsLimitedSparkline}{\sparklineTikzBars{14.15, 15.91, 17.45, 18.08, 16.58, 20.41, 21.78, 31.01}}
\newcommand{\VeryAllFileswithcommitscontributorsConsistentSparkline}{\sparklineTikzBars{15.29, 18.66, 18.55, 24.29, 23.68, 23.51, 25.41, 32.36}}
\newcommand{\VeryAllFileswithcommitscontributorsPervasiveSparkline}{\sparklineTikzBars{63.61, 55.80, 55.03, 48.78, 49.21, 45.48, 43.92, 27.52}}
\newcommand{\VeryAllFileswithcommitsageyearsNoneSparkline}{\sparklineTikzBars{0.00, 0.00, 0.00, 0.00, 0.00, 0.00}}
\newcommand{\VeryAllFileswithcommitsageyearsExperimentalSparkline}{\sparklineTikzBars{6.31, 8.18, 9.44, 11.14, 10.51, 11.80}}
\newcommand{\VeryAllFileswithcommitsageyearsLimitedSparkline}{\sparklineTikzBars{14.56, 16.36, 19.48, 19.33, 24.57, 23.61}}
\newcommand{\VeryAllFileswithcommitsageyearsConsistentSparkline}{\sparklineTikzBars{18.85, 19.36, 21.87, 23.94, 22.26, 26.23}}
\newcommand{\VeryAllFileswithcommitsageyearsPervasiveSparkline}{\sparklineTikzBars{60.28, 56.09, 49.20, 45.58, 42.66, 38.36}}
\newcommand{\VeryFilelevelcodelinesNoneSparkline}{\sparklineTikzBars{0.00, 0.00, 0.00, 0.00, 0.00, 0.00, 0.00, 0.00, 0.00, 0.00}}
\newcommand{\VeryFilelevelcodelinesExperimentalSparkline}{\sparklineTikzBars{8.46, 6.36, 7.20, 7.42, 8.69, 9.96, 11.65, 9.96, 10.59, 8.69}}
\newcommand{\VeryFilelevelcodelinesLimitedSparkline}{\sparklineTikzBars{20.30, 20.55, 13.77, 15.89, 15.04, 15.89, 19.92, 17.37, 16.10, 20.34}}
\newcommand{\VeryFilelevelcodelinesConsistentSparkline}{\sparklineTikzBars{19.24, 20.34, 20.97, 21.82, 17.80, 19.70, 18.64, 21.82, 22.88, 19.49}}
\newcommand{\VeryFilelevelcodelinesPervasiveSparkline}{\sparklineTikzBars{52.01, 52.75, 58.05, 54.87, 58.47, 54.45, 49.79, 50.85, 50.42, 51.48}}
\newcommand{\VeryFilelevelcontributorsNoneSparkline}{\sparklineTikzBars{0.00, 0.00, 0.00, 0.00, 0.00, 0.00, 0.00, 0.00}}
\newcommand{\VeryFilelevelcontributorsExperimentalSparkline}{\sparklineTikzBars{7.40, 9.55, 8.90, 9.29, 9.17, 10.69, 8.74, 9.11}}
\newcommand{\VeryFilelevelcontributorsLimitedSparkline}{\sparklineTikzBars{14.31, 15.28, 16.36, 17.48, 15.90, 17.92, 20.68, 29.11}}
\newcommand{\VeryFilelevelcontributorsConsistentSparkline}{\sparklineTikzBars{14.12, 18.29, 17.36, 22.57, 23.85, 22.33, 24.31, 32.44}}
\newcommand{\VeryFilelevelcontributorsPervasiveSparkline}{\sparklineTikzBars{64.17, 56.88, 57.39, 50.66, 51.07, 49.06, 46.27, 29.33}}
\newcommand{\VeryFilelevelageyearsNoneSparkline}{\sparklineTikzBars{0.00, 0.00, 0.00, 0.00, 0.00, 0.00}}
\newcommand{\VeryFilelevelageyearsExperimentalSparkline}{\sparklineTikzBars{6.26, 8.26, 10.51, 10.64, 9.47, 12.02}}
\newcommand{\VeryFilelevelageyearsLimitedSparkline}{\sparklineTikzBars{14.44, 16.01, 16.49, 19.00, 24.95, 21.71}}
\newcommand{\VeryFilelevelageyearsConsistentSparkline}{\sparklineTikzBars{18.29, 18.19, 21.98, 22.19, 21.31, 26.74}}
\newcommand{\VeryFilelevelageyearsPervasiveSparkline}{\sparklineTikzBars{61.02, 57.54, 51.02, 48.18, 44.26, 39.53}}
\newcommand{\VeryFilesignoredonlycodelinesNoneSparkline}{\sparklineTikzBars{0.00, 0.00, 0.00, 0.00, 0.00, 0.00, 0.00, 0.00, 0.00, 0.00}}
\newcommand{\VeryFilesignoredonlycodelinesExperimentalSparkline}{\sparklineTikzBars{5.81, 4.71, 4.65, 8.24, 6.98, 5.88, 7.06, 12.79, 14.12, 17.44}}
\newcommand{\VeryFilesignoredonlycodelinesLimitedSparkline}{\sparklineTikzBars{19.77, 15.29, 19.77, 21.18, 19.77, 32.94, 29.41, 24.42, 21.18, 22.09}}
\newcommand{\VeryFilesignoredonlycodelinesConsistentSparkline}{\sparklineTikzBars{33.72, 34.12, 27.91, 21.18, 27.91, 21.18, 31.76, 15.12, 15.29, 25.58}}
\newcommand{\VeryFilesignoredonlycodelinesPervasiveSparkline}{\sparklineTikzBars{40.70, 45.88, 47.67, 49.41, 45.35, 40.00, 31.76, 47.67, 49.41, 34.88}}
\newcommand{\VeryFilesignoredonlycontributorsNoneSparkline}{\sparklineTikzBars{0.00, 0.00, 0.00, 0.00, 0.00, 0.00, 0.00, 0.00}}
\newcommand{\VeryFilesignoredonlycontributorsExperimentalSparkline}{\sparklineTikzBars{4.76, 10.12, 9.38, 6.33, 18.87, 10.14, 7.35, 11.25}}
\newcommand{\VeryFilesignoredonlycontributorsLimitedSparkline}{\sparklineTikzBars{13.33, 19.64, 23.44, 21.52, 20.75, 31.88, 29.41, 40.00}}
\newcommand{\VeryFilesignoredonlycontributorsConsistentSparkline}{\sparklineTikzBars{20.95, 20.83, 25.00, 34.18, 22.64, 28.99, 30.88, 32.50}}
\newcommand{\VeryFilesignoredonlycontributorsPervasiveSparkline}{\sparklineTikzBars{60.95, 49.40, 42.19, 37.97, 37.74, 28.99, 32.35, 16.25}}
\newcommand{\VeryFilesignoredonlyageyearsNoneSparkline}{\sparklineTikzBars{0.00, 0.00, 0.00, 0.00, 0.00, 0.00}}
\newcommand{\VeryFilesignoredonlyageyearsExperimentalSparkline}{\sparklineTikzBars{6.60, 7.69, 4.14, 13.82, 16.33, 10.64}}
\newcommand{\VeryFilesignoredonlyageyearsLimitedSparkline}{\sparklineTikzBars{15.23, 18.55, 34.32, 21.14, 22.45, 34.04}}
\newcommand{\VeryFilesignoredonlyageyearsConsistentSparkline}{\sparklineTikzBars{21.83, 26.70, 21.30, 33.33, 27.55, 23.40}}
\newcommand{\VeryFilesignoredonlyageyearsPervasiveSparkline}{\sparklineTikzBars{56.35, 47.06, 40.24, 31.71, 33.67, 31.91}}
\newcommand{\VeryCommitsonlycodelinesNoneSparkline}{\sparklineTikzBars{0.00, 0.00, 0.00, 0.00, 0.00, 0.00, 0.00, 0.00, 0.00, 0.00}}
\newcommand{\VeryCommitsonlycodelinesExperimentalSparkline}{\sparklineTikzBars{5.84, 5.11, 8.03, 4.38, 8.03, 5.11, 10.95, 13.14, 16.06, 18.25}}
\newcommand{\VeryCommitsonlycodelinesLimitedSparkline}{\sparklineTikzBars{31.39, 22.63, 27.74, 32.85, 25.55, 32.85, 24.82, 29.93, 29.93, 28.47}}
\newcommand{\VeryCommitsonlycodelinesConsistentSparkline}{\sparklineTikzBars{24.09, 25.55, 32.12, 31.39, 21.90, 28.47, 24.09, 24.09, 26.28, 21.17}}
\newcommand{\VeryCommitsonlycodelinesPervasiveSparkline}{\sparklineTikzBars{38.69, 46.72, 32.12, 31.39, 44.53, 33.58, 40.15, 32.85, 27.74, 32.12}}
\newcommand{\VeryCommitsonlycontributorsNoneSparkline}{\sparklineTikzBars{0.00, 0.00, 0.00, 0.00, 0.00, 0.00, 0.00, 0.00}}
\newcommand{\VeryCommitsonlycontributorsExperimentalSparkline}{\sparklineTikzBars{7.48, 8.37, 5.85, 7.88, 8.33, 6.19, 12.12, 23.36}}
\newcommand{\VeryCommitsonlycontributorsLimitedSparkline}{\sparklineTikzBars{16.93, 23.90, 35.64, 28.08, 37.96, 30.93, 36.36, 33.58}}
\newcommand{\VeryCommitsonlycontributorsConsistentSparkline}{\sparklineTikzBars{16.54, 21.12, 20.21, 29.06, 33.33, 38.14, 37.12, 29.93}}
\newcommand{\VeryCommitsonlycontributorsPervasiveSparkline}{\sparklineTikzBars{59.06, 46.61, 38.30, 34.98, 20.37, 24.74, 14.39, 13.14}}
\newcommand{\VeryCommitsonlyageyearsNoneSparkline}{\sparklineTikzBars{0.00, 0.00, 0.00, 0.00, 0.00, 0.00}}
\newcommand{\VeryCommitsonlyageyearsExperimentalSparkline}{\sparklineTikzBars{7.76, 7.97, 12.16, 7.23, 10.62, 13.22}}
\newcommand{\VeryCommitsonlyageyearsLimitedSparkline}{\sparklineTikzBars{19.83, 26.58, 28.63, 34.04, 29.65, 38.02}}
\newcommand{\VeryCommitsonlyageyearsConsistentSparkline}{\sparklineTikzBars{21.55, 23.59, 23.92, 29.36, 31.86, 26.45}}
\newcommand{\VeryCommitsonlyageyearsPervasiveSparkline}{\sparklineTikzBars{50.86, 41.86, 35.29, 29.36, 27.88, 22.31}}
\newcommand{\VeryAllFilesNoneValue}{19.00\%}
\newcommand{\VeryAllFilesExperimentalValue}{7.19\%}
\newcommand{\VeryAllFilesLimitedValue}{14.82\%}
\newcommand{\VeryAllFilesConsistentValue}{17.05\%}
\newcommand{\VeryAllFilesPervasiveValue}{41.94\%}
\newcommand{\VeryAllFileswithcommitsNoneValue}{0.00\%}
\newcommand{\VeryAllFileswithcommitsExperimentalValue}{8.88\%}
\newcommand{\VeryAllFileswithcommitsLimitedValue}{18.29\%}
\newcommand{\VeryAllFileswithcommitsConsistentValue}{21.05\%}
\newcommand{\VeryAllFileswithcommitsPervasiveValue}{51.78\%}
\newcommand{\VeryFilelevelNoneValue}{0.00\%}
\newcommand{\VeryFilelevelExperimentalValue}{8.90\%}
\newcommand{\VeryFilelevelLimitedValue}{17.52\%}
\newcommand{\VeryFilelevelConsistentValue}{20.27\%}
\newcommand{\VeryFilelevelPervasiveValue}{53.31\%}
\newcommand{\VeryFilesignoredonlyNoneValue}{0.00\%}
\newcommand{\VeryFilesignoredonlyExperimentalValue}{8.77\%}
\newcommand{\VeryFilesignoredonlyLimitedValue}{22.57\%}
\newcommand{\VeryFilesignoredonlyConsistentValue}{25.38\%}
\newcommand{\VeryFilesignoredonlyPervasiveValue}{43.27\%}
\newcommand{\VeryCommitsonlyNoneValue}{0.00\%}
\newcommand{\VeryCommitsonlyExperimentalValue}{9.49\%}
\newcommand{\VeryCommitsonlyLimitedValue}{28.61\%}
\newcommand{\VeryCommitsonlyConsistentValue}{25.91\%}
\newcommand{\VeryCommitsonlyPervasiveValue}{35.99\%}

\newcommand{\VerybinfiletotalfilesSparkline}{\sparklineTikzBars{74.34, 72.59, 19.71, 12.55, 20.81}}
\newcommand{\VerybinfiletotallinesSparkline}{\sparklineTikzBars{6.48, 18.48, 23.15, 45.78, 37.57, 26.07, 42.47}}
\newcommand{\VerybinfiletotalchangesSparkline}{\sparklineTikzBars{16.66, 44.30, 33.40, 36.47, 69.16}}
\newcommand{\VerybinfileadoptionratioSparkline}{\sparklineTikzBars{46.10, 13.57, 26.72, 31.26, 82.36}}
\newcommand{\VerybinfiletotalfileszeroValue}{37.2\%}
\newcommand{\VerybinfiletotalfilesoneValue}{36.3\%}
\newcommand{\VerybinfiletotalfilestwoValue}{9.9\%}
\newcommand{\VerybinfiletotalfilesthreeValue}{6.3\%}
\newcommand{\VerybinfiletotalfilesfourValue}{10.4\%}
\newcommand{\VerybinfiletotallineszeroValue}{3.2\%}
\newcommand{\VerybinfiletotallinesoneValue}{9.2\%}
\newcommand{\VerybinfiletotallinestwoValue}{11.6\%}
\newcommand{\VerybinfiletotallinesthreeValue}{22.9\%}
\newcommand{\VerybinfiletotallinesfourValue}{18.8\%}
\newcommand{\VerybinfiletotallinesfiveValue}{13.0\%}
\newcommand{\VerybinfiletotallinessixValue}{21.2\%}
\newcommand{\VerybinfiletotalchangeszeroValue}{8.3\%}
\newcommand{\VerybinfiletotalchangesoneValue}{22.1\%}
\newcommand{\VerybinfiletotalchangestwoValue}{16.7\%}
\newcommand{\VerybinfiletotalchangesthreeValue}{18.2\%}
\newcommand{\VerybinfiletotalchangesfourValue}{34.6\%}
\newcommand{\VerybinfileadoptionratiozeroValue}{23.0\%}
\newcommand{\VerybinfileadoptionratiooneValue}{6.8\%}
\newcommand{\VerybinfileadoptionratiotwoValue}{13.4\%}
\newcommand{\VerybinfileadoptionratiothreeValue}{15.6\%}
\newcommand{\VerybinfileadoptionratiofourValue}{41.2\%}

\newcommand{\VerycodelinestooluseSparkline}{\sparklineTikzBars{112.38, 160.63, 182.53, 186.48, 218.10, 238.08, 238.44, 259.97, 263.53, 240.90}}
\newcommand{\VerycontributorstooluseSparkline}{\sparklineTikzBars{192.88, 217.64, 223.41, 225.28, 223.60, 207.04, 205.06}}
\newcommand{\VerycommitstooluseSparkline}{\sparklineTikzBars{173.16, 181.15, 192.03, 200.16, 210.72, 213.18, 222.45, 222.06, 245.60, 241.82}}
\newcommand{\VerytotalissuestooluseSparkline}{\sparklineTikzBars{198.56, 205.19, 202.91, 207.03, 207.84, 208.00, 223.29, 255.24}}
\newcommand{\VerytotalpullrequeststooluseSparkline}{\sparklineTikzBars{168.56, 192.55, 203.64, 211.16, 206.23, 220.60, 229.85, 236.44, 273.66}}
\newcommand{\VeryageyearstooluseSparkline}{\sparklineTikzBars{250.61, 242.09, 205.95, 185.81, 169.97, 154.83}}
\newcommand{\VerycodelinestooluseDecilezeroBoundary}{522}
\newcommand{\VerycodelinestooluseDecileoneBoundary}{9.3K}
\newcommand{\VerycodelinestooluseDeciletwoBoundary}{15K}
\newcommand{\VerycodelinestooluseDecilethreeBoundary}{21K}
\newcommand{\VerycodelinestooluseDecilefourBoundary}{28K}
\newcommand{\VerycodelinestooluseDecilefiveBoundary}{38K}
\newcommand{\VerycodelinestooluseDecilesixBoundary}{52K}
\newcommand{\VerycodelinestooluseDecilesevenBoundary}{74K}
\newcommand{\VerycodelinestooluseDecileeightBoundary}{113K}
\newcommand{\VerycodelinestooluseDecilenineBoundary}{220K}
\newcommand{\VerycodelinestooluseDecilezero}{32.11\%}
\newcommand{\VerycodelinestooluseDecileone}{45.90\%}
\newcommand{\VerycodelinestooluseDeciletwo}{52.15\%}
\newcommand{\VerycodelinestooluseDecilethree}{53.28\%}
\newcommand{\VerycodelinestooluseDecilefour}{62.31\%}
\newcommand{\VerycodelinestooluseDecilefive}{68.02\%}
\newcommand{\VerycodelinestooluseDecilesix}{68.12\%}
\newcommand{\VerycodelinestooluseDecileseven}{74.28\%}
\newcommand{\VerycodelinestooluseDecileeight}{75.29\%}
\newcommand{\VerycodelinestooluseDecilenine}{68.83\%}
\newcommand{\VerycontributorstooluseDecilezeroBoundary}{0.0}
\newcommand{\VerycontributorstooluseDecileoneBoundary}{1.0}
\newcommand{\VerycontributorstooluseDeciletwoBoundary}{2.0}
\newcommand{\VerycontributorstooluseDecilethreeBoundary}{3.0}
\newcommand{\VerycontributorstooluseDecilefourBoundary}{4.0}
\newcommand{\VerycontributorstooluseDecilefiveBoundary}{6.0}
\newcommand{\VerycontributorstooluseDecilesixBoundary}{12}
\newcommand{\VerycontributorstooluseDecilezero}{55.11\%}
\newcommand{\VerycontributorstooluseDecileone}{62.18\%}
\newcommand{\VerycontributorstooluseDeciletwo}{63.83\%}
\newcommand{\VerycontributorstooluseDecilethree}{64.37\%}
\newcommand{\VerycontributorstooluseDecilefour}{63.89\%}
\newcommand{\VerycontributorstooluseDecilefive}{59.15\%}
\newcommand{\VerycontributorstooluseDecilesix}{58.59\%}
\newcommand{\VerycommitstooluseDecilezeroBoundary}{100}
\newcommand{\VerycommitstooluseDecileoneBoundary}{116}
\newcommand{\VerycommitstooluseDeciletwoBoundary}{136}
\newcommand{\VerycommitstooluseDecilethreeBoundary}{160}
\newcommand{\VerycommitstooluseDecilefourBoundary}{191}
\newcommand{\VerycommitstooluseDecilefiveBoundary}{232}
\newcommand{\VerycommitstooluseDecilesixBoundary}{287}
\newcommand{\VerycommitstooluseDecilesevenBoundary}{376}
\newcommand{\VerycommitstooluseDecileeightBoundary}{528}
\newcommand{\VerycommitstooluseDecilenineBoundary}{892}
\newcommand{\VerycommitstooluseDecilezero}{49.47\%}
\newcommand{\VerycommitstooluseDecileone}{51.76\%}
\newcommand{\VerycommitstooluseDeciletwo}{54.86\%}
\newcommand{\VerycommitstooluseDecilethree}{57.19\%}
\newcommand{\VerycommitstooluseDecilefour}{60.21\%}
\newcommand{\VerycommitstooluseDecilefive}{60.91\%}
\newcommand{\VerycommitstooluseDecilesix}{63.56\%}
\newcommand{\VerycommitstooluseDecileseven}{63.45\%}
\newcommand{\VerycommitstooluseDecileeight}{70.17\%}
\newcommand{\VerycommitstooluseDecilenine}{69.09\%}
\newcommand{\VerytotalissuestooluseDecilezeroBoundary}{0.0}
\newcommand{\VerytotalissuestooluseDecileoneBoundary}{1.0}
\newcommand{\VerytotalissuestooluseDeciletwoBoundary}{2.0}
\newcommand{\VerytotalissuestooluseDecilethreeBoundary}{5.0}
\newcommand{\VerytotalissuestooluseDecilefourBoundary}{9.0}
\newcommand{\VerytotalissuestooluseDecilefiveBoundary}{17}
\newcommand{\VerytotalissuestooluseDecilesixBoundary}{32}
\newcommand{\VerytotalissuestooluseDecilesevenBoundary}{71}
\newcommand{\VerytotalissuestooluseDecilezero}{56.73\%}
\newcommand{\VerytotalissuestooluseDecileone}{58.62\%}
\newcommand{\VerytotalissuestooluseDeciletwo}{57.98\%}
\newcommand{\VerytotalissuestooluseDecilethree}{59.15\%}
\newcommand{\VerytotalissuestooluseDecilefour}{59.38\%}
\newcommand{\VerytotalissuestooluseDecilefive}{59.43\%}
\newcommand{\VerytotalissuestooluseDecilesix}{63.80\%}
\newcommand{\VerytotalissuestooluseDecileseven}{72.93\%}
\newcommand{\VerytotalpullrequeststooluseDecilezeroBoundary}{0.0}
\newcommand{\VerytotalpullrequeststooluseDecileoneBoundary}{1.0}
\newcommand{\VerytotalpullrequeststooluseDeciletwoBoundary}{3.0}
\newcommand{\VerytotalpullrequeststooluseDecilethreeBoundary}{7.0}
\newcommand{\VerytotalpullrequeststooluseDecilefourBoundary}{16}
\newcommand{\VerytotalpullrequeststooluseDecilefiveBoundary}{28}
\newcommand{\VerytotalpullrequeststooluseDecilesixBoundary}{46}
\newcommand{\VerytotalpullrequeststooluseDecilesevenBoundary}{77}
\newcommand{\VerytotalpullrequeststooluseDecileeightBoundary}{150}
\newcommand{\VerytotalpullrequeststooluseDecilezero}{48.16\%}
\newcommand{\VerytotalpullrequeststooluseDecileone}{55.02\%}
\newcommand{\VerytotalpullrequeststooluseDeciletwo}{58.18\%}
\newcommand{\VerytotalpullrequeststooluseDecilethree}{60.33\%}
\newcommand{\VerytotalpullrequeststooluseDecilefour}{58.92\%}
\newcommand{\VerytotalpullrequeststooluseDecilefive}{63.03\%}
\newcommand{\VerytotalpullrequeststooluseDecilesix}{65.67\%}
\newcommand{\VerytotalpullrequeststooluseDecileseven}{67.55\%}
\newcommand{\VerytotalpullrequeststooluseDecileeight}{78.19\%}
\newcommand{\VeryageyearstooluseDecilezeroBoundary}{0.0}
\newcommand{\VeryageyearstooluseDecileoneBoundary}{0.1}
\newcommand{\VeryageyearstooluseDeciletwoBoundary}{0.2}
\newcommand{\VeryageyearstooluseDecilethreeBoundary}{0.3}
\newcommand{\VeryageyearstooluseDecilefourBoundary}{0.4}
\newcommand{\VeryageyearstooluseDecilefiveBoundary}{0.5}
\newcommand{\VeryageyearstooluseDecilezero}{71.60\%}
\newcommand{\VeryageyearstooluseDecileone}{69.17\%}
\newcommand{\VeryageyearstooluseDeciletwo}{58.84\%}
\newcommand{\VeryageyearstooluseDecilethree}{53.09\%}
\newcommand{\VeryageyearstooluseDecilefour}{48.56\%}
\newcommand{\VeryageyearstooluseDecilefive}{44.24\%}

\newcommand{\AllAllFilescodelinesNoneSparkline}{\sparklineTikzBars{28.90, 28.45, 26.56, 25.40, 24.13, 21.80, 21.84, 17.08, 19.23, 21.98}}
\newcommand{\AllAllFilescodelinesExperimentalSparkline}{\sparklineTikzBars{4.37, 3.56, 5.78, 7.12, 8.00, 8.61, 10.08, 11.16, 12.64, 13.37}}
\newcommand{\AllAllFilescodelinesLimitedSparkline}{\sparklineTikzBars{15.52, 15.74, 17.75, 20.16, 21.17, 22.54, 22.45, 26.43, 28.24, 28.02}}
\newcommand{\AllAllFilescodelinesConsistentSparkline}{\sparklineTikzBars{23.86, 26.23, 23.47, 25.54, 26.48, 26.99, 28.49, 29.72, 24.95, 24.87}}
\newcommand{\AllAllFilescodelinesPervasiveSparkline}{\sparklineTikzBars{27.35, 26.03, 26.43, 21.77, 20.23, 20.05, 17.14, 15.60, 14.93, 11.76}}
\newcommand{\AllAllFilescontributorsNoneSparkline}{\sparklineTikzBars{39.81, 34.03, 27.41, 24.33, 22.69, 21.22, 18.68, 17.52, 14.77, 10.11}}
\newcommand{\AllAllFilescontributorsExperimentalSparkline}{\sparklineTikzBars{6.21, 7.14, 7.49, 7.28, 8.58, 8.27, 9.34, 7.43, 9.58, 14.02}}
\newcommand{\AllAllFilescontributorsLimitedSparkline}{\sparklineTikzBars{12.33, 15.70, 16.86, 20.53, 21.64, 22.37, 23.68, 25.22, 28.86, 34.10}}
\newcommand{\AllAllFilescontributorsConsistentSparkline}{\sparklineTikzBars{17.75, 18.66, 23.39, 25.22, 26.32, 27.19, 28.16, 32.79, 31.02, 31.87}}
\newcommand{\AllAllFilescontributorsPervasiveSparkline}{\sparklineTikzBars{23.91, 24.48, 24.84, 22.64, 20.78, 20.94, 20.15, 17.04, 15.78, 9.91}}
\newcommand{\AllAllFilesageyearsNoneSparkline}{\sparklineTikzBars{19.96, 24.23, 22.52, 25.20, 23.33, 24.79, 24.22, 24.43, 22.28, 25.07}}
\newcommand{\AllAllFilesageyearsExperimentalSparkline}{\sparklineTikzBars{9.32, 9.12, 9.91, 8.06, 8.78, 6.77, 7.98, 6.75, 7.65, 10.26}}
\newcommand{\AllAllFilesageyearsLimitedSparkline}{\sparklineTikzBars{21.89, 24.78, 23.00, 22.88, 22.61, 19.48, 21.08, 22.60, 21.29, 18.39}}
\newcommand{\AllAllFilesageyearsConsistentSparkline}{\sparklineTikzBars{25.62, 25.32, 26.59, 23.75, 26.47, 26.93, 27.22, 28.41, 26.76, 23.35}}
\newcommand{\AllAllFilesageyearsPervasiveSparkline}{\sparklineTikzBars{23.21, 16.54, 17.98, 20.12, 18.81, 22.03, 19.51, 17.81, 22.02, 22.93}}
\newcommand{\AllAllFileswithcommitscodelinesNoneSparkline}{\sparklineTikzBars{0.00, 0.00, 0.00, 0.00, 0.00, 0.00, 0.00, 0.00, 0.00, 0.00}}
\newcommand{\AllAllFileswithcommitscodelinesExperimentalSparkline}{\sparklineTikzBars{5.88, 5.90, 7.56, 10.11, 10.54, 11.96, 12.49, 13.19, 15.83, 17.31}}
\newcommand{\AllAllFileswithcommitscodelinesLimitedSparkline}{\sparklineTikzBars{21.77, 22.01, 25.07, 27.09, 28.65, 27.70, 30.08, 31.31, 35.80, 35.68}}
\newcommand{\AllAllFileswithcommitscodelinesConsistentSparkline}{\sparklineTikzBars{33.71, 36.44, 31.22, 34.92, 34.45, 35.53, 35.71, 36.41, 30.52, 31.90}}
\newcommand{\AllAllFileswithcommitscodelinesPervasiveSparkline}{\sparklineTikzBars{38.63, 35.65, 36.15, 27.88, 26.36, 24.80, 21.72, 19.09, 17.85, 15.11}}
\newcommand{\AllAllFileswithcommitscontributorsNoneSparkline}{\sparklineTikzBars{0.00, 0.00, 0.00, 0.00, 0.00, 0.00, 0.00, 0.00, 0.00, 0.00}}
\newcommand{\AllAllFileswithcommitscontributorsExperimentalSparkline}{\sparklineTikzBars{10.32, 10.77, 9.90, 10.36, 11.98, 10.52, 9.77, 9.63, 11.47, 16.23}}
\newcommand{\AllAllFileswithcommitscontributorsLimitedSparkline}{\sparklineTikzBars{20.48, 24.89, 22.50, 29.10, 26.59, 29.00, 28.09, 32.84, 34.25, 38.54}}
\newcommand{\AllAllFileswithcommitscontributorsConsistentSparkline}{\sparklineTikzBars{29.48, 28.33, 34.67, 32.38, 35.75, 33.39, 37.95, 39.27, 36.28, 34.66}}
\newcommand{\AllAllFileswithcommitscontributorsPervasiveSparkline}{\sparklineTikzBars{39.72, 36.01, 32.93, 28.15, 25.68, 27.09, 24.19, 18.26, 18.01, 10.58}}
\newcommand{\AllAllFileswithcommitsageyearsNoneSparkline}{\sparklineTikzBars{0.00, 0.00, 0.00, 0.00, 0.00, 0.00, 0.00, 0.00, 0.00, 0.00}}
\newcommand{\AllAllFileswithcommitsageyearsExperimentalSparkline}{\sparklineTikzBars{11.65, 12.04, 13.27, 10.63, 11.72, 8.95, 10.53, 8.93, 9.88, 13.50}}
\newcommand{\AllAllFileswithcommitsageyearsLimitedSparkline}{\sparklineTikzBars{27.35, 32.70, 29.48, 30.63, 29.12, 26.52, 27.81, 29.91, 27.34, 24.71}}
\newcommand{\AllAllFileswithcommitsageyearsConsistentSparkline}{\sparklineTikzBars{32.01, 33.42, 34.14, 32.24, 34.70, 35.56, 35.91, 37.59, 34.74, 30.98}}
\newcommand{\AllAllFileswithcommitsageyearsPervasiveSparkline}{\sparklineTikzBars{29.00, 21.83, 23.10, 26.50, 24.47, 28.97, 25.74, 23.57, 28.04, 30.80}}
\newcommand{\AllFilelevelcodelinesNoneSparkline}{\sparklineTikzBars{0.00, 0.00, 0.00, 0.00, 0.00, 0.00, 0.00, 0.00, 0.00, 0.00}}
\newcommand{\AllFilelevelcodelinesExperimentalSparkline}{\sparklineTikzBars{5.77, 5.67, 7.36, 9.75, 10.35, 11.85, 11.14, 13.13, 15.22, 16.72}}
\newcommand{\AllFilelevelcodelinesLimitedSparkline}{\sparklineTikzBars{21.19, 21.89, 24.08, 25.77, 27.16, 28.78, 29.75, 30.65, 34.83, 35.42}}
\newcommand{\AllFilelevelcodelinesConsistentSparkline}{\sparklineTikzBars{32.84, 36.32, 30.25, 35.42, 34.33, 33.96, 37.21, 37.01, 31.24, 32.14}}
\newcommand{\AllFilelevelcodelinesPervasiveSparkline}{\sparklineTikzBars{40.20, 36.12, 38.31, 29.05, 28.16, 25.40, 21.89, 19.20, 18.71, 15.72}}
\newcommand{\AllFilelevelcontributorsNoneSparkline}{\sparklineTikzBars{0.00, 0.00, 0.00, 0.00, 0.00, 0.00, 0.00, 0.00, 0.00, 0.00}}
\newcommand{\AllFilelevelcontributorsExperimentalSparkline}{\sparklineTikzBars{9.95, 11.16, 9.78, 10.39, 10.93, 9.74, 9.18, 9.70, 11.21, 15.04}}
\newcommand{\AllFilelevelcontributorsLimitedSparkline}{\sparklineTikzBars{19.81, 23.12, 22.11, 28.63, 27.22, 28.85, 26.79, 32.51, 33.03, 38.55}}
\newcommand{\AllFilelevelcontributorsConsistentSparkline}{\sparklineTikzBars{29.05, 27.94, 34.25, 31.96, 36.10, 33.30, 38.61, 38.91, 36.84, 35.16}}
\newcommand{\AllFilelevelcontributorsPervasiveSparkline}{\sparklineTikzBars{41.20, 37.79, 33.86, 29.02, 25.74, 28.11, 25.42, 18.89, 18.92, 11.25}}
\newcommand{\AllFilelevelageyearsNoneSparkline}{\sparklineTikzBars{0.00, 0.00, 0.00, 0.00, 0.00, 0.00, 0.00, 0.00, 0.00, 0.00}}
\newcommand{\AllFilelevelageyearsExperimentalSparkline}{\sparklineTikzBars{11.90, 12.03, 11.76, 9.88, 10.73, 8.54, 10.88, 8.17, 9.66, 13.54}}
\newcommand{\AllFilelevelageyearsLimitedSparkline}{\sparklineTikzBars{26.25, 31.43, 29.55, 30.23, 29.15, 25.23, 27.62, 30.37, 26.53, 23.69}}
\newcommand{\AllFilelevelageyearsConsistentSparkline}{\sparklineTikzBars{32.20, 32.99, 34.27, 32.44, 33.79, 36.21, 35.73, 37.13, 35.40, 30.67}}
\newcommand{\AllFilelevelageyearsPervasiveSparkline}{\sparklineTikzBars{29.66, 23.55, 24.42, 27.45, 26.33, 30.02, 25.77, 24.32, 28.40, 32.10}}
\newcommand{\AllFilesignoredonlycodelinesNoneSparkline}{\sparklineTikzBars{0.00, 0.00, 0.00, 0.00, 0.00, 0.00, 0.00, 0.00, 0.00, 0.00}}
\newcommand{\AllFilesignoredonlycodelinesExperimentalSparkline}{\sparklineTikzBars{6.77, 9.09, 9.85, 9.77, 12.12, 18.94, 12.78, 20.45, 18.94, 21.05}}
\newcommand{\AllFilesignoredonlycodelinesLimitedSparkline}{\sparklineTikzBars{26.32, 22.73, 34.09, 33.83, 34.09, 29.55, 26.32, 37.88, 43.18, 39.85}}
\newcommand{\AllFilesignoredonlycodelinesConsistentSparkline}{\sparklineTikzBars{39.10, 37.88, 38.64, 33.83, 36.36, 34.09, 42.11, 24.24, 25.00, 30.08}}
\newcommand{\AllFilesignoredonlycodelinesPervasiveSparkline}{\sparklineTikzBars{27.82, 30.30, 17.42, 22.56, 17.42, 17.42, 18.80, 17.42, 12.88, 9.02}}
\newcommand{\AllFilesignoredonlycontributorsNoneSparkline}{\sparklineTikzBars{0.00, 0.00, 0.00, 0.00, 0.00, 0.00, 0.00, 0.00, 0.00, 0.00}}
\newcommand{\AllFilesignoredonlycontributorsExperimentalSparkline}{\sparklineTikzBars{12.88, 7.97, 12.37, 8.90, 18.64, 16.67, 15.94, 8.59, 15.04, 23.66}}
\newcommand{\AllFilesignoredonlycontributorsLimitedSparkline}{\sparklineTikzBars{25.15, 37.68, 22.68, 33.56, 20.34, 30.30, 37.68, 35.94, 42.86, 38.93}}
\newcommand{\AllFilesignoredonlycontributorsConsistentSparkline}{\sparklineTikzBars{32.52, 31.16, 41.24, 31.51, 37.29, 34.09, 32.61, 41.41, 31.58, 31.30}}
\newcommand{\AllFilesignoredonlycontributorsPervasiveSparkline}{\sparklineTikzBars{29.45, 23.19, 23.71, 26.03, 23.73, 18.94, 13.77, 14.06, 10.53, 6.11}}
\newcommand{\AllFilesignoredonlyageyearsNoneSparkline}{\sparklineTikzBars{0.00, 0.00, 0.00, 0.00, 0.00, 0.00, 0.00, 0.00, 0.00, 0.00}}
\newcommand{\AllFilesignoredonlyageyearsExperimentalSparkline}{\sparklineTikzBars{9.68, 11.71, 18.94, 15.04, 14.08, 18.32, 15.08, 11.45, 9.77, 16.15}}
\newcommand{\AllFilesignoredonlyageyearsLimitedSparkline}{\sparklineTikzBars{31.61, 40.54, 39.39, 28.57, 30.28, 32.82, 30.95, 28.24, 36.84, 30.00}}
\newcommand{\AllFilesignoredonlyageyearsConsistentSparkline}{\sparklineTikzBars{31.61, 36.04, 31.06, 36.84, 38.03, 32.06, 35.71, 35.88, 29.32, 35.38}}
\newcommand{\AllFilesignoredonlyageyearsPervasiveSparkline}{\sparklineTikzBars{27.10, 11.71, 10.61, 19.55, 17.61, 16.79, 18.25, 24.43, 24.06, 18.46}}
\newcommand{\AllCommitsonlycodelinesNoneSparkline}{\sparklineTikzBars{0.00, 0.00, 0.00, 0.00, 0.00, 0.00, 0.00, 0.00, 0.00, 0.00}}
\newcommand{\AllCommitsonlycodelinesExperimentalSparkline}{\sparklineTikzBars{4.66, 6.39, 8.49, 8.31, 9.59, 11.70, 13.42, 16.62, 17.81, 24.84}}
\newcommand{\AllCommitsonlycodelinesLimitedSparkline}{\sparklineTikzBars{31.14, 30.87, 29.59, 32.69, 35.16, 35.74, 36.99, 37.17, 38.54, 37.63}}
\newcommand{\AllCommitsonlycodelinesConsistentSparkline}{\sparklineTikzBars{41.00, 39.09, 39.54, 38.90, 36.44, 36.75, 32.79, 32.15, 31.87, 26.48}}
\newcommand{\AllCommitsonlycodelinesPervasiveSparkline}{\sparklineTikzBars{23.20, 23.65, 22.37, 20.09, 18.81, 15.81, 16.80, 14.06, 11.78, 11.05}}
\newcommand{\AllCommitsonlycontributorsNoneSparkline}{\sparklineTikzBars{0.00, 0.00, 0.00, 0.00, 0.00, 0.00, 0.00, 0.00, 0.00, 0.00}}
\newcommand{\AllCommitsonlycontributorsExperimentalSparkline}{\sparklineTikzBars{8.65, 10.36, 9.49, 10.65, 12.30, 11.24, 11.46, 10.86, 13.79, 23.30}}
\newcommand{\AllCommitsonlycontributorsLimitedSparkline}{\sparklineTikzBars{30.37, 33.00, 32.03, 34.48, 33.79, 34.98, 33.64, 36.47, 36.76, 40.61}}
\newcommand{\AllCommitsonlycontributorsConsistentSparkline}{\sparklineTikzBars{32.49, 35.01, 37.32, 36.37, 36.72, 37.89, 37.49, 36.74, 36.21, 29.37}}
\newcommand{\AllCommitsonlycontributorsPervasiveSparkline}{\sparklineTikzBars{28.49, 21.63, 21.17, 18.50, 17.19, 15.89, 17.42, 15.93, 13.24, 6.72}}
\newcommand{\AllCommitsonlyageyearsNoneSparkline}{\sparklineTikzBars{0.00, 0.00, 0.00, 0.00, 0.00, 0.00, 0.00, 0.00, 0.00, 0.00}}
\newcommand{\AllCommitsonlyageyearsExperimentalSparkline}{\sparklineTikzBars{16.33, 14.35, 12.62, 12.72, 11.35, 9.67, 10.04, 8.70, 11.06, 14.89}}
\newcommand{\AllCommitsonlyageyearsLimitedSparkline}{\sparklineTikzBars{38.86, 34.80, 34.35, 33.68, 34.48, 35.17, 31.36, 34.36, 37.14, 30.90}}
\newcommand{\AllCommitsonlyageyearsConsistentSparkline}{\sparklineTikzBars{30.26, 34.44, 35.17, 37.70, 36.59, 35.73, 37.67, 38.89, 33.09, 35.77}}
\newcommand{\AllCommitsonlyageyearsPervasiveSparkline}{\sparklineTikzBars{14.55, 16.41, 17.86, 15.90, 17.59, 19.43, 20.94, 18.04, 18.71, 18.45}}
\newcommand{\AllAllFilesNoneValue}{23.54\%}
\newcommand{\AllAllFilesExperimentalValue}{8.47\%}
\newcommand{\AllAllFilesLimitedValue}{21.80\%}
\newcommand{\AllAllFilesConsistentValue}{26.06\%}
\newcommand{\AllAllFilesPervasiveValue}{20.13\%}
\newcommand{\AllAllFileswithcommitsNoneValue}{0.00\%}
\newcommand{\AllAllFileswithcommitsExperimentalValue}{11.08\%}
\newcommand{\AllAllFileswithcommitsLimitedValue}{28.51\%}
\newcommand{\AllAllFileswithcommitsConsistentValue}{34.08\%}
\newcommand{\AllAllFileswithcommitsPervasiveValue}{26.33\%}
\newcommand{\AllFilelevelNoneValue}{0.00\%}
\newcommand{\AllFilelevelExperimentalValue}{10.70\%}
\newcommand{\AllFilelevelLimitedValue}{27.95\%}
\newcommand{\AllFilelevelConsistentValue}{34.07\%}
\newcommand{\AllFilelevelPervasiveValue}{27.28\%}
\newcommand{\AllFilesignoredonlyNoneValue}{0.00\%}
\newcommand{\AllFilesignoredonlyExperimentalValue}{13.97\%}
\newcommand{\AllFilesignoredonlyLimitedValue}{32.78\%}
\newcommand{\AllFilesignoredonlyConsistentValue}{34.14\%}
\newcommand{\AllFilesignoredonlyPervasiveValue}{19.11\%}
\newcommand{\AllCommitsonlyNoneValue}{0.00\%}
\newcommand{\AllCommitsonlyExperimentalValue}{12.18\%}
\newcommand{\AllCommitsonlyLimitedValue}{34.55\%}
\newcommand{\AllCommitsonlyConsistentValue}{35.50\%}
\newcommand{\AllCommitsonlyPervasiveValue}{17.76\%}

\newcommand{\AllbinfiletotalfilesSparkline}{\sparklineTikzBars{98.18, 66.90, 16.59, 9.55, 8.77}}
\newcommand{\AllbinfiletotallinesSparkline}{\sparklineTikzBars{10.14, 26.52, 30.50, 57.85, 31.23, 18.87, 24.89}}
\newcommand{\AllbinfiletotalchangesSparkline}{\sparklineTikzBars{41.39, 71.14, 33.51, 24.34, 29.63}}
\newcommand{\AllbinfileadoptionratioSparkline}{\sparklineTikzBars{63.01, 13.81, 35.84, 45.22, 42.12}}
\newcommand{\AllbinfiletotalfileszeroValue}{49.1\%}
\newcommand{\AllbinfiletotalfilesoneValue}{33.5\%}
\newcommand{\AllbinfiletotalfilestwoValue}{8.3\%}
\newcommand{\AllbinfiletotalfilesthreeValue}{4.8\%}
\newcommand{\AllbinfiletotalfilesfourValue}{4.4\%}
\newcommand{\AllbinfiletotallineszeroValue}{5.1\%}
\newcommand{\AllbinfiletotallinesoneValue}{13.3\%}
\newcommand{\AllbinfiletotallinestwoValue}{15.3\%}
\newcommand{\AllbinfiletotallinesthreeValue}{28.9\%}
\newcommand{\AllbinfiletotallinesfourValue}{15.6\%}
\newcommand{\AllbinfiletotallinesfiveValue}{9.4\%}
\newcommand{\AllbinfiletotallinessixValue}{12.4\%}
\newcommand{\AllbinfiletotalchangeszeroValue}{20.7\%}
\newcommand{\AllbinfiletotalchangesoneValue}{35.6\%}
\newcommand{\AllbinfiletotalchangestwoValue}{16.8\%}
\newcommand{\AllbinfiletotalchangesthreeValue}{12.2\%}
\newcommand{\AllbinfiletotalchangesfourValue}{14.8\%}
\newcommand{\AllbinfileadoptionratiozeroValue}{31.5\%}
\newcommand{\AllbinfileadoptionratiooneValue}{6.9\%}
\newcommand{\AllbinfileadoptionratiotwoValue}{17.9\%}
\newcommand{\AllbinfileadoptionratiothreeValue}{22.6\%}
\newcommand{\AllbinfileadoptionratiofourValue}{21.1\%}

\newcommand{\AllcodelinestooluseSparkline}{\sparklineTikzBars{29.69, 35.39, 39.84, 42.63, 49.54, 53.94, 61.39, 64.81, 68.56, 68.73}}
\newcommand{\AllcontributorstooluseSparkline}{\sparklineTikzBars{38.86, 47.26, 48.29, 46.58, 46.72, 49.38, 50.96, 53.52, 56.81, 78.18}}
\newcommand{\AllcommitstooluseSparkline}{\sparklineTikzBars{40.69, 41.02, 41.65, 42.62, 45.93, 47.56, 49.35, 57.41, 65.67, 82.66}}
\newcommand{\AlltotalissuestooluseSparkline}{\sparklineTikzBars{40.20, 43.26, 45.00, 43.67, 46.90, 51.14, 54.32, 61.36, 88.86}}
\newcommand{\AlltotalpullrequeststooluseSparkline}{\sparklineTikzBars{26.98, 32.68, 37.24, 42.72, 44.58, 47.62, 52.40, 57.32, 66.01, 107.38}}
\newcommand{\AllageyearstooluseSparkline}{\sparklineTikzBars{101.60, 60.93, 55.70, 47.11, 44.94, 44.40, 43.49, 39.16, 38.09, 37.69}}
\newcommand{\AllcodelinestooluseDecilezeroBoundary}{5.0K}
\newcommand{\AllcodelinestooluseDecileoneBoundary}{7.5K}
\newcommand{\AllcodelinestooluseDeciletwoBoundary}{11K}
\newcommand{\AllcodelinestooluseDecilethreeBoundary}{15K}
\newcommand{\AllcodelinestooluseDecilefourBoundary}{21K}
\newcommand{\AllcodelinestooluseDecilefiveBoundary}{30K}
\newcommand{\AllcodelinestooluseDecilesixBoundary}{44K}
\newcommand{\AllcodelinestooluseDecilesevenBoundary}{70K}
\newcommand{\AllcodelinestooluseDecileeightBoundary}{128K}
\newcommand{\AllcodelinestooluseDecilenineBoundary}{320K}
\newcommand{\AllcodelinestooluseDecilezero}{8.48\%}
\newcommand{\AllcodelinestooluseDecileone}{10.11\%}
\newcommand{\AllcodelinestooluseDeciletwo}{11.38\%}
\newcommand{\AllcodelinestooluseDecilethree}{12.18\%}
\newcommand{\AllcodelinestooluseDecilefour}{14.15\%}
\newcommand{\AllcodelinestooluseDecilefive}{15.41\%}
\newcommand{\AllcodelinestooluseDecilesix}{17.54\%}
\newcommand{\AllcodelinestooluseDecileseven}{18.52\%}
\newcommand{\AllcodelinestooluseDecileeight}{19.59\%}
\newcommand{\AllcodelinestooluseDecilenine}{19.64\%}
\newcommand{\AllcontributorstooluseDecilezeroBoundary}{0.0}
\newcommand{\AllcontributorstooluseDecileoneBoundary}{1.0}
\newcommand{\AllcontributorstooluseDeciletwoBoundary}{3.0}
\newcommand{\AllcontributorstooluseDecilethreeBoundary}{4.0}
\newcommand{\AllcontributorstooluseDecilefourBoundary}{6.0}
\newcommand{\AllcontributorstooluseDecilefiveBoundary}{10}
\newcommand{\AllcontributorstooluseDecilesixBoundary}{14}
\newcommand{\contributorstooluseDecilesevenBoundary}{21}
\newcommand{\contributorstooluseDecileeightBoundary}{34}
\newcommand{\contributorstooluseDecilenineBoundary}{68}
\newcommand{\AllcontributorstooluseDecilezero}{11.10\%}
\newcommand{\AllcontributorstooluseDecileone}{13.50\%}
\newcommand{\AllcontributorstooluseDeciletwo}{13.80\%}
\newcommand{\AllcontributorstooluseDecilethree}{13.31\%}
\newcommand{\AllcontributorstooluseDecilefour}{13.35\%}
\newcommand{\AllcontributorstooluseDecilefive}{14.11\%}
\newcommand{\AllcontributorstooluseDecilesix}{14.56\%}
\newcommand{\contributorstooluseDecileseven}{15.29\%}
\newcommand{\contributorstooluseDecileeight}{16.23\%}
\newcommand{\contributorstooluseDecilenine}{22.34\%}
\newcommand{\AllcommitstooluseDecilezeroBoundary}{100}
\newcommand{\AllcommitstooluseDecileoneBoundary}{159}
\newcommand{\AllcommitstooluseDeciletwoBoundary}{233}
\newcommand{\AllcommitstooluseDecilethreeBoundary}{324}
\newcommand{\AllcommitstooluseDecilefourBoundary}{446}
\newcommand{\AllcommitstooluseDecilefiveBoundary}{610}
\newcommand{\AllcommitstooluseDecilesixBoundary}{849}
\newcommand{\AllcommitstooluseDecilesevenBoundary}{1.2K}
\newcommand{\AllcommitstooluseDecileeightBoundary}{1.9K}
\newcommand{\AllcommitstooluseDecilenineBoundary}{3.9K}
\newcommand{\AllcommitstooluseDecilezero}{11.62\%}
\newcommand{\AllcommitstooluseDecileone}{11.72\%}
\newcommand{\AllcommitstooluseDeciletwo}{11.90\%}
\newcommand{\AllcommitstooluseDecilethree}{12.18\%}
\newcommand{\AllcommitstooluseDecilefour}{13.12\%}
\newcommand{\AllcommitstooluseDecilefive}{13.59\%}
\newcommand{\AllcommitstooluseDecilesix}{14.10\%}
\newcommand{\AllcommitstooluseDecileseven}{16.40\%}
\newcommand{\AllcommitstooluseDecileeight}{18.76\%}
\newcommand{\AllcommitstooluseDecilenine}{23.62\%}
\newcommand{\AlltotalissuestooluseDecilezeroBoundary}{0.0}
\newcommand{\AlltotalissuestooluseDecileoneBoundary}{3.0}
\newcommand{\AlltotalissuestooluseDeciletwoBoundary}{9.0}
\newcommand{\AlltotalissuestooluseDecilethreeBoundary}{19}
\newcommand{\AlltotalissuestooluseDecilefourBoundary}{35}
\newcommand{\AlltotalissuestooluseDecilefiveBoundary}{60}
\newcommand{\AlltotalissuestooluseDecilesixBoundary}{105}
\newcommand{\AlltotalissuestooluseDecilesevenBoundary}{195}
\newcommand{\totalissuestooluseDecileeightBoundary}{458}
\newcommand{\AlltotalissuestooluseDecilezero}{11.49\%}
\newcommand{\AlltotalissuestooluseDecileone}{12.36\%}
\newcommand{\AlltotalissuestooluseDeciletwo}{12.86\%}
\newcommand{\AlltotalissuestooluseDecilethree}{12.48\%}
\newcommand{\AlltotalissuestooluseDecilefour}{13.40\%}
\newcommand{\AlltotalissuestooluseDecilefive}{14.61\%}
\newcommand{\AlltotalissuestooluseDecilesix}{15.52\%}
\newcommand{\AlltotalissuestooluseDecileseven}{17.53\%}
\newcommand{\totalissuestooluseDecileeight}{25.39\%}
\newcommand{\AlltotalpullrequeststooluseDecilezeroBoundary}{0.0}
\newcommand{\AlltotalpullrequeststooluseDecileoneBoundary}{1.0}
\newcommand{\AlltotalpullrequeststooluseDeciletwoBoundary}{8.0}
\newcommand{\AlltotalpullrequeststooluseDecilethreeBoundary}{22}
\newcommand{\AlltotalpullrequeststooluseDecilefourBoundary}{46}
\newcommand{\AlltotalpullrequeststooluseDecilefiveBoundary}{85}
\newcommand{\AlltotalpullrequeststooluseDecilesixBoundary}{145}
\newcommand{\AlltotalpullrequeststooluseDecilesevenBoundary}{244}
\newcommand{\AlltotalpullrequeststooluseDecileeightBoundary}{434}
\newcommand{\totalpullrequeststooluseDecilenineBoundary}{930}
\newcommand{\AlltotalpullrequeststooluseDecilezero}{7.71\%}
\newcommand{\AlltotalpullrequeststooluseDecileone}{9.34\%}
\newcommand{\AlltotalpullrequeststooluseDeciletwo}{10.64\%}
\newcommand{\AlltotalpullrequeststooluseDecilethree}{12.20\%}
\newcommand{\AlltotalpullrequeststooluseDecilefour}{12.74\%}
\newcommand{\AlltotalpullrequeststooluseDecilefive}{13.60\%}
\newcommand{\AlltotalpullrequeststooluseDecilesix}{14.97\%}
\newcommand{\AlltotalpullrequeststooluseDecileseven}{16.38\%}
\newcommand{\AlltotalpullrequeststooluseDecileeight}{18.86\%}
\newcommand{\totalpullrequeststooluseDecilenine}{30.68\%}
\newcommand{\AllageyearstooluseDecilezeroBoundary}{0.0}
\newcommand{\AllageyearstooluseDecileoneBoundary}{1.1}
\newcommand{\AllageyearstooluseDeciletwoBoundary}{2.0}
\newcommand{\AllageyearstooluseDecilethreeBoundary}{2.9}
\newcommand{\AllageyearstooluseDecilefourBoundary}{3.8}
\newcommand{\AllageyearstooluseDecilefiveBoundary}{4.8}
\newcommand{\ageyearstooluseDecilesixBoundary}{5.8}
\newcommand{\ageyearstooluseDecilesevenBoundary}{7.1}
\newcommand{\ageyearstooluseDecileeightBoundary}{8.7}
\newcommand{\ageyearstooluseDecilenineBoundary}{11}
\newcommand{\AllageyearstooluseDecilezero}{29.03\%}
\newcommand{\AllageyearstooluseDecileone}{17.41\%}
\newcommand{\AllageyearstooluseDeciletwo}{15.91\%}
\newcommand{\AllageyearstooluseDecilethree}{13.46\%}
\newcommand{\AllageyearstooluseDecilefour}{12.84\%}
\newcommand{\AllageyearstooluseDecilefive}{12.69\%}
\newcommand{\ageyearstooluseDecilesix}{12.42\%}
\newcommand{\ageyearstooluseDecileseven}{11.19\%}
\newcommand{\ageyearstooluseDecileeight}{10.88\%}
\newcommand{\ageyearstooluseDecilenine}{10.77\%}

We analyze the distribution of the various adoption metrics 
to understand the magnitude of adoption changes. We analyze adoption metrics characterizing both file-level adoption and commit-level adoption.

\subsection{Distribution of adoption metrics}

\begin{table}[ht]
\scriptsize
\centering
\caption{Binned distributions of file and commit level adoption metrics}
\label{tab:binned_adoption_metrics}
\begin{tabular}{l|l|rrrrrrr}
\hline
\toprule
\textbf{Metric} & \textbf{Distribution} & \textbf{Bin 1} & \textbf{Bin 2} & \textbf{Bin 3} & \textbf{Bin 4} & \textbf{Bin 5} & \textbf{Bin 6} & \textbf{Bin 7} \\
\midrule
Total Files &  & 1 & 2-5 & 6-10 & 11-20 & 21+ & & \\
Newer projects & \quad \VerybinfiletotalfilesSparkline & \VerybinfiletotalfileszeroValue & \VerybinfiletotalfilesoneValue & \VerybinfiletotalfilestwoValue & \VerybinfiletotalfilesthreeValue & \VerybinfiletotalfilesfourValue & & \\
\rowcolor{lightgray!50} Older projects & \quad \AllbinfiletotalfilesSparkline & \AllbinfiletotalfileszeroValue & \AllbinfiletotalfilesoneValue & \AllbinfiletotalfilestwoValue & \AllbinfiletotalfilesthreeValue & \AllbinfiletotalfilesfourValue & & \\
\midrule
Total Lines & & 0-10 & 11-50 & 51-100 & 101-250 & 251-500 & 501-1000 & 1001+ \\
Newer projects & \quad \VerybinfiletotallinesSparkline & \VerybinfiletotallineszeroValue & \VerybinfiletotallinesoneValue & \VerybinfiletotallinestwoValue & \VerybinfiletotallinesthreeValue & \VerybinfiletotallinesfourValue & \VerybinfiletotallinesfiveValue & \VerybinfiletotallinessixValue \\
\rowcolor{lightgray!50} Older projects & \quad \AllbinfiletotallinesSparkline & \AllbinfiletotallineszeroValue & \AllbinfiletotallinesoneValue & \AllbinfiletotallinestwoValue & \AllbinfiletotallinesthreeValue & \AllbinfiletotallinesfourValue & \AllbinfiletotallinesfiveValue & \AllbinfiletotallinessixValue \\
\midrule
Total Changes & & 1 & 2-5 & 6-10 & 11-20 & 21+ & & \\
Newer projects & \quad \VerybinfiletotalchangesSparkline & \VerybinfiletotalchangeszeroValue & \VerybinfiletotalchangesoneValue & \VerybinfiletotalchangestwoValue & \VerybinfiletotalchangesthreeValue & \VerybinfiletotalchangesfourValue & & \\
\rowcolor{lightgray!50} Older projects & \quad \AllbinfiletotalchangesSparkline & \AllbinfiletotalchangeszeroValue & \AllbinfiletotalchangesoneValue & \AllbinfiletotalchangestwoValue & \AllbinfiletotalchangesthreeValue & \AllbinfiletotalchangesfourValue & & \\
\midrule
Adoption Ratio & & None & Experimental & Limited & Consistent & Pervasive & & \\
Newer projects & \quad \VerybinfileadoptionratioSparkline & \VerybinfileadoptionratiozeroValue & \VerybinfileadoptionratiooneValue & \VerybinfileadoptionratiotwoValue & \VerybinfileadoptionratiothreeValue & \VerybinfileadoptionratiofourValue & & \\
\rowcolor{lightgray!50} Older projects & \quad \AllbinfileadoptionratioSparkline & \AllbinfileadoptionratiozeroValue & \AllbinfileadoptionratiooneValue & \AllbinfileadoptionratiotwoValue & \AllbinfileadoptionratiothreeValue & \AllbinfileadoptionratiofourValue & & \\
\bottomrule
\end{tabular}
\end{table}

\Cref{tab:binned_adoption_metrics} shows the distribution of the adoption metrics we measure for each type of projects, using the bins we defined in the analysis. For comparison, the rows representing the metrics of older projects have a gray background. For the commit ratio, we define the bins as follow:

\begin{itemize}
    \item ``None'': there are no AI-assisted commits that we could identify.
    \item ``Experimental'': there are less than 1\% of AI-assisted commits we could identify.
    \item ``Limited'': the AI-assisted commit ratio is between 1 and 5\%.
    \item ``Consistent'': the AI-assisted commit ratio is between 5 and 20\%.
    \item ``Pervasive'': the AI-assisted commit ratio is more than 20\%.
\end{itemize}

Based on the data, we can make the following observations:
\begin{itemize}
    \item In terms of number of files, we see a marked tendency for the newer projects to have more files than the older: the smallest bin (just one file) is still the largest one, but is smaller in the newer projects. Conversely, the largest bins are higher. In particular, the proportion of projects with more than 20 agent configuration or guidance files is more than twice as large in newer projects.
    \item For total lines, we see a similar behavior. The first four bins (lines of code less than 250 lines) are lower in newer projects. On the other hand, the categories with more than 250 lines are higher, with the highest (1000+ lines) showing the largest difference: it is almost twice as large for newer projects; indeed, the largest bin is also the second most common.
    \item In terms of changes to agent configuration files, we see an even bigger difference. The largest bin for the older projects is the one grouping 2 to 5 changes. On the other hand, for the newer projects, the largest bin is the one with the most extreme values, covering projects with more than 20 commits changing agent configuration files. This bin is more than twice as common for newest projects, and covers more than 1/3rd of the newest projects. 
    \item Finally, the AI-assisted commit ratio follows a similar trend. The lowest bin (``None'') is smaller in newest projects, while the highest bin (``Pervasive'') is almost twice as more common in newer projects, and is by far the most common in the newest projects. This bin is covers more than 40\% of all newest projects. 
\end{itemize}

\begin{figure}[htbp]
    \centering
    \includegraphics[width=0.95\textwidth]{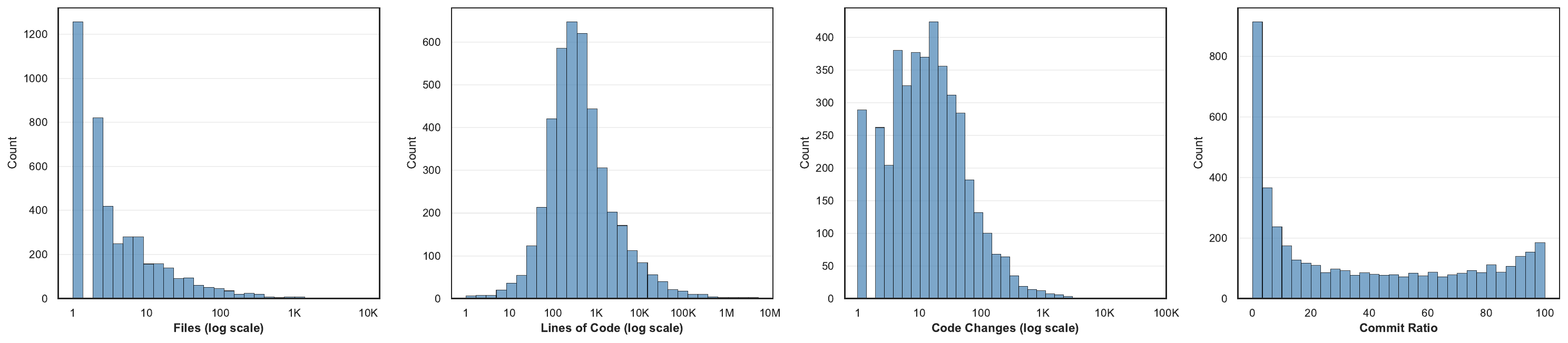}
    \caption{Distribution of adoption metrics for newer projects}
    \label{fig:metrics-distribution-very}

    \vspace{1cm} 

    \includegraphics[width=0.95\textwidth]{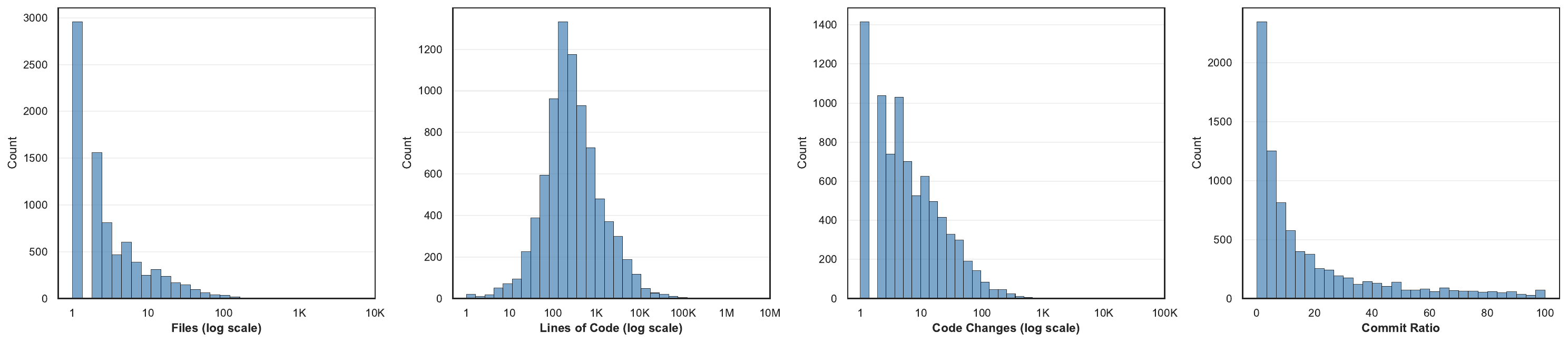}
    \caption{Distribution of adoption metrics for older projects}
    \label{fig:metrics-distribution-all}
\end{figure}

To get a more precise view of the metrics, \Cref{fig:metrics-distribution-very} and \Cref{fig:metrics-distribution-all} show the distribution of the metrics of both types of project, without pre-defined bins. The Y axes differ, but the X axes are aligned. The graphs show a similar trend as the bin-level analysis, but for code changes and commit ratio, there is some interesting nuances.

\begin{itemize}
    \item The shape of the curves is very different for the number of changes, where we clearly see that the relative proportion of projects with higher number of changes is clearly larger in the newest projects.  
    \item for commit ratio, we see a completely different behavior in the distribution of the ``pervasive'' category. ``Pervasive'' adopters are both more common, and also show significantly more extreme values. For older projects, there is a tendency for the curve to decrease: large commit ratios are very uncommon. For newer projects, the curve is flat, and even rises at the end (projects with more than 80\% of AI assisted commits). The proportion of extreme adopters of coding agents is thus much higher in newer projects.
    \item In both cases (older and newer projects), there is a high proportion of projects with a commit ratio of 0, raising questions as to whether these projects have no AI activity, or rather hidden AI activity.
\end{itemize}

\subsection{File-level adoption and project metrics}

\Cref{tab:file_adoption} shows the distribution of file-level adoption over project metrics, for both newer projects and older projects. The rows of older projects have a gray background. In each case, we split the distribution in deciles of the metric of interest, and we compute the adoption for the specific decile. We focus on two aspects: the evolution of the adoption over deciles, and the differences in decile values between older and newer projects. Particularly, we expect older projects to be larger in all metrics, and we discuss the differences with this in mind. We do not compare adoption rates between both kinds of projects, as we did that globally in the previous section. 

\begin{table}[ht]
\centering
\caption{File adoption statistics versus project-level metrics, by deciles}
\label{tab:file_adoption}
\tiny
\begin{tabular}{l|l|l|cccccccccc}
\hline
\toprule\textbf{Metric} & \textbf{Aspect} &  \textbf{Sparkline} & \textbf{D1} & \textbf{D2} & \textbf{D3} & \textbf{D4} & \textbf{D5} & \textbf{D6} & \textbf{D7} & \textbf{D8} & \textbf{D9} & \textbf{D10} \\
\midrule
\midrule
\multirow{4}{*}{LOC} & Deciles &  \multirow{2}{*}{} & \VerycodelinestooluseDecilezeroBoundary & \VerycodelinestooluseDecileoneBoundary & \VerycodelinestooluseDeciletwoBoundary & \VerycodelinestooluseDecilethreeBoundary & \VerycodelinestooluseDecilefourBoundary & \VerycodelinestooluseDecilefiveBoundary & \VerycodelinestooluseDecilesixBoundary & \VerycodelinestooluseDecilesevenBoundary & \VerycodelinestooluseDecileeightBoundary & \VerycodelinestooluseDecilenineBoundary \\
  & File &  \VerycodelinestooluseSparkline & \VerycodelinestooluseDecilezero & \VerycodelinestooluseDecileone & \VerycodelinestooluseDeciletwo & \VerycodelinestooluseDecilethree & \VerycodelinestooluseDecilefour & \VerycodelinestooluseDecilefive & \VerycodelinestooluseDecilesix & \VerycodelinestooluseDecileseven & \VerycodelinestooluseDecileeight & \VerycodelinestooluseDecilenine \\
\rowcolor{lightgray!50} & Deciles &  \multirow{2}{*}{} & \AllcodelinestooluseDecilezeroBoundary & \AllcodelinestooluseDecileoneBoundary & \AllcodelinestooluseDeciletwoBoundary & \AllcodelinestooluseDecilethreeBoundary & \AllcodelinestooluseDecilefourBoundary & \AllcodelinestooluseDecilefiveBoundary & \AllcodelinestooluseDecilesixBoundary & \AllcodelinestooluseDecilesevenBoundary & \AllcodelinestooluseDecileeightBoundary & \AllcodelinestooluseDecilenineBoundary \\
\rowcolor{lightgray!50}  & File &  \AllcodelinestooluseSparkline & \AllcodelinestooluseDecilezero & \AllcodelinestooluseDecileone & \AllcodelinestooluseDeciletwo & \AllcodelinestooluseDecilethree & \AllcodelinestooluseDecilefour & \AllcodelinestooluseDecilefive & \AllcodelinestooluseDecilesix & \AllcodelinestooluseDecileseven & \AllcodelinestooluseDecileeight & \AllcodelinestooluseDecilenine \\
 \midrule
\multirow{4}{*}{Contributors} & Deciles &  \multirow{2}{*}{} & \VerycontributorstooluseDecilezeroBoundary & \VerycontributorstooluseDecileoneBoundary & \VerycontributorstooluseDeciletwoBoundary & \VerycontributorstooluseDecilethreeBoundary & \VerycontributorstooluseDecilefourBoundary & \VerycontributorstooluseDecilefiveBoundary & \VerycontributorstooluseDecilesixBoundary &  &  &  \\
  & File &  \VerycontributorstooluseSparkline & \VerycontributorstooluseDecilezero & \VerycontributorstooluseDecileone & \VerycontributorstooluseDeciletwo & \VerycontributorstooluseDecilethree & \VerycontributorstooluseDecilefour & \VerycontributorstooluseDecilefive & \VerycontributorstooluseDecilesix &  &  &  \\
\rowcolor{lightgray!50} & Deciles &  \multirow{2}{*}{} & \AllcontributorstooluseDecilezeroBoundary & \AllcontributorstooluseDecileoneBoundary & \AllcontributorstooluseDeciletwoBoundary & \AllcontributorstooluseDecilethreeBoundary & \AllcontributorstooluseDecilefourBoundary & \AllcontributorstooluseDecilefiveBoundary & \AllcontributorstooluseDecilesixBoundary & \contributorstooluseDecilesevenBoundary & \contributorstooluseDecileeightBoundary & \contributorstooluseDecilenineBoundary \\
\rowcolor{lightgray!50}  & File &  \AllcontributorstooluseSparkline & \AllcontributorstooluseDecilezero & \AllcontributorstooluseDecileone & \AllcontributorstooluseDeciletwo & \AllcontributorstooluseDecilethree & \AllcontributorstooluseDecilefour & \AllcontributorstooluseDecilefive & \AllcontributorstooluseDecilesix & \contributorstooluseDecileseven & \contributorstooluseDecileeight & \contributorstooluseDecilenine \\
 \midrule
\multirow{4}{*}{Commits} & Deciles &  \multirow{2}{*}{} & \VerycommitstooluseDecilezeroBoundary & \VerycommitstooluseDecileoneBoundary & \VerycommitstooluseDeciletwoBoundary & \VerycommitstooluseDecilethreeBoundary & \VerycommitstooluseDecilefourBoundary & \VerycommitstooluseDecilefiveBoundary & \VerycommitstooluseDecilesixBoundary & \VerycommitstooluseDecilesevenBoundary & \VerycommitstooluseDecileeightBoundary & \VerycommitstooluseDecilenineBoundary \\
  & File &  \VerycommitstooluseSparkline & \VerycommitstooluseDecilezero & \VerycommitstooluseDecileone & \VerycommitstooluseDeciletwo & \VerycommitstooluseDecilethree & \VerycommitstooluseDecilefour & \VerycommitstooluseDecilefive & \VerycommitstooluseDecilesix & \VerycommitstooluseDecileseven & \VerycommitstooluseDecileeight & \VerycommitstooluseDecilenine \\
\rowcolor{lightgray!50} & Deciles &  \multirow{2}{*}{} & \AllcommitstooluseDecilezeroBoundary & \AllcommitstooluseDecileoneBoundary & \AllcommitstooluseDeciletwoBoundary & \AllcommitstooluseDecilethreeBoundary & \AllcommitstooluseDecilefourBoundary & \AllcommitstooluseDecilefiveBoundary & \AllcommitstooluseDecilesixBoundary & \AllcommitstooluseDecilesevenBoundary & \AllcommitstooluseDecileeightBoundary & \AllcommitstooluseDecilenineBoundary \\
\rowcolor{lightgray!50}  & File &  \AllcommitstooluseSparkline & \AllcommitstooluseDecilezero & \AllcommitstooluseDecileone & \AllcommitstooluseDeciletwo & \AllcommitstooluseDecilethree & \AllcommitstooluseDecilefour & \AllcommitstooluseDecilefive & \AllcommitstooluseDecilesix & \AllcommitstooluseDecileseven & \AllcommitstooluseDecileeight & \AllcommitstooluseDecilenine \\
 \midrule
\multirow{4}{*}{Issues} & Deciles &  \multirow{2}{*}{} & \VerytotalissuestooluseDecilezeroBoundary & \VerytotalissuestooluseDecileoneBoundary & \VerytotalissuestooluseDeciletwoBoundary & \VerytotalissuestooluseDecilethreeBoundary & \VerytotalissuestooluseDecilefourBoundary & \VerytotalissuestooluseDecilefiveBoundary & \VerytotalissuestooluseDecilesixBoundary & \VerytotalissuestooluseDecilesevenBoundary &  &  \\
  & File &  \VerytotalissuestooluseSparkline & \VerytotalissuestooluseDecilezero & \VerytotalissuestooluseDecileone & \VerytotalissuestooluseDeciletwo & \VerytotalissuestooluseDecilethree & \VerytotalissuestooluseDecilefour & \VerytotalissuestooluseDecilefive & \VerytotalissuestooluseDecilesix & \VerytotalissuestooluseDecileseven &  &  \\
\rowcolor{lightgray!50} & Deciles &  \multirow{2}{*}{} & \AlltotalissuestooluseDecilezeroBoundary & \AlltotalissuestooluseDecileoneBoundary & \AlltotalissuestooluseDeciletwoBoundary & \AlltotalissuestooluseDecilethreeBoundary & \AlltotalissuestooluseDecilefourBoundary & \AlltotalissuestooluseDecilefiveBoundary & \AlltotalissuestooluseDecilesixBoundary & \AlltotalissuestooluseDecilesevenBoundary & \totalissuestooluseDecileeightBoundary &  \\
\rowcolor{lightgray!50}  & File &  \AlltotalissuestooluseSparkline & \AlltotalissuestooluseDecilezero & \AlltotalissuestooluseDecileone & \AlltotalissuestooluseDeciletwo & \AlltotalissuestooluseDecilethree & \AlltotalissuestooluseDecilefour & \AlltotalissuestooluseDecilefive & \AlltotalissuestooluseDecilesix & \AlltotalissuestooluseDecileseven & \totalissuestooluseDecileeight &  \\
 \midrule
\multirow{4}{*}{Pull Request} & Deciles &  \multirow{2}{*}{} & \VerytotalpullrequeststooluseDecilezeroBoundary & \VerytotalpullrequeststooluseDecileoneBoundary & \VerytotalpullrequeststooluseDeciletwoBoundary & \VerytotalpullrequeststooluseDecilethreeBoundary & \VerytotalpullrequeststooluseDecilefourBoundary & \VerytotalpullrequeststooluseDecilefiveBoundary & \VerytotalpullrequeststooluseDecilesixBoundary & \VerytotalpullrequeststooluseDecilesevenBoundary & \VerytotalpullrequeststooluseDecileeightBoundary &  \\
  & File &  \VerytotalpullrequeststooluseSparkline & \VerytotalpullrequeststooluseDecilezero & \VerytotalpullrequeststooluseDecileone & \VerytotalpullrequeststooluseDeciletwo & \VerytotalpullrequeststooluseDecilethree & \VerytotalpullrequeststooluseDecilefour & \VerytotalpullrequeststooluseDecilefive & \VerytotalpullrequeststooluseDecilesix & \VerytotalpullrequeststooluseDecileseven & \VerytotalpullrequeststooluseDecileeight &  \\
\rowcolor{lightgray!50} & Deciles &  \multirow{2}{*}{} & \AlltotalpullrequeststooluseDecilezeroBoundary & \AlltotalpullrequeststooluseDecileoneBoundary & \AlltotalpullrequeststooluseDeciletwoBoundary & \AlltotalpullrequeststooluseDecilethreeBoundary & \AlltotalpullrequeststooluseDecilefourBoundary & \AlltotalpullrequeststooluseDecilefiveBoundary & \AlltotalpullrequeststooluseDecilesixBoundary & \AlltotalpullrequeststooluseDecilesevenBoundary & \AlltotalpullrequeststooluseDecileeightBoundary & \totalpullrequeststooluseDecilenineBoundary \\
\rowcolor{lightgray!50}  & File &  \AlltotalpullrequeststooluseSparkline & \AlltotalpullrequeststooluseDecilezero & \AlltotalpullrequeststooluseDecileone & \AlltotalpullrequeststooluseDeciletwo & \AlltotalpullrequeststooluseDecilethree & \AlltotalpullrequeststooluseDecilefour & \AlltotalpullrequeststooluseDecilefive & \AlltotalpullrequeststooluseDecilesix & \AlltotalpullrequeststooluseDecileseven & \AlltotalpullrequeststooluseDecileeight & \totalpullrequeststooluseDecilenine \\
 \midrule
\multirow{4}{*}{Age (years)} & Deciles &  \multirow{2}{*}{} & \VeryageyearstooluseDecilezeroBoundary & \VeryageyearstooluseDecileoneBoundary & \VeryageyearstooluseDeciletwoBoundary & \VeryageyearstooluseDecilethreeBoundary & \VeryageyearstooluseDecilefourBoundary & \VeryageyearstooluseDecilefiveBoundary &  &  &  &  \\
  & File &  \VeryageyearstooluseSparkline & \VeryageyearstooluseDecilezero & \VeryageyearstooluseDecileone & \VeryageyearstooluseDeciletwo & \VeryageyearstooluseDecilethree & \VeryageyearstooluseDecilefour & \VeryageyearstooluseDecilefive &  &  &  &  \\
\rowcolor{lightgray!50} & Deciles &  \multirow{2}{*}{} & \AllageyearstooluseDecilezeroBoundary & \AllageyearstooluseDecileoneBoundary & \AllageyearstooluseDeciletwoBoundary & \AllageyearstooluseDecilethreeBoundary & \AllageyearstooluseDecilefourBoundary & \AllageyearstooluseDecilefiveBoundary & \ageyearstooluseDecilesixBoundary & \ageyearstooluseDecilesevenBoundary & \ageyearstooluseDecileeightBoundary & \ageyearstooluseDecilenineBoundary \\
\rowcolor{lightgray!50}  & File &  \AllageyearstooluseSparkline & \AllageyearstooluseDecilezero & \AllageyearstooluseDecileone & \AllageyearstooluseDeciletwo & \AllageyearstooluseDecilethree & \AllageyearstooluseDecilefour & \AllageyearstooluseDecilefive & \ageyearstooluseDecilesix & \ageyearstooluseDecileseven & \ageyearstooluseDecileeight & \ageyearstooluseDecilenine \\
 \bottomrule
\end{tabular}
\end{table}

\paragraph{Lines of code}

For lines of code, looking at the deciles, we see that newer projects are paradoxically \emph{larger} in the first deciles (deciles two to eight); for instance, the boundary of decile 8 is 74,000 lines of code for newer projects, while it is only 70,000 lines of code for older projects. The trend reverts for deciles 9 and 10, where newer projects are smaller. The boundary of the largest decile is 320,000 lines of code for older projects, and 220,000 for newer projects. While the newer projects are smaller, they are not markedly smaller, especially since the oldest projects are a few months old at best.

In terms of adoption, we see the same behavior for older and newer projects: larger projects have more adoption, with the exception of the slight drop of the last decile.

\paragraph{Contributors}

Newer projects are overall significantly smaller in terms of contributors than the older projects, across most deciles. In terms of adoption, we see that the curve is slowly rising acrosss deciles for older projects, while it is more flat for newer projects (there is even a middle bump).


\paragraph{Commits}
Newer projects have significantly less commits than older ones. Both the curve of older and newer projects rises with deciles,with the exception of the largest decile for commits.

\paragraph{Issues and Pull requests}
As with commits, issues and pull requests are less numerous for newer projects. In both cases the curve for issues rises, with a notable increase in the largest decile. The curves are more steep for pull requests, but the behaviour is similar.

\paragraph{Age}

Newer projects are younger by construction. We see that even in this samples, the very youngest projects have a higher adoption than the marginally older ones.


\subsection{Commit-level adoption and project-level metrics}

\Cref{tab:overall_commit_ratios} focus on the commit-level metric, the commit ratio. Once again, we compare the two categories of projects. Beyond the increase in intensity of adoption (higher commit ratios) of the newer projects, we can make several observations.

\paragraph{Lines of code} The relationship between commit ratio and lines of code is different. Notably, for older projects, the largest projects tended to have lower commit ratios. This is not the case anymore: the adoption across deciles is more or less flat when we look at the more intensive category, the "pervasive" projects (more than 20\% of commits detected as AI-assisted). Conversely, the lowest categories, "experimental" and "limited", tend to rise for larger, older projects; this is not the case for newer projects.

\paragraph{Contributors} The relationship between commit ratio and lines of code is more similar in this category. Both older and newer projects have decreases in the "none" category, and increases in the "limited" and "consistent" categories. For the "pervasive" category, there is a regular drop for older projects, but the curse is more flat for newer projects, excep the last decile.

\paragraph{Age} For age, we once again see differences. For older projects, there is no clear trend. For newer projects, we see a decrease of intensity in the "pervasive" category: the youngest of the newer projects have a more intense adoption.


\begin{table}[ht]
\scriptsize
\centering
\caption{Commit adoption across different project categories}
\label{tab:overall_commit_ratios}
\begin{tabular}{l|r|rrrrr}
\hline
\toprule
\textbf{Category} & \textbf{N} & \textbf{None} & \textbf{Experimental} & \textbf{Limited} & \textbf{Consistent} & \textbf{Pervasive} \\
\midrule
New Projects -- All Files & 6,884 & \VeryAllFilesNoneValue & \VeryAllFilesExperimentalValue & \VeryAllFilesLimitedValue & \VeryAllFilesConsistentValue & \VeryAllFilesPervasiveValue \\
\quad by LOC & & \VeryAllFilescodelinesNoneSparkline & \VeryAllFilescodelinesExperimentalSparkline & \VeryAllFilescodelinesLimitedSparkline & \VeryAllFilescodelinesConsistentSparkline & \VeryAllFilescodelinesPervasiveSparkline \\
\quad by Contributors & & \VeryAllFilescontributorsNoneSparkline & \VeryAllFilescontributorsExperimentalSparkline & \VeryAllFilescontributorsLimitedSparkline & \VeryAllFilescontributorsConsistentSparkline & \VeryAllFilescontributorsPervasiveSparkline \\
\quad by Age & & \VeryAllFilesageyearsNoneSparkline & \VeryAllFilesageyearsExperimentalSparkline & \VeryAllFilesageyearsLimitedSparkline & \VeryAllFilesageyearsConsistentSparkline & \VeryAllFilesageyearsPervasiveSparkline \\
\midrule
\rowcolor{lightgray!50} Older Projects -- All Files & 14,874 & \AllAllFilesNoneValue & \AllAllFilesExperimentalValue & \AllAllFilesLimitedValue & \AllAllFilesConsistentValue & \AllAllFilesPervasiveValue \\
\quad by LOC & & \AllAllFilescodelinesNoneSparkline & \AllAllFilescodelinesExperimentalSparkline & \AllAllFilescodelinesLimitedSparkline & \AllAllFilescodelinesConsistentSparkline & \AllAllFilescodelinesPervasiveSparkline \\
\quad by Contributors & & \AllAllFilescontributorsNoneSparkline & \AllAllFilescontributorsExperimentalSparkline & \AllAllFilescontributorsLimitedSparkline & \AllAllFilescontributorsConsistentSparkline & \AllAllFilescontributorsPervasiveSparkline \\
\quad by Age & & \AllAllFilesageyearsNoneSparkline & \AllAllFilesageyearsExperimentalSparkline & \AllAllFilesageyearsLimitedSparkline & \AllAllFilesageyearsConsistentSparkline & \AllAllFilesageyearsPervasiveSparkline \\
\bottomrule
\end{tabular}
\end{table}

\begin{table}[ht]
\tiny
\centering
\caption{Commit adoption across different project categories}
\label{tab:detailled_commit_ratios}
\begin{tabular}{l|rrrr|rrrr}
\hline
\toprule
\textbf{Category} & \textbf{Experimental} & \textbf{Limited} & \textbf{Consistent} & \textbf{Pervasive} & \textbf{Experimental} & \textbf{Limited} & \textbf{Consistent} & \textbf{Pervasive} \\
\midrule
& \multicolumn{4}{c}{\textbf{Newer projects}}  & \multicolumn{4}{c}{\cellcolor{lightgray!50} \textbf{Older projects}} \\
All Files with commits & \VeryAllFileswithcommitsExperimentalValue & \VeryAllFileswithcommitsLimitedValue & \VeryAllFileswithcommitsConsistentValue & \VeryAllFileswithcommitsPervasiveValue 
& \AllAllFileswithcommitsExperimentalValue & \AllAllFileswithcommitsLimitedValue & \AllAllFileswithcommitsConsistentValue & \AllAllFileswithcommitsPervasiveValue \\
\quad by LOC & \VeryAllFileswithcommitscodelinesExperimentalSparkline & \VeryAllFileswithcommitscodelinesLimitedSparkline & \VeryAllFileswithcommitscodelinesConsistentSparkline & \VeryAllFileswithcommitscodelinesPervasiveSparkline 
& \AllAllFileswithcommitscodelinesExperimentalSparkline & \AllAllFileswithcommitscodelinesLimitedSparkline & \AllAllFileswithcommitscodelinesConsistentSparkline & \AllAllFileswithcommitscodelinesPervasiveSparkline \\
\quad by Contributors & \VeryAllFileswithcommitscontributorsExperimentalSparkline & \VeryAllFileswithcommitscontributorsLimitedSparkline & \VeryAllFileswithcommitscontributorsConsistentSparkline & \VeryAllFileswithcommitscontributorsPervasiveSparkline & \AllAllFileswithcommitscontributorsExperimentalSparkline & \AllAllFileswithcommitscontributorsLimitedSparkline & \AllAllFileswithcommitscontributorsConsistentSparkline & \AllAllFileswithcommitscontributorsPervasiveSparkline \\
\quad by Age & \VeryAllFileswithcommitsageyearsExperimentalSparkline & \VeryAllFileswithcommitsageyearsLimitedSparkline & \VeryAllFileswithcommitsageyearsConsistentSparkline & \VeryAllFileswithcommitsageyearsPervasiveSparkline & \AllAllFileswithcommitsageyearsExperimentalSparkline & \AllAllFileswithcommitsageyearsLimitedSparkline & \AllAllFileswithcommitsageyearsConsistentSparkline & \AllAllFileswithcommitsageyearsPervasiveSparkline \\
\midrule
File level & \VeryFilelevelExperimentalValue & \VeryFilelevelLimitedValue & \VeryFilelevelConsistentValue & \VeryFilelevelPervasiveValue & \AllFilelevelExperimentalValue & \AllFilelevelLimitedValue & \AllFilelevelConsistentValue & \AllFilelevelPervasiveValue \\
\quad by LOC & \VeryFilelevelcodelinesExperimentalSparkline & \VeryFilelevelcodelinesLimitedSparkline & \VeryFilelevelcodelinesConsistentSparkline & \VeryFilelevelcodelinesPervasiveSparkline & \AllFilelevelcodelinesExperimentalSparkline & \AllFilelevelcodelinesLimitedSparkline & \AllFilelevelcodelinesConsistentSparkline & \AllFilelevelcodelinesPervasiveSparkline \\
\quad by Contributors & \VeryFilelevelcontributorsExperimentalSparkline & \VeryFilelevelcontributorsLimitedSparkline & \VeryFilelevelcontributorsConsistentSparkline & \VeryFilelevelcontributorsPervasiveSparkline & \AllFilelevelcontributorsExperimentalSparkline & \AllFilelevelcontributorsLimitedSparkline & \AllFilelevelcontributorsConsistentSparkline & \AllFilelevelcontributorsPervasiveSparkline \\
\quad by Age & \VeryFilelevelageyearsExperimentalSparkline & \VeryFilelevelageyearsLimitedSparkline & \VeryFilelevelageyearsConsistentSparkline & \VeryFilelevelageyearsPervasiveSparkline & \AllFilelevelageyearsExperimentalSparkline & \AllFilelevelageyearsLimitedSparkline & \AllFilelevelageyearsConsistentSparkline & \AllFilelevelageyearsPervasiveSparkline \\
\midrule
Files ignored only & \VeryFilesignoredonlyExperimentalValue & \VeryFilesignoredonlyLimitedValue & \VeryFilesignoredonlyConsistentValue & \VeryFilesignoredonlyPervasiveValue & \AllFilesignoredonlyExperimentalValue & \AllFilesignoredonlyLimitedValue & \AllFilesignoredonlyConsistentValue & \AllFilesignoredonlyPervasiveValue \\
\quad by LOC & \VeryFilesignoredonlycodelinesExperimentalSparkline & \VeryFilesignoredonlycodelinesLimitedSparkline & \VeryFilesignoredonlycodelinesConsistentSparkline & \VeryFilesignoredonlycodelinesPervasiveSparkline & \AllFilesignoredonlycodelinesExperimentalSparkline & \AllFilesignoredonlycodelinesLimitedSparkline & \AllFilesignoredonlycodelinesConsistentSparkline & \AllFilesignoredonlycodelinesPervasiveSparkline \\
\quad by Contributors & \VeryFilesignoredonlycontributorsExperimentalSparkline & \VeryFilesignoredonlycontributorsLimitedSparkline & \VeryFilesignoredonlycontributorsConsistentSparkline & \VeryFilesignoredonlycontributorsPervasiveSparkline & \AllFilesignoredonlycontributorsExperimentalSparkline & \AllFilesignoredonlycontributorsLimitedSparkline & \AllFilesignoredonlycontributorsConsistentSparkline & \AllFilesignoredonlycontributorsPervasiveSparkline \\
\quad by Age & \VeryFilesignoredonlyageyearsExperimentalSparkline & \VeryFilesignoredonlyageyearsLimitedSparkline & \VeryFilesignoredonlyageyearsConsistentSparkline & \VeryFilesignoredonlyageyearsPervasiveSparkline & \AllFilesignoredonlyageyearsExperimentalSparkline & \AllFilesignoredonlyageyearsLimitedSparkline & \AllFilesignoredonlyageyearsConsistentSparkline & \AllFilesignoredonlyageyearsPervasiveSparkline \\
\midrule
Commits only & \VeryCommitsonlyExperimentalValue & \VeryCommitsonlyLimitedValue & \VeryCommitsonlyConsistentValue & \VeryCommitsonlyPervasiveValue & \AllCommitsonlyExperimentalValue & \AllCommitsonlyLimitedValue & \AllCommitsonlyConsistentValue & \AllCommitsonlyPervasiveValue \\
\quad by LOC & \VeryCommitsonlycodelinesExperimentalSparkline & \VeryCommitsonlycodelinesLimitedSparkline & \VeryCommitsonlycodelinesConsistentSparkline & \VeryCommitsonlycodelinesPervasiveSparkline & \AllCommitsonlycodelinesExperimentalSparkline & \AllCommitsonlycodelinesLimitedSparkline & \AllCommitsonlycodelinesConsistentSparkline & \AllCommitsonlycodelinesPervasiveSparkline \\
\quad by Contributors & \VeryCommitsonlycontributorsExperimentalSparkline & \VeryCommitsonlycontributorsLimitedSparkline & \VeryCommitsonlycontributorsConsistentSparkline & \VeryCommitsonlycontributorsPervasiveSparkline & \AllCommitsonlycontributorsExperimentalSparkline & \AllCommitsonlycontributorsLimitedSparkline & \AllCommitsonlycontributorsConsistentSparkline & \AllCommitsonlycontributorsPervasiveSparkline \\
\quad by Age & \VeryCommitsonlyageyearsExperimentalSparkline & \VeryCommitsonlyageyearsLimitedSparkline & \VeryCommitsonlyageyearsConsistentSparkline & \VeryCommitsonlyageyearsPervasiveSparkline & \AllCommitsonlyageyearsExperimentalSparkline & \AllCommitsonlyageyearsLimitedSparkline & \AllCommitsonlyageyearsConsistentSparkline & \AllCommitsonlyageyearsPervasiveSparkline \\
\bottomrule
\end{tabular}
\end{table}

\paragraph{Adoption categories and commit ratio} \Cref{tab:detailled_commit_ratios} zooms in further on the commit ratio, exploring it on several categories of adopters. Overall, commit-level adoption is higher in projects that also show file-level adoption; it is lower in projects that have file-level adoption in \texttt{.gitignore} files; and it is lowest in projects that have no file-level adoption at all, only commits.

\section{RQ3: Contexts of Use of Coding Agents}

We examine repository properties and categorize adoption patterns across different dimensions.

\newcommand{\VeryTopOrganizationsMacroToolUse}{65.13\%}
\newcommand{\VeryTopOrganizationsMicroToolUse}{63.77\%}
\newcommand{\VeryTopOrganizationsRelativeDiffMacro}{-9.32\%}
\newcommand{\VeryTopOrganizationsRelativeDiffMicro}{-11.23\%}

\newcommand{\VeryMicrosoftToolUse}{63.46\%}
\newcommand{\VeryMicrosoftRepoCount}{52}
\newcommand{\VeryAmazonToolUse}{75.00\%}
\newcommand{\VeryAmazonRepoCount}{24}
\newcommand{\VeryNvidiaToolUse}{56.25\%}
\newcommand{\VeryNvidiaRepoCount}{16}
\newcommand{\VeryGoogleToolUse}{44.44\%}
\newcommand{\VeryGoogleRepoCount}{9}
\newcommand{\VeryJetbrainsToolUse}{62.50\%}
\newcommand{\VeryJetbrainsRepoCount}{8}
\newcommand{\VeryAlibabaToolUse}{57.14\%}
\newcommand{\VeryAlibabaRepoCount}{7}
\newcommand{\CloudflareToolUse}{83.33\%}
\newcommand{\CloudflareRepoCount}{6}
\newcommand{\VeryHuggingfaceToolUse}{50.00\%}
\newcommand{\VeryHuggingfaceRepoCount}{6}
\newcommand{\VeryAutomatticToolUse}{50.00\%}
\newcommand{\VeryAutomatticRepoCount}{2}
\newcommand{\VeryMetaToolUse}{100.00\%}
\newcommand{\VeryMetaRepoCount}{2}
\newcommand{\VeryApacheToolUse}{100.00\%}
\newcommand{\VeryApacheRepoCount}{1}
\newcommand{\VeryElasticToolUse}{0.00\%}
\newcommand{\VeryElasticRepoCount}{1}
\newcommand{\VeryGrafanaToolUse}{100.00\%}
\newcommand{\VeryGrafanaRepoCount}{1}
\newcommand{\VeryMozillaToolUse}{100.00\%}
\newcommand{\VeryMozillaRepoCount}{1}
\newcommand{\VeryShopifyToolUse}{0.00\%}
\newcommand{\VeryShopifyRepoCount}{1}
\newcommand{\VeryTencentToolUse}{100.00\%}
\newcommand{\VeryTencentRepoCount}{1}

\begin{figure}[htbp]
    \centering
    \includegraphics[width=0.8\textwidth]{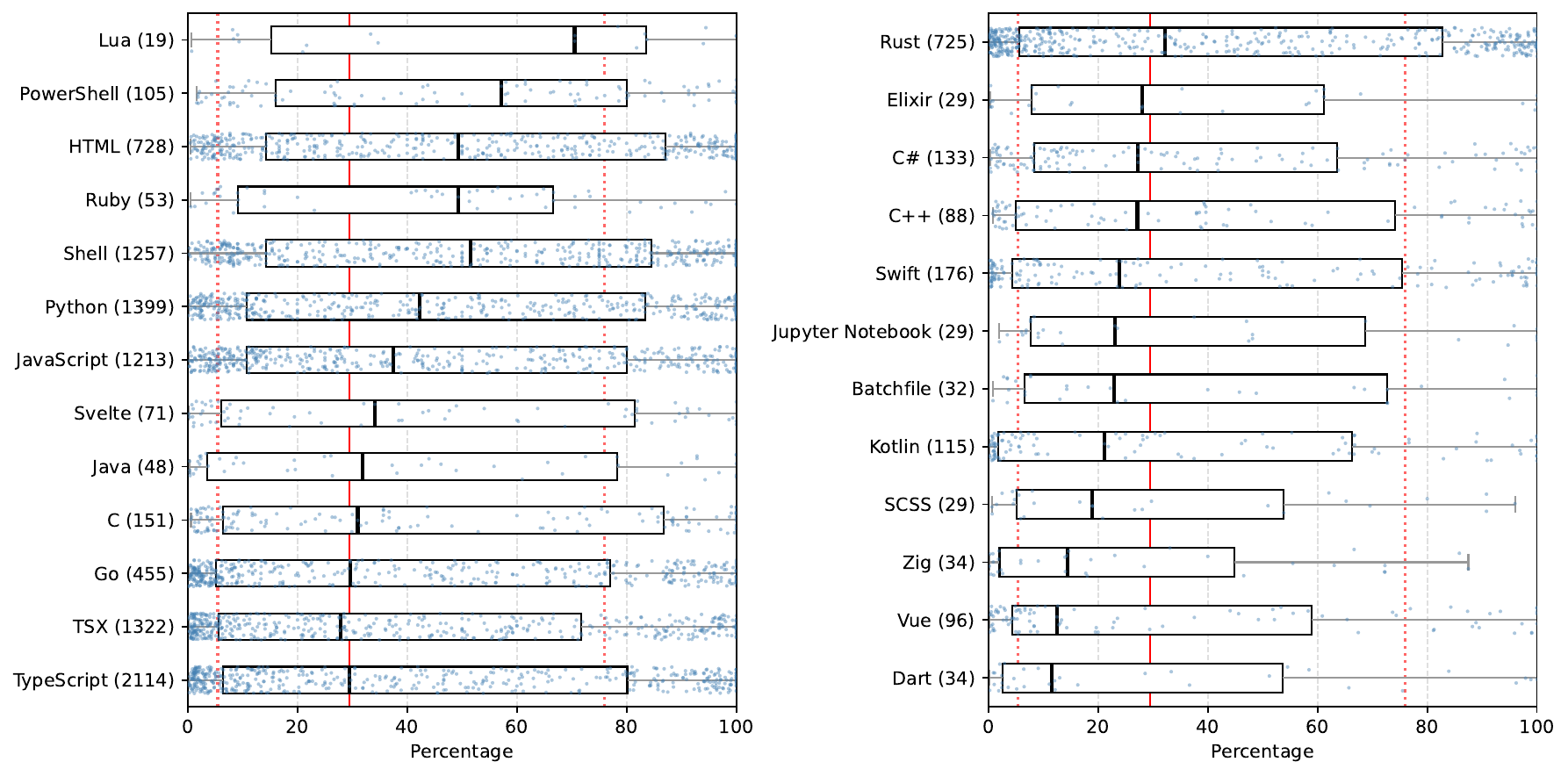}
    \caption{Distribution of commit ratio by language for newer projects}
    \label{fig:languages-adoption-frequency-very}

    \vspace{1cm} 

    \includegraphics[width=0.8\textwidth]{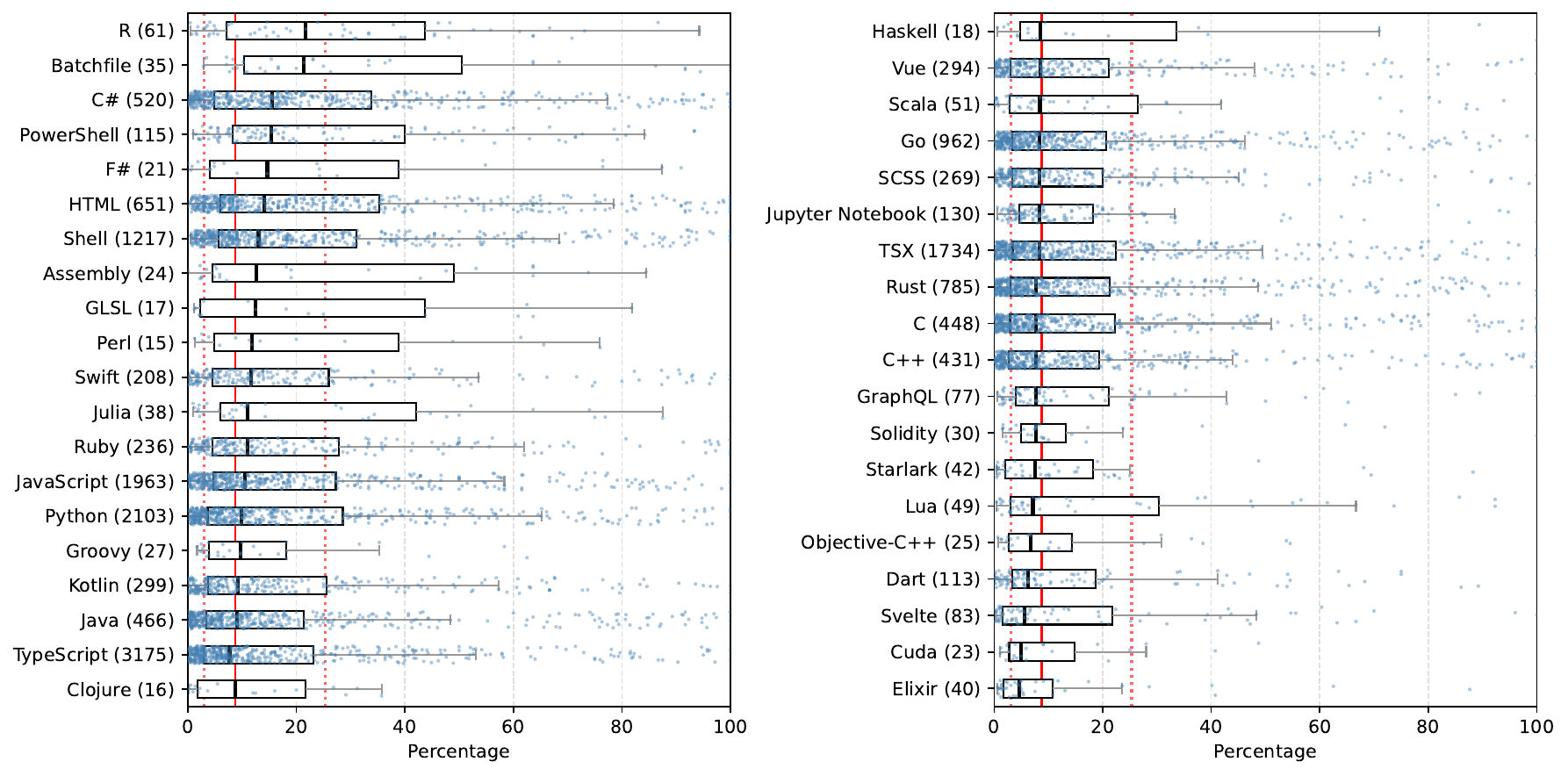}
    \caption{Distribution of commit ratio by language for older projects}
    \label{fig:languages-adoption-frequency-all}
\end{figure}

\Cref{fig:languages-adoption-frequency-very} and \Cref{fig:languages-adoption-frequency-all} contrast the commit ratio of projects by programming language, for newer and older projects. The red full line shows the median commit ratio for each set of projects. The dotted red lines show the upper and lower quartiles. The first thing we clearly see is that the overall commit ratio is much higher in the newer projects. The median is close to 30\% vs ~10\% for older projects, while the upper quartile is close to 25\% for the older projects, whereas it is around 75\% for the newer projects. In both cases, the newer projects seem to have \emph{triple} the commit ratio.

Comparing the programming languages is not straightforward, as both the amount of languages that cross the minimum threshold, and their size in terms of projects that use a programming language are different. Overall, there are more variations in commit ratio per programming language in the newer projects.

One interesting comparison aspect is the number of projects per language in both sets, which shows some revealing differences. If we compare the percentage of adopters that use a given language in the newer projects with the ones in the older project, there are clear trends in terms of programming languages. Table ~\ref{tab:language_evolution} shows this data.

\begin{table}[h]
\centering
\caption{Language Evolution}\label{tab:language_evolution}
\begin{tabular}{|l|r|r|l|r|r|}
\hline
\textbf{Language} & \textbf{New} & \textbf{\% New} & \textbf{Old} & \textbf{\% Old} & \textbf{Evolution (\%)} \\ \hline
Shell & 1257 & 18.26\% & 1217 & 8.18\% & 223.17\% \\ \hline
Rust & 725 & 10.53\% & 785 & 5.28\% & 199.55\% \\ \hline
Batchfile & 32 & 0.46\% & 35 & 0.24\% & 197.55\% \\ \hline
Powershell & 105 & 1.53\% & 115 & 0.77\% & 197.28\% \\ \hline
HTML & 651 & 9.46\% & 728 & 4.89\% & 193.21\% \\ \hline
Svelte & 71 & 1.03\% & 83 & 0.56\% & 184.83\% \\ \hline
Swift & 176 & 2.56\% & 208 & 1.40\% & 182.83\% \\ \hline
TSX & 1322 & 19.20\% & 1734 & 11.66\% & 164.73\% \\ \hline
Elixir & 29 & 0.42\% & 40 & 0.27\% & 156.65\% \\ \hline
Typescript & 2114 & 30.71\% & 3175 & 21.35\% & 143.86\% \\ \hline
Python & 1399 & 20.32\% & 2103 & 14.14\% & 143.74\% \\ \hline
Javascript & 1213 & 17.62\% & 1963 & 13.20\% & 133.51\% \\ \hline
Go & 455 & 6.61\% & 962 & 6.47\% & 102.19\% \\ \hline
Lua & 19 & 0.28\% & 49 & 0.33\% & 83.78\% \\ \hline
Kotlin & 115 & 1.67\% & 299 & 2.01\% & 83.10\% \\ \hline
C & 151 & 2.19\% & 448 & 3.01\% & 72.83\% \\ \hline
Vue & 96 & 1.39\% & 294 & 1.98\% & 70.55\% \\ \hline
Dart & 34 & 0.49\% & 113 & 0.76\% & 65.01\% \\ \hline
C\# & 133 & 1.93\% & 520 & 3.50\% & 55.26\% \\ \hline
Jupyter & 29 & 0.42\% & 130 & 0.87\% & 48.20\% \\ \hline
Ruby & 53 & 0.77\% & 256 & 1.72\% & 44.73\% \\ \hline
C++ & 88 & 1.28\% & 431 & 2.90\% & 44.12\% \\ \hline
SCCS & 29 & 0.42\% & 269 & 1.81\% & 23.29\% \\ \hline
Java & 48 & 0.70\% & 466 & 3.13\% & 22.26\% \\ \hline
\end{tabular}
\end{table}

Some languages are over-represented in newer projects. We see that in particular for shell scripting languages (Shell, Batchfile, Powershell), and some markup languages such as HTML. For programming languages, Rust is the most over-represented in newer projects. Other languages with sizeable usage that are over-represented include Javascript and its variants (Typescript, TSX), and Python.

On the other hand, several established languages are under-represented in newer projects. These include systems languages such as C and C++ (newer projects seem to favor Rust), as well as general-purpose languages such as C\#, Kotlin, or, to an even larger extent, Java.

\begin{figure}[h]
\centering
\includegraphics[width=\textwidth]{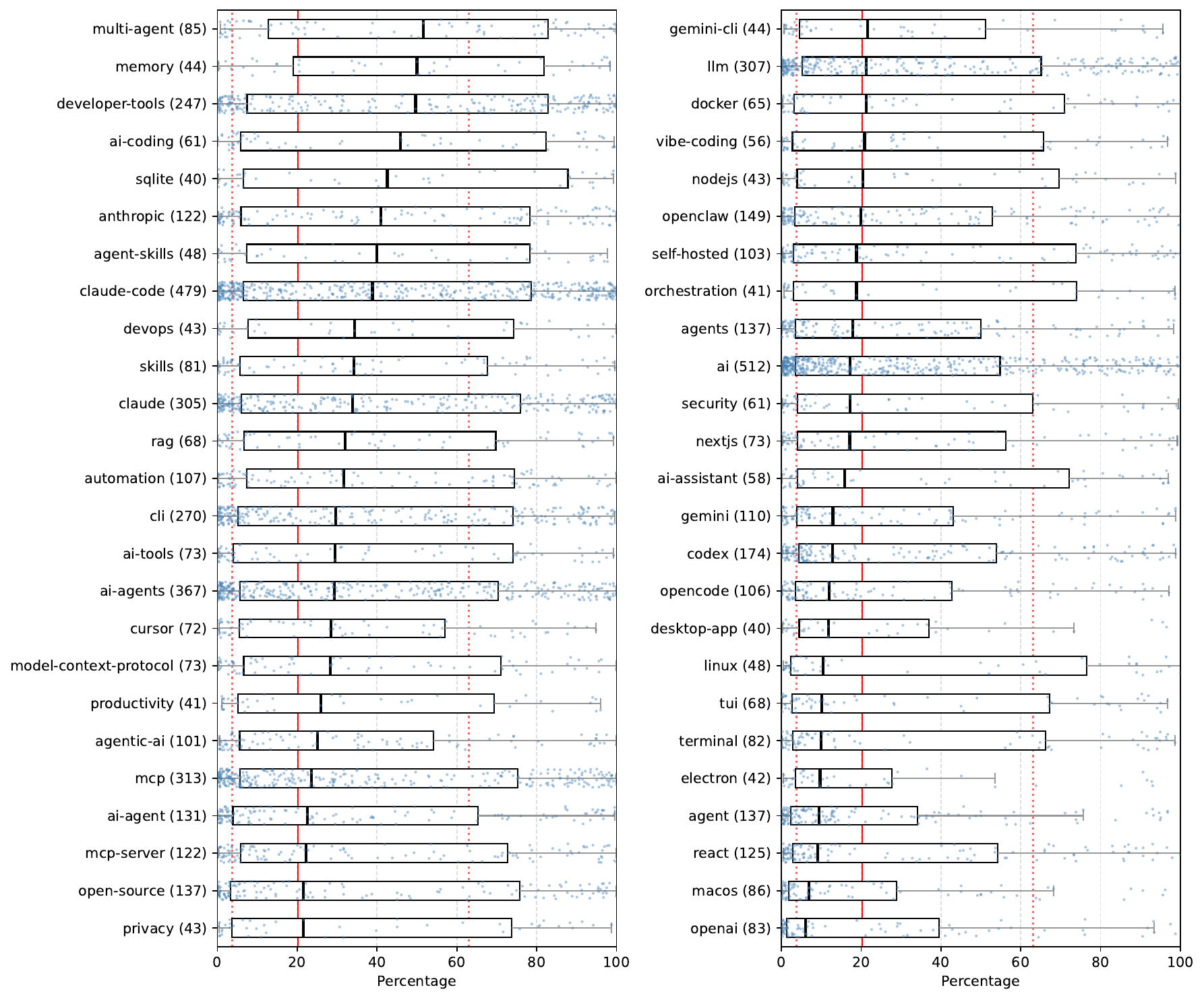}
\caption{Topic-stratified analysis of commit adoption ratios, comparing adoption patterns across repository categorizations, for newer projects.}
\label{fig:topic-adoption-distribution-very}
\end{figure}

\begin{figure}[h]
\centering
\includegraphics[width=\textwidth]{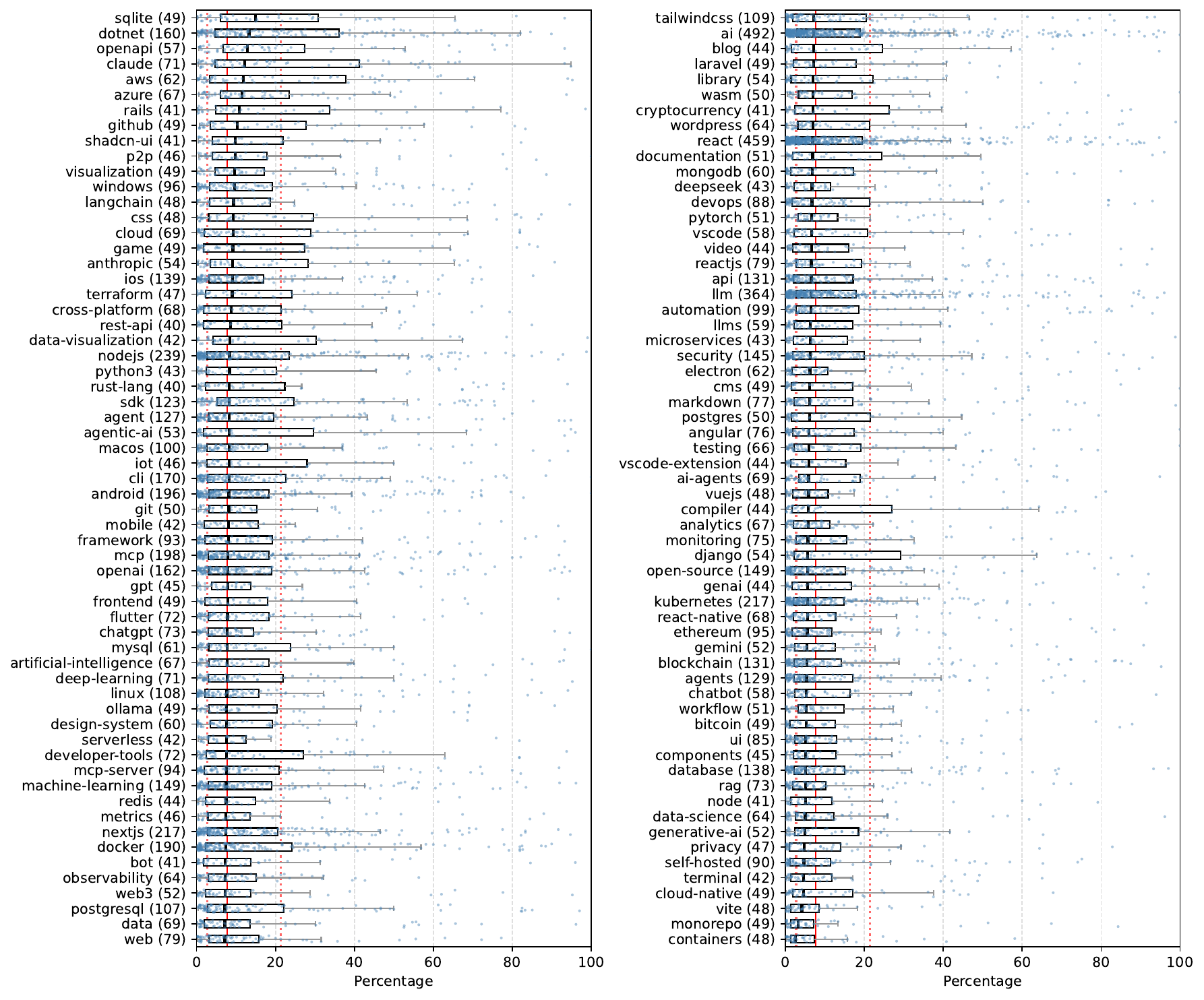}
\caption{Topic-stratified analysis of commit adoption ratios, comparing adoption patterns across repository categorizations, for older projects.}
\label{fig:topic-adoption-distribution-all}
\end{figure}

\Cref{fig:topic-adoption-distribution-very} and \Cref{fig:topic-adoption-distribution-all} contrast the commit ratio of projects by topics, for newer and older projects. We can clearly see a very different distribution of topics. While for the older projects, there is a broad distribution of topics, with some over-representation of machine learning and artifical intelligence topics, we can clearly see that for the newer projects, the over-representation is overwhelming, with a particular focus on agentic AI (e.g. mcp, ai-agents, openclaw, agent-skills, or simply agents) and explicit mentions of coding agents (claude-code, cursor, opencode). The topics that are not explicitely about artificial intelligence, such as developer tools, cli, productivity, or orchestration, can be implicitly linked to the broad phenomenon of agentic coding tools, that is vastly present in the other topics. In short, there is a large amount of ``dogfooding'': coding agents used to drive improvements to AI coding in general.

\newcommand{\AllTopOrganizationsMacroToolUse}{24.18\%}
\newcommand{\AllTopOrganizationsMicroToolUse}{20.99\%}
\newcommand{\AllTopOrganizationsRelativeDiffMacro}{-8.61\%}
\newcommand{\AllTopOrganizationsRelativeDiffMicro}{-20.69\%}

\newcommand{\AllMicrosoftToolUse}{30.73\%}
\newcommand{\AllMicrosoftRepoCount}{934}
\newcommand{\AllGoogleToolUse}{10.54\%}
\newcommand{\AllGoogleRepoCount}{626}
\newcommand{\AllApacheToolUse}{12.71\%}
\newcommand{\AllApacheRepoCount}{535}
\newcommand{\AllAmazonToolUse}{16.63\%}
\newcommand{\AllAmazonRepoCount}{415}
\newcommand{\HashicorpToolUse}{2.29\%}
\newcommand{\HashicorpRepoCount}{131}
\newcommand{\AllMetaToolUse}{21.30\%}
\newcommand{\AllMetaRepoCount}{108}
\newcommand{\AllNvidiaToolUse}{26.17\%}
\newcommand{\AllNvidiaRepoCount}{107}
\newcommand{\AllGrafanaToolUse}{38.30\%}
\newcommand{\AllGrafanaRepoCount}{94}
\newcommand{\AllJetbrainsToolUse}{21.59\%}
\newcommand{\AllJetbrainsRepoCount}{88}
\newcommand{\AllElasticToolUse}{20.73\%}
\newcommand{\AllElasticRepoCount}{82}
\newcommand{\AllMozillaToolUse}{10.96\%}
\newcommand{\AllMozillaRepoCount}{73}
\newcommand{\AllAutomatticToolUse}{63.08\%}
\newcommand{\AllAutomatticRepoCount}{65}
\newcommand{\AllShopifyToolUse}{19.35\%}
\newcommand{\AllShopifyRepoCount}{62}
\newcommand{\AllAlibabaToolUse}{24.56\%}
\newcommand{\AllAlibabaRepoCount}{57}
\newcommand{\RustToolUse}{7.69\%}
\newcommand{\RustRepoCount}{52}
\newcommand{\KubernetesToolUse}{6.00\%}
\newcommand{\KubernetesRepoCount}{50}
\newcommand{\AllHuggingfaceToolUse}{24.49\%}
\newcommand{\AllHuggingfaceRepoCount}{49}
\newcommand{\SymfonyToolUse}{10.20\%}
\newcommand{\SymfonyRepoCount}{49}
\newcommand{\AllTencentToolUse}{16.33\%}
\newcommand{\AllTencentRepoCount}{49}
\newcommand{\IobrokerToolUse}{100.00\%}
\newcommand{\IobrokerRepoCount}{48}

\begin{table}[ht]
\scriptsize
\centering
\caption{File-based adoption for top 20 organizations}
\label{tab:organization_adoption}
\begin{tabular}{l|rr@{\hspace{0.5cm}}|rr}
\toprule
\textbf{Organization} & \textbf{New Adoption pct} & \textbf{New Repositories} & \textbf{Old Adoption pct} & \textbf{Old Repositories} \\
\midrule
Microsoft & \VeryMicrosoftToolUse & \VeryMicrosoftRepoCount & \AllMicrosoftToolUse & \AllMicrosoftRepoCount \\
Amazon & \VeryAmazonToolUse & \VeryAmazonRepoCount & \AllAmazonToolUse & \AllAmazonRepoCount \\
Nvidia & \VeryNvidiaToolUse & \VeryNvidiaRepoCount & \AllNvidiaToolUse & \AllNvidiaRepoCount \\
Google & \VeryGoogleToolUse & \VeryGoogleRepoCount & \AllGoogleToolUse & \AllGoogleRepoCount \\
Jetbrains & \VeryJetbrainsToolUse & \VeryJetbrainsRepoCount & \AllJetbrainsToolUse & \AllJetbrainsRepoCount \\
Alibaba & \VeryAlibabaToolUse & \VeryAlibabaRepoCount & \AllAlibabaToolUse & \AllAlibabaRepoCount \\
Huggingface & \VeryHuggingfaceToolUse & \VeryHuggingfaceRepoCount & \AllHuggingfaceToolUse & \AllHuggingfaceRepoCount \\
\bottomrule
\end{tabular}
\end{table}

\Cref{tab:organization_adoption} shows the differences in adoption for newer vs older projects in some of the top 100 organizations. The list is shorter, as few such organizations created a large number of repositories in that period, that matched our sampling criteria. We see that their adoption has risen in line with the broad trends. Interestingly, while the organizations had a higher level of adoption among the older projects (14.70\% adoption overall), this is no longer the case for the newer projects (60\% of adoption overall).


\section{RQ4: Evolution of Adoption Over Time, and Tool-Specific Adoption}

We present visualizations of tool-specific adoption patterns and co-adoption relationships. \Cref{fig:adoption-very} and \Cref{fig:adoption-all} contrast the adoption timeline for both newer and older projects. Unsurprisingly, the adoption timeline is much sharper for the newer projects. We do see a similar inflection point, in that the number of adopters grows very significantly in early 2026. The spike is of course less sharp, but still visible, for the older projects.

\newcommand{\VerytotalAdoptingProjects}{8\,562}
\newcommand{\VeryfirstAdoptionDate}{2025-01-05}
\newcommand{\VerylastAdoptionDate}{2026-04-08}
\newcommand{\VeryadoptionTimeSpanDays}{458}

\newcommand{\VerytotalToolAdoptions}{16\,225}
\newcommand{\VeryuniqueTools}{47}
\newcommand{\VerydisplayedTools}{47}
\newcommand{\VerytopToolName}{Claude\_Code}
\newcommand{\VerytopToolCount}{6\,443}
\newcommand{\VeryClaudeCodeCount}{6\,443}
\newcommand{\VeryGenericCount}{3\,092}
\newcommand{\VeryCopilotCount}{2\,177}
\newcommand{\VeryCodexCount}{1\,231}
\newcommand{\VeryCursorCount}{1\,064}
\newcommand{\VeryGeminiCount}{499}
\newcommand{\VeryOpencodeCount}{298}
\newcommand{\VeryAmpCount}{230}
\newcommand{\VeryJulesCount}{227}
\newcommand{\VeryCoderabbitCount}{196}
\newcommand{\VeryWindsurfCount}{71}
\newcommand{\VeryKiroCount}{69}
\newcommand{\VeryDevinCount}{59}
\newcommand{\VeryClineCount}{59}
\newcommand{\SpecKitCount}{55}
\newcommand{\VerySerenaCount}{53}
\newcommand{\PiCount}{52}
\newcommand{\VeryQwenCoderCount}{45}
\newcommand{\VeryAiderCount}{35}
\newcommand{\TraeCount}{31}
\newcommand{\VeryotherToolsCount}{239}

\newcommand{\VerytotalOrganizationAdoptions}{16\,225}
\newcommand{\VeryuniqueOrganizations}{7\,418}
\newcommand{\VerydisplayedOrganizations}{35}
\newcommand{\VerytopOrganizationName}{dicklesworthstone}
\newcommand{\VerytopOrganizationCount}{110}
\newcommand{\dicklesworthstoneCount}{110}
\newcommand{\VerymicrosoftCount}{97}
\newcommand{\nirholasCount}{37}
\newcommand{\vercellabsCount}{37}
\newcommand{\VerynvidiaCount}{34}
\newcommand{\VeryamazonCount}{31}
\newcommand{\VerygetsentryCount}{28}
\newcommand{\bearcoveCount}{26}
\newcommand{\kiloorgCount}{26}
\newcommand{\snowdreamtechCount}{25}
\newcommand{\steipeteCount}{25}
\newcommand{\openclawCount}{25}
\newcommand{\avifeneshCount}{23}
\newcommand{\blockCount}{20}
\newcommand{\hmbownCount}{20}
\newcommand{\VeryvercelCount}{19}
\newcommand{\jordanhubbardCount}{19}
\newcommand{\VeryalibabaCount}{18}
\newcommand{\hkudsCount}{18}
\newcommand{\boostcampwmtwozerotwofiveCount}{17}
\newcommand{\martianengineeringCount}{17}
\newcommand{\cyberfabricCount}{17}
\newcommand{\jfiveiktwooCount}{16}
\newcommand{\firstflukeCount}{16}
\newcommand{\VerycloudflareCount}{16}

\newcommand{\VeryminSubsetSize}{40}
\newcommand{\VeryminFrequency}{250}
\newcommand{\VerytotalProjects}{8\,562}
\newcommand{\VeryprojectsUsingoneTools}{4\,247}
\newcommand{\VeryprojectsUsingtwoTools}{2\,432}
\newcommand{\VeryprojectsUsingthreeTools}{1\,082}
\newcommand{\VeryprojectsUsingfourTools}{454}
\newcommand{\VeryprojectsUsingfiveTools}{192}
\newcommand{\VeryprojectsUsingsixTools}{83}
\newcommand{\VeryprojectsUsingsevenTools}{30}
\newcommand{\VeryprojectsUsingeightTools}{21}
\newcommand{\VeryprojectsUsingnineTools}{13}
\newcommand{\VeryprojectsUsingonezeroTools}{3}
\newcommand{\VeryprojectsUsingoneoneTools}{4}
\newcommand{\projectsUsingtwofiveTools}{1}
\newcommand{\VerymeanToolsPerProject}{1.8950011679514132}
\newcommand{\VerymedianToolsPerProject}{2.0}
\newcommand{\VerymaxToolsPerProject}{25}
\newcommand{\VeryminToolsPerProject}{1}
\newcommand{\VeryprojectsAfterFiltering}{8\,407}
\newcommand{\VeryprojectsFilteredOut}{155}
\newcommand{\VerytoolsAnalyzed}{7}
\newcommand{\VerytotalIntersections}{91}
\newcommand{\VeryprojectsInShownIntersections}{7\,701}
\newcommand{\VeryprojectsInFilteredIntersections}{706}
\newcommand{\VeryprojectsWithMultipleTools}{24}
\newcommand{\VerylargestIntersectionSize}{2\,925}
\newcommand{\VerylargestIntersectionTools}{Claude\_Code}
\newcommand{\VeryintersectionClaudeCode}{Claude\_Code (2925)}
\newcommand{\VeryintersectionClaudeCodeSize}{2\,925}
\newcommand{\VeryintersectionClaudeCodeToolCount}{1}
\newcommand{\VeryintersectionClaudeCodeGeneric}{Claude\_Code $\cap$ Generic (1010)}
\newcommand{\VeryintersectionClaudeCodeGenericSize}{1\,010}
\newcommand{\VeryintersectionClaudeCodeGenericToolCount}{2}
\newcommand{\VeryintersectionCopilot}{Copilot (672)}
\newcommand{\VeryintersectionCopilotSize}{672}
\newcommand{\VeryintersectionCopilotToolCount}{1}
\newcommand{\VeryintersectionGeneric}{Generic (469)}
\newcommand{\VeryintersectionGenericSize}{469}
\newcommand{\VeryintersectionGenericToolCount}{1}
\newcommand{\VeryintersectionClaudeCodeCopilot}{Claude\_Code $\cap$ Copilot (463)}
\newcommand{\VeryintersectionClaudeCodeCopilotSize}{463}
\newcommand{\VeryintersectionClaudeCodeCopilotToolCount}{2}
\newcommand{\VeryintersectionClaudeCodeCodexGeneric}{Claude\_Code $\cap$ Codex $\cap$ Generic (234)}
\newcommand{\VeryintersectionClaudeCodeCodexGenericSize}{234}
\newcommand{\VeryintersectionClaudeCodeCodexGenericToolCount}{3}
\newcommand{\VeryintersectionClaudeCodeCodex}{Claude\_Code $\cap$ Codex (222)}
\newcommand{\VeryintersectionClaudeCodeCodexSize}{222}
\newcommand{\VeryintersectionClaudeCodeCodexToolCount}{2}
\newcommand{\VeryintersectionClaudeCodeCopilotGeneric}{Claude\_Code $\cap$ Copilot $\cap$ Generic (217)}
\newcommand{\VeryintersectionClaudeCodeCopilotGenericSize}{217}
\newcommand{\VeryintersectionClaudeCodeCopilotGenericToolCount}{3}
\newcommand{\VeryintersectionClaudeCodeCursor}{Claude\_Code $\cap$ Cursor (216)}
\newcommand{\VeryintersectionClaudeCodeCursorSize}{216}
\newcommand{\VeryintersectionClaudeCodeCursorToolCount}{2}
\newcommand{\VeryintersectionCursor}{Cursor (168)}
\newcommand{\VeryintersectionCursorSize}{168}
\newcommand{\VeryintersectionCursorToolCount}{1}
\newcommand{\VeryintersectionClaudeCodeCursorGeneric}{Claude\_Code $\cap$ Cursor $\cap$ Generic (137)}
\newcommand{\VeryintersectionClaudeCodeCursorGenericSize}{137}
\newcommand{\VeryintersectionClaudeCodeCursorGenericToolCount}{3}
\newcommand{\VeryintersectionCodex}{Codex (130)}
\newcommand{\VeryintersectionCodexSize}{130}
\newcommand{\VeryintersectionCodexToolCount}{1}
\newcommand{\VeryintersectionCopilotGeneric}{Copilot $\cap$ Generic (122)}
\newcommand{\VeryintersectionCopilotGenericSize}{122}
\newcommand{\VeryintersectionCopilotGenericToolCount}{2}
\newcommand{\intersectionClaudeCodeCodexCopilotGeneric}{Claude\_Code $\cap$ Codex $\cap$ Copilot $\cap$ Generic (94)}
\newcommand{\intersectionClaudeCodeCodexCopilotGenericSize}{94}
\newcommand{\intersectionClaudeCodeCodexCopilotGenericToolCount}{4}
\newcommand{\VeryintersectionClaudeCodeGeminiGeneric}{Claude\_Code $\cap$ Gemini $\cap$ Generic (79)}
\newcommand{\VeryintersectionClaudeCodeGeminiGenericSize}{79}
\newcommand{\VeryintersectionClaudeCodeGeminiGenericToolCount}{3}
\newcommand{\VeryintersectionCodexGeneric}{Codex $\cap$ Generic (75)}
\newcommand{\VeryintersectionCodexGenericSize}{75}
\newcommand{\VeryintersectionCodexGenericToolCount}{2}
\newcommand{\VeryintersectionClaudeCodeCopilotCursorGeneric}{Claude\_Code $\cap$ Copilot $\cap$ Cursor $\cap$ Generic (73)}
\newcommand{\VeryintersectionClaudeCodeCopilotCursorGenericSize}{73}
\newcommand{\VeryintersectionClaudeCodeCopilotCursorGenericToolCount}{4}
\newcommand{\VeryintersectionClaudeCodeGemini}{Claude\_Code $\cap$ Gemini (68)}
\newcommand{\VeryintersectionClaudeCodeGeminiSize}{68}
\newcommand{\VeryintersectionClaudeCodeGeminiToolCount}{2}
\newcommand{\intersectionClaudeCodeCodexCopilot}{Claude\_Code $\cap$ Codex $\cap$ Copilot (66)}
\newcommand{\intersectionClaudeCodeCodexCopilotSize}{66}
\newcommand{\intersectionClaudeCodeCodexCopilotToolCount}{3}
\newcommand{\VeryintersectionClaudeCodeCopilotCursor}{Claude\_Code $\cap$ Copilot $\cap$ Cursor (60)}
\newcommand{\VeryintersectionClaudeCodeCopilotCursorSize}{60}
\newcommand{\VeryintersectionClaudeCodeCopilotCursorToolCount}{3}
\newcommand{\VeryintersectionClaudeCodeCodexCursorGeneric}{Claude\_Code $\cap$ Codex $\cap$ Cursor $\cap$ Generic (56)}
\newcommand{\VeryintersectionClaudeCodeCodexCursorGenericSize}{56}
\newcommand{\VeryintersectionClaudeCodeCodexCursorGenericToolCount}{4}
\newcommand{\intersectionClaudeCodeGenericOpencode}{Claude\_Code $\cap$ Generic $\cap$ Opencode (50)}
\newcommand{\intersectionClaudeCodeGenericOpencodeSize}{50}
\newcommand{\intersectionClaudeCodeGenericOpencodeToolCount}{3}
\newcommand{\VeryintersectionCursorGeneric}{Cursor $\cap$ Generic (48)}
\newcommand{\VeryintersectionCursorGenericSize}{48}
\newcommand{\VeryintersectionCursorGenericToolCount}{2}
\newcommand{\VeryintersectionGemini}{Gemini (47)}
\newcommand{\VeryintersectionGeminiSize}{47}
\newcommand{\VeryintersectionGeminiToolCount}{1}
\newcommand{\VerymeanIntersectionSize}{320}
\newcommand{\VerymedianIntersectionSize}{126}
\newcommand{\VeryshownIntersectionsoneTools}{6}
\newcommand{\VeryshownIntersectionstwoTools}{8}
\newcommand{\VeryshownIntersectionsthreeTools}{7}
\newcommand{\shownIntersectionsfourTools}{3}


\newcommand{\AlltotalAdoptingProjects}{32\,134}
\newcommand{\AllfirstAdoptionDate}{2025-01-01}
\newcommand{\AlllastAdoptionDate}{2026-04-05}
\newcommand{\AlladoptionTimeSpanDays}{459}

\newcommand{\AlltotalToolAdoptions}{52\,556}
\newcommand{\AlluniqueTools}{56}
\newcommand{\AlldisplayedTools}{56}
\newcommand{\AlltopToolName}{Claude\_Code}
\newcommand{\AlltopToolCount}{16\,897}
\newcommand{\AllClaudeCodeCount}{16\,897}
\newcommand{\AllCopilotCount}{15\,794}
\newcommand{\AllGenericCount}{6\,234}
\newcommand{\AllCursorCount}{3\,436}
\newcommand{\AllCodexCount}{2\,938}
\newcommand{\AllGeminiCount}{1\,408}
\newcommand{\AllCoderabbitCount}{1\,372}
\newcommand{\AllJulesCount}{931}
\newcommand{\AllAiderCount}{529}
\newcommand{\AllDevinCount}{464}
\newcommand{\AllAmpCount}{330}
\newcommand{\AllOpencodeCount}{308}
\newcommand{\AllClineCount}{241}
\newcommand{\SourceryCount}{203}
\newcommand{\AllWindsurfCount}{178}
\newcommand{\JunieCount}{172}
\newcommand{\OpenHandsCount}{123}
\newcommand{\AllSerenaCount}{111}
\newcommand{\AllKiroCount}{102}
\newcommand{\AllQwenCoderCount}{87}
\newcommand{\AllotherToolsCount}{690}

\newcommand{\AlltotalOrganizationAdoptions}{52\,556}
\newcommand{\AlluniqueOrganizations}{21\,374}
\newcommand{\AlldisplayedOrganizations}{35}
\newcommand{\AlltopOrganizationName}{microsoft}
\newcommand{\AlltopOrganizationCount}{740}
\newcommand{\AllmicrosoftCount}{740}
\newcommand{\apacheCount}{330}
\newcommand{\openshiftCount}{175}
\newcommand{\googleCount}{174}
\newcommand{\grafanaCount}{131}
\newcommand{\AllamazonCount}{126}
\newcommand{\automatticCount}{126}
\newcommand{\AllgetsentryCount}{119}
\newcommand{\datadogCount}{117}
\newcommand{\AllnvidiaCount}{111}
\newcommand{\elasticCount}{104}
\newcommand{\pulumiCount}{98}
\newcommand{\micronautCount}{71}
\newcommand{\iobrokerCount}{71}
\newcommand{\huggingfaceCount}{63}
\newcommand{\mattermostCount}{63}
\newcommand{\langchainaiCount}{58}
\newcommand{\mongodbCount}{57}
\newcommand{\opentelemetryCount}{55}
\newcommand{\adobeCount}{50}
\newcommand{\tencentCount}{49}
\newcommand{\jetbrainsCount}{48}
\newcommand{\AllalibabaCount}{47}
\newcommand{\AllcloudflareCount}{47}
\newcommand{\AllvercelCount}{45}

\newcommand{\AllminSubsetSize}{40}
\newcommand{\AllminFrequency}{250}
\newcommand{\AlltotalProjects}{32\,134}
\newcommand{\AllprojectsUsingoneTools}{20\,108}
\newcommand{\AllprojectsUsingtwoTools}{7\,150}
\newcommand{\AllprojectsUsingthreeTools}{2\,806}
\newcommand{\AllprojectsUsingfourTools}{1\,232}
\newcommand{\AllprojectsUsingfiveTools}{499}
\newcommand{\AllprojectsUsingsixTools}{192}
\newcommand{\AllprojectsUsingsevenTools}{89}
\newcommand{\AllprojectsUsingeightTools}{37}
\newcommand{\AllprojectsUsingnineTools}{13}
\newcommand{\AllprojectsUsingonezeroTools}{3}
\newcommand{\AllprojectsUsingoneoneTools}{1}
\newcommand{\projectsUsingonethreeTools}{1}
\newcommand{\projectsUsingonefourTools}{1}
\newcommand{\projectsUsingonesevenTools}{2}
\newcommand{\AllmeanToolsPerProject}{1.6349971992282317}
\newcommand{\AllmedianToolsPerProject}{1.0}
\newcommand{\AllmaxToolsPerProject}{17}
\newcommand{\AllminToolsPerProject}{1}
\newcommand{\AllprojectsAfterFiltering}{31\,731}
\newcommand{\AllprojectsFilteredOut}{403}
\newcommand{\AlltoolsAnalyzed}{12}
\newcommand{\AlltotalIntersections}{485}
\newcommand{\AllprojectsInShownIntersections}{29\,595}
\newcommand{\AllprojectsInFilteredIntersections}{2\,136}
\newcommand{\AllprojectsWithMultipleTools}{50}
\newcommand{\AlllargestIntersectionSize}{8\,471}
\newcommand{\AlllargestIntersectionTools}{Copilot}
\newcommand{\AllintersectionCopilot}{Copilot (8471)}
\newcommand{\AllintersectionCopilotSize}{8\,471}
\newcommand{\AllintersectionCopilotToolCount}{1}
\newcommand{\AllintersectionClaudeCode}{Claude\_Code (7872)}
\newcommand{\AllintersectionClaudeCodeSize}{7\,872}
\newcommand{\AllintersectionClaudeCodeToolCount}{1}
\newcommand{\AllintersectionClaudeCodeCopilot}{Claude\_Code $\cap$ Copilot (2574)}
\newcommand{\AllintersectionClaudeCodeCopilotSize}{2\,574}
\newcommand{\AllintersectionClaudeCodeCopilotToolCount}{2}
\newcommand{\AllintersectionClaudeCodeGeneric}{Claude\_Code $\cap$ Generic (1113)}
\newcommand{\AllintersectionClaudeCodeGenericSize}{1\,113}
\newcommand{\AllintersectionClaudeCodeGenericToolCount}{2}
\newcommand{\AllintersectionGeneric}{Generic (1101)}
\newcommand{\AllintersectionGenericSize}{1\,101}
\newcommand{\AllintersectionGenericToolCount}{1}
\newcommand{\AllintersectionCursor}{Cursor (798)}
\newcommand{\AllintersectionCursorSize}{798}
\newcommand{\AllintersectionCursorToolCount}{1}
\newcommand{\AllintersectionClaudeCodeCopilotGeneric}{Claude\_Code $\cap$ Copilot $\cap$ Generic (716)}
\newcommand{\AllintersectionClaudeCodeCopilotGenericSize}{716}
\newcommand{\AllintersectionClaudeCodeCopilotGenericToolCount}{3}
\newcommand{\AllintersectionCodex}{Codex (653)}
\newcommand{\AllintersectionCodexSize}{653}
\newcommand{\AllintersectionCodexToolCount}{1}
\newcommand{\AllintersectionCopilotGeneric}{Copilot $\cap$ Generic (554)}
\newcommand{\AllintersectionCopilotGenericSize}{554}
\newcommand{\AllintersectionCopilotGenericToolCount}{2}
\newcommand{\AllintersectionClaudeCodeCursor}{Claude\_Code $\cap$ Cursor (499)}
\newcommand{\AllintersectionClaudeCodeCursorSize}{499}
\newcommand{\AllintersectionClaudeCodeCursorToolCount}{2}
\newcommand{\AllintersectionGemini}{Gemini (326)}
\newcommand{\AllintersectionGeminiSize}{326}
\newcommand{\AllintersectionGeminiToolCount}{1}
\newcommand{\intersectionCoderabbit}{Coderabbit (315)}
\newcommand{\intersectionCoderabbitSize}{315}
\newcommand{\intersectionCoderabbitToolCount}{1}
\newcommand{\intersectionJules}{Jules (287)}
\newcommand{\intersectionJulesSize}{287}
\newcommand{\intersectionJulesToolCount}{1}
\newcommand{\AllintersectionClaudeCodeCodex}{Claude\_Code $\cap$ Codex (279)}
\newcommand{\AllintersectionClaudeCodeCodexSize}{279}
\newcommand{\AllintersectionClaudeCodeCodexToolCount}{2}
\newcommand{\AllintersectionClaudeCodeCopilotCursor}{Claude\_Code $\cap$ Copilot $\cap$ Cursor (278)}
\newcommand{\AllintersectionClaudeCodeCopilotCursorSize}{278}
\newcommand{\AllintersectionClaudeCodeCopilotCursorToolCount}{3}
\newcommand{\intersectionCopilotCursor}{Copilot $\cap$ Cursor (238)}
\newcommand{\intersectionCopilotCursorSize}{238}
\newcommand{\intersectionCopilotCursorToolCount}{2}
\newcommand{\intersectionClaudeCodeCopilotCodex}{Claude\_Code $\cap$ Copilot $\cap$ Codex (219)}
\newcommand{\intersectionClaudeCodeCopilotCodexSize}{219}
\newcommand{\intersectionClaudeCodeCopilotCodexToolCount}{3}
\newcommand{\AllintersectionClaudeCodeCursorGeneric}{Claude\_Code $\cap$ Cursor $\cap$ Generic (214)}
\newcommand{\AllintersectionClaudeCodeCursorGenericSize}{214}
\newcommand{\AllintersectionClaudeCodeCursorGenericToolCount}{3}
\newcommand{\AllintersectionClaudeCodeCopilotCursorGeneric}{Claude\_Code $\cap$ Copilot $\cap$ Cursor $\cap$ Generic (210)}
\newcommand{\AllintersectionClaudeCodeCopilotCursorGenericSize}{210}
\newcommand{\AllintersectionClaudeCodeCopilotCursorGenericToolCount}{4}
\newcommand{\intersectionCopilotCodex}{Copilot $\cap$ Codex (204)}
\newcommand{\intersectionCopilotCodexSize}{204}
\newcommand{\intersectionCopilotCodexToolCount}{2}
\newcommand{\intersectionClaudeCodeCopilotCodexGeneric}{Claude\_Code $\cap$ Copilot $\cap$ Codex $\cap$ Generic (192)}
\newcommand{\intersectionClaudeCodeCopilotCodexGenericSize}{192}
\newcommand{\intersectionClaudeCodeCopilotCodexGenericToolCount}{4}
\newcommand{\intersectionAider}{Aider (188)}
\newcommand{\intersectionAiderSize}{188}
\newcommand{\intersectionAiderToolCount}{1}
\newcommand{\AllintersectionClaudeCodeCodexGeneric}{Claude\_Code $\cap$ Codex $\cap$ Generic (185)}
\newcommand{\AllintersectionClaudeCodeCodexGenericSize}{185}
\newcommand{\AllintersectionClaudeCodeCodexGenericToolCount}{3}
\newcommand{\intersectionCopilotCoderabbit}{Copilot $\cap$ Coderabbit (156)}
\newcommand{\intersectionCopilotCoderabbitSize}{156}
\newcommand{\intersectionCopilotCoderabbitToolCount}{2}
\newcommand{\AllintersectionCodexGeneric}{Codex $\cap$ Generic (144)}
\newcommand{\AllintersectionCodexGenericSize}{144}
\newcommand{\AllintersectionCodexGenericToolCount}{2}
\newcommand{\intersectionCopilotGemini}{Copilot $\cap$ Gemini (133)}
\newcommand{\intersectionCopilotGeminiSize}{133}
\newcommand{\intersectionCopilotGeminiToolCount}{2}
\newcommand{\intersectionClaudeCodeCoderabbit}{Claude\_Code $\cap$ Coderabbit (125)}
\newcommand{\intersectionClaudeCodeCoderabbitSize}{125}
\newcommand{\intersectionClaudeCodeCoderabbitToolCount}{2}
\newcommand{\AllintersectionClaudeCodeGemini}{Claude\_Code $\cap$ Gemini (114)}
\newcommand{\AllintersectionClaudeCodeGeminiSize}{114}
\newcommand{\AllintersectionClaudeCodeGeminiToolCount}{2}
\newcommand{\intersectionClaudeCodeCopilotCoderabbit}{Claude\_Code $\cap$ Copilot $\cap$ Coderabbit (98)}
\newcommand{\intersectionClaudeCodeCopilotCoderabbitSize}{98}
\newcommand{\intersectionClaudeCodeCopilotCoderabbitToolCount}{3}
\newcommand{\intersectionClaudeCodeCopilotCodexCursorGeneric}{Claude\_Code $\cap$ Copilot $\cap$ Codex $\cap$ Cursor $\cap$ Generic (95)}
\newcommand{\intersectionClaudeCodeCopilotCodexCursorGenericSize}{95}
\newcommand{\intersectionClaudeCodeCopilotCodexCursorGenericToolCount}{5}
\newcommand{\intersectionCopilotCodexGeneric}{Copilot $\cap$ Codex $\cap$ Generic (92)}
\newcommand{\intersectionCopilotCodexGenericSize}{92}
\newcommand{\intersectionCopilotCodexGenericToolCount}{3}
\newcommand{\intersectionDevin}{Devin (91)}
\newcommand{\intersectionDevinSize}{91}
\newcommand{\intersectionDevinToolCount}{1}
\newcommand{\AllintersectionCursorGeneric}{Cursor $\cap$ Generic (85)}
\newcommand{\AllintersectionCursorGenericSize}{85}
\newcommand{\AllintersectionCursorGenericToolCount}{2}
\newcommand{\intersectionClaudeCodeCopilotGeminiGeneric}{Claude\_Code $\cap$ Copilot $\cap$ Gemini $\cap$ Generic (79)}
\newcommand{\intersectionClaudeCodeCopilotGeminiGenericSize}{79}
\newcommand{\intersectionClaudeCodeCopilotGeminiGenericToolCount}{4}
\newcommand{\intersectionCopilotJules}{Copilot $\cap$ Jules (78)}
\newcommand{\intersectionCopilotJulesSize}{78}
\newcommand{\intersectionCopilotJulesToolCount}{2}
\newcommand{\intersectionClaudeCodeCopilotGemini}{Claude\_Code $\cap$ Copilot $\cap$ Gemini (72)}
\newcommand{\intersectionClaudeCodeCopilotGeminiSize}{72}
\newcommand{\intersectionClaudeCodeCopilotGeminiToolCount}{3}
\newcommand{\intersectionJulesGemini}{Jules $\cap$ Gemini (71)}
\newcommand{\intersectionJulesGeminiSize}{71}
\newcommand{\intersectionJulesGeminiToolCount}{2}
\newcommand{\AllintersectionClaudeCodeGeminiGeneric}{Claude\_Code $\cap$ Gemini $\cap$ Generic (68)}
\newcommand{\AllintersectionClaudeCodeGeminiGenericSize}{68}
\newcommand{\AllintersectionClaudeCodeGeminiGenericToolCount}{3}
\newcommand{\intersectionClaudeCodeJules}{Claude\_Code $\cap$ Jules (67)}
\newcommand{\intersectionClaudeCodeJulesSize}{67}
\newcommand{\intersectionClaudeCodeJulesToolCount}{2}
\newcommand{\intersectionClaudeCodeAider}{Claude\_Code $\cap$ Aider (66)}
\newcommand{\intersectionClaudeCodeAiderSize}{66}
\newcommand{\intersectionClaudeCodeAiderToolCount}{2}
\newcommand{\AllintersectionClaudeCodeCodexCursorGeneric}{Claude\_Code $\cap$ Codex $\cap$ Cursor $\cap$ Generic (59)}
\newcommand{\AllintersectionClaudeCodeCodexCursorGenericSize}{59}
\newcommand{\AllintersectionClaudeCodeCodexCursorGenericToolCount}{4}
\newcommand{\intersectionAmp}{Amp (53)}
\newcommand{\intersectionAmpSize}{53}
\newcommand{\intersectionAmpToolCount}{1}
\newcommand{\intersectionClaudeCodeDevin}{Claude\_Code $\cap$ Devin (51)}
\newcommand{\intersectionClaudeCodeDevinSize}{51}
\newcommand{\intersectionClaudeCodeDevinToolCount}{2}
\newcommand{\intersectionClaudeCodeAmp}{Claude\_Code $\cap$ Amp (49)}
\newcommand{\intersectionClaudeCodeAmpSize}{49}
\newcommand{\intersectionClaudeCodeAmpToolCount}{2}
\newcommand{\intersectionClaudeCodeCodexCursor}{Claude\_Code $\cap$ Codex $\cap$ Cursor (48)}
\newcommand{\intersectionClaudeCodeCodexCursorSize}{48}
\newcommand{\intersectionClaudeCodeCodexCursorToolCount}{3}
\newcommand{\intersectionClaudeCodeCopilotCoderabbitGeneric}{Claude\_Code $\cap$ Copilot $\cap$ Coderabbit $\cap$ Generic (47)}
\newcommand{\intersectionClaudeCodeCopilotCoderabbitGenericSize}{47}
\newcommand{\intersectionClaudeCodeCopilotCoderabbitGenericToolCount}{4}
\newcommand{\intersectionCopilotCursorGeneric}{Copilot $\cap$ Cursor $\cap$ Generic (44)}
\newcommand{\intersectionCopilotCursorGenericSize}{44}
\newcommand{\intersectionCopilotCursorGenericToolCount}{3}
\newcommand{\intersectionClaudeCodeCoderabbitGeneric}{Claude\_Code $\cap$ Coderabbit $\cap$ Generic (42)}
\newcommand{\intersectionClaudeCodeCoderabbitGenericSize}{42}
\newcommand{\intersectionClaudeCodeCoderabbitGenericToolCount}{3}
\newcommand{\intersectionClaudeCodeCopilotJules}{Claude\_Code $\cap$ Copilot $\cap$ Jules (41)}
\newcommand{\intersectionClaudeCodeCopilotJulesSize}{41}
\newcommand{\intersectionClaudeCodeCopilotJulesToolCount}{3}
\newcommand{\intersectionClaudeCodeCopilotCodexCursor}{Claude\_Code $\cap$ Copilot $\cap$ Codex $\cap$ Cursor (41)}
\newcommand{\intersectionClaudeCodeCopilotCodexCursorSize}{41}
\newcommand{\intersectionClaudeCodeCopilotCodexCursorToolCount}{4}
\newcommand{\AllmeanIntersectionSize}{591}
\newcommand{\AllmedianIntersectionSize}{138}
\newcommand{\AllshownIntersectionsoneTools}{11}
\newcommand{\AllshownIntersectionstwoTools}{19}
\newcommand{\AllshownIntersectionsthreeTools}{13}
\newcommand{\AllshownIntersectionsfourTools}{6}
\newcommand{\AllshownIntersectionsfiveTools}{1}

\begin{figure}[htbp]
    \centering
    \includegraphics[width=0.8\textwidth]{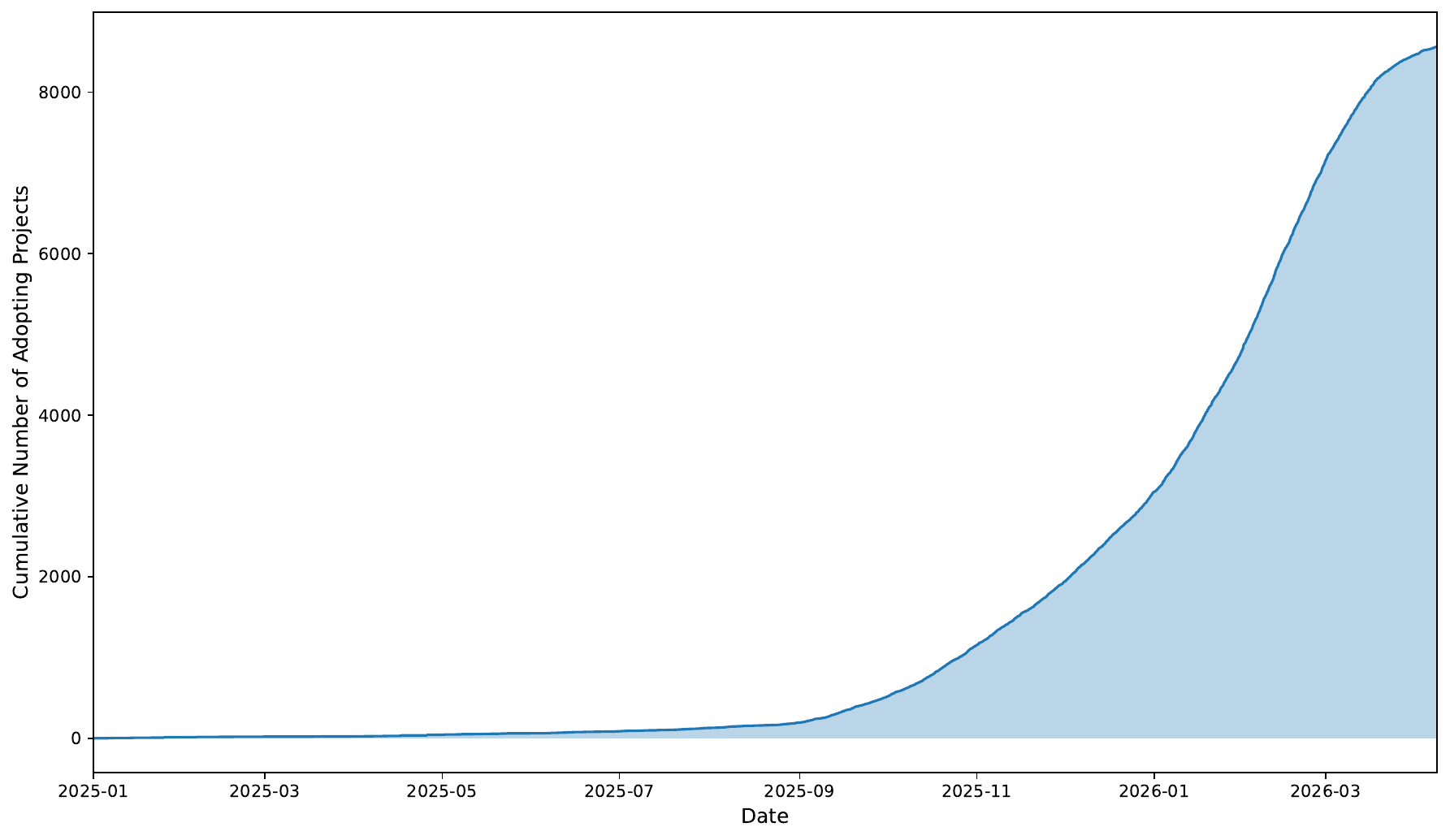}
    \caption{Cumulative adoption of any tool across all \VerytotalAdoptingProjects\ projects from \VeryfirstAdoptionDate\ to \VerylastAdoptionDate.}
    \label{fig:adoption-very}

    \vspace{1cm} 

    \includegraphics[width=0.8\textwidth]{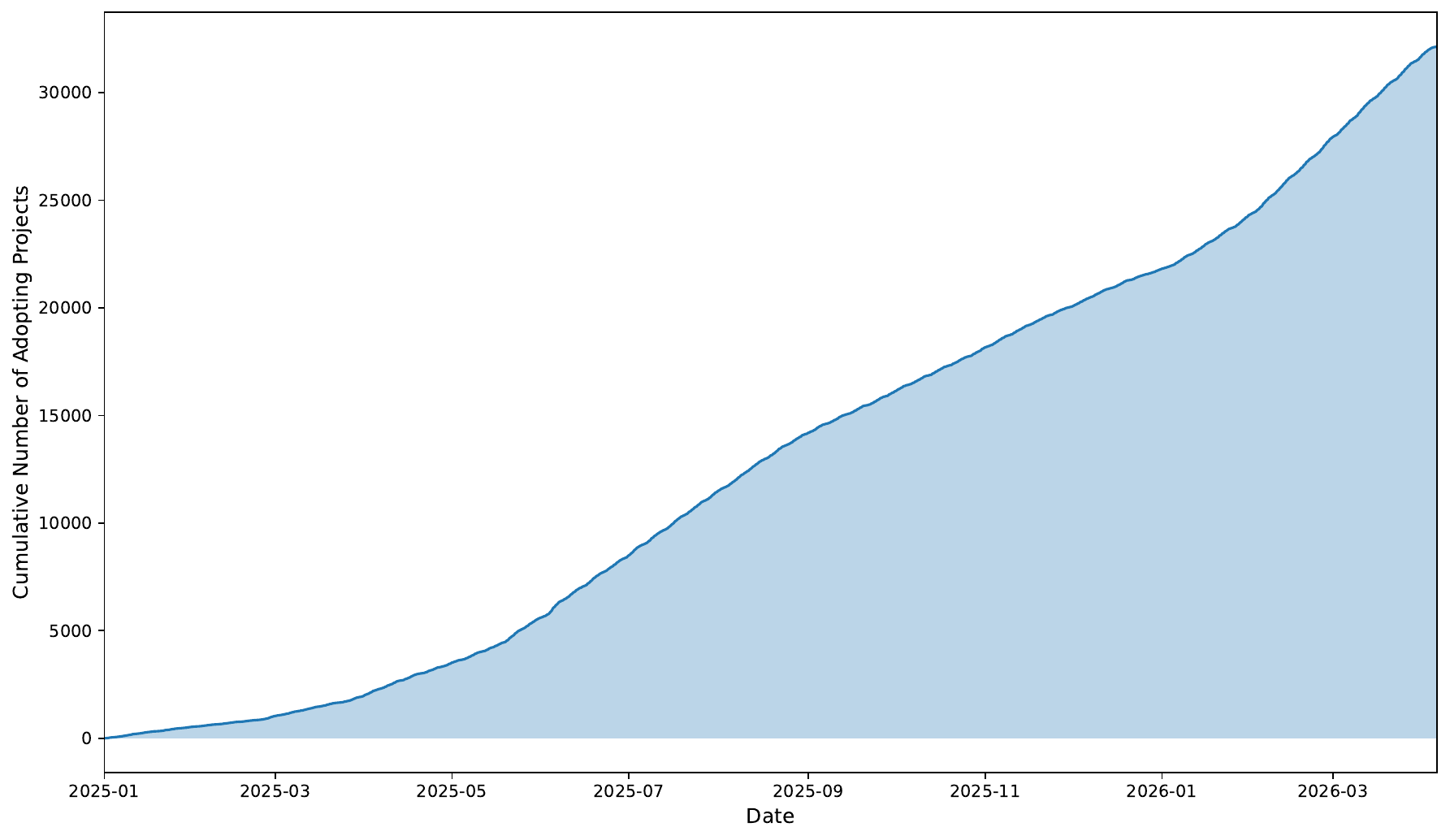}
    \caption{Cumulative adoption of any tool across all \AlltotalAdoptingProjects\ projects from \AllfirstAdoptionDate\ to \AlllastAdoptionDate.}
    \label{fig:adoption-all}
\end{figure}

\begin{figure}[htbp]
    \centering
    \includegraphics[width=0.8\textwidth]{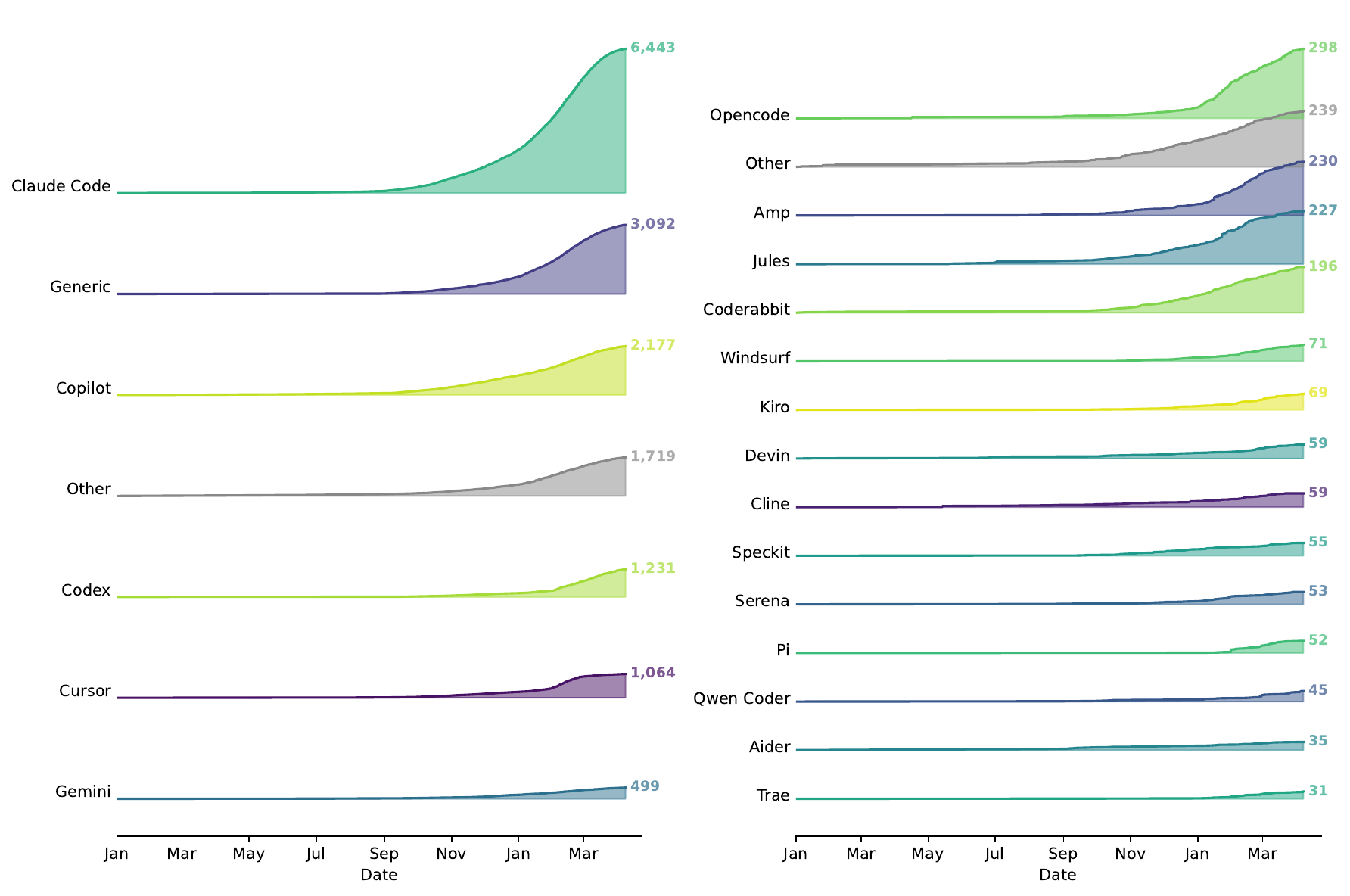}
    \caption{Ridgeline plot showing adoption patterns for the \VerydisplayedTools\ tools, with \VerytopToolName\ being the most adopted (\VerytopToolCount\ projects).}
    \label{fig:tooladoption-very}

    \vspace{1cm} 

    \includegraphics[width=0.8\textwidth]{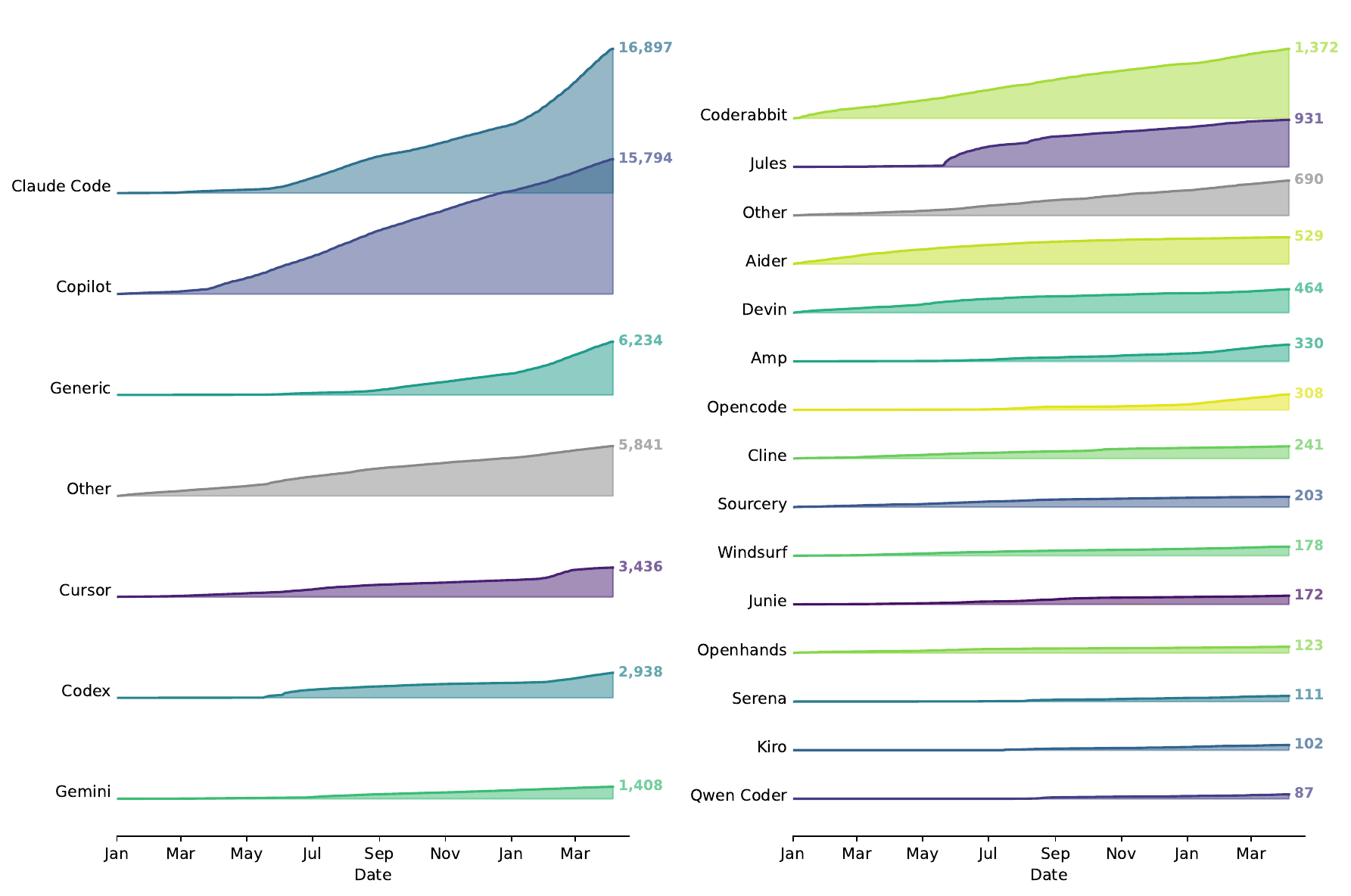}
    \caption{Ridgeline plot showing adoption patterns for the \AlldisplayedTools\ tools, with \AlltopToolName\ being the most adopted (\AlltopToolCount\ projects).}
    \label{fig:tooladoption-all}
\end{figure}

\Cref{fig:tooladoption-very} and \Cref{fig:tooladoption-all} contrast the adoption timeline for both newer and older projects, where the adoption is plotted for each tool. We can clearly see that newer projects have some differences in terms of tool preferences. In particular, Claude Code is the most popular tool in both dataset, but it is by a short margin ($\approx$ 6\%) over Copilot in the older projects, while Copilot it a distant third for newer projects. In this corpus, Claude Code is three times more popular. The second category for newer projects is the ``Generic'' category (mostly identified by their use of AGENTS.md).  This includes a sizeable portion of Codex users. Considering the explicit uses of Codex as well, it is likely that Codex is the second most popular tool for the newer projects, although a distant second to Claude Code.

Beyond Claude Code and Codex, other tools that see a comparatively higher usage in newer projects include OpenCode, Amp, and Kiro.

\begin{figure}[htbp]
    \centering
    \includegraphics[width=0.8\textwidth]{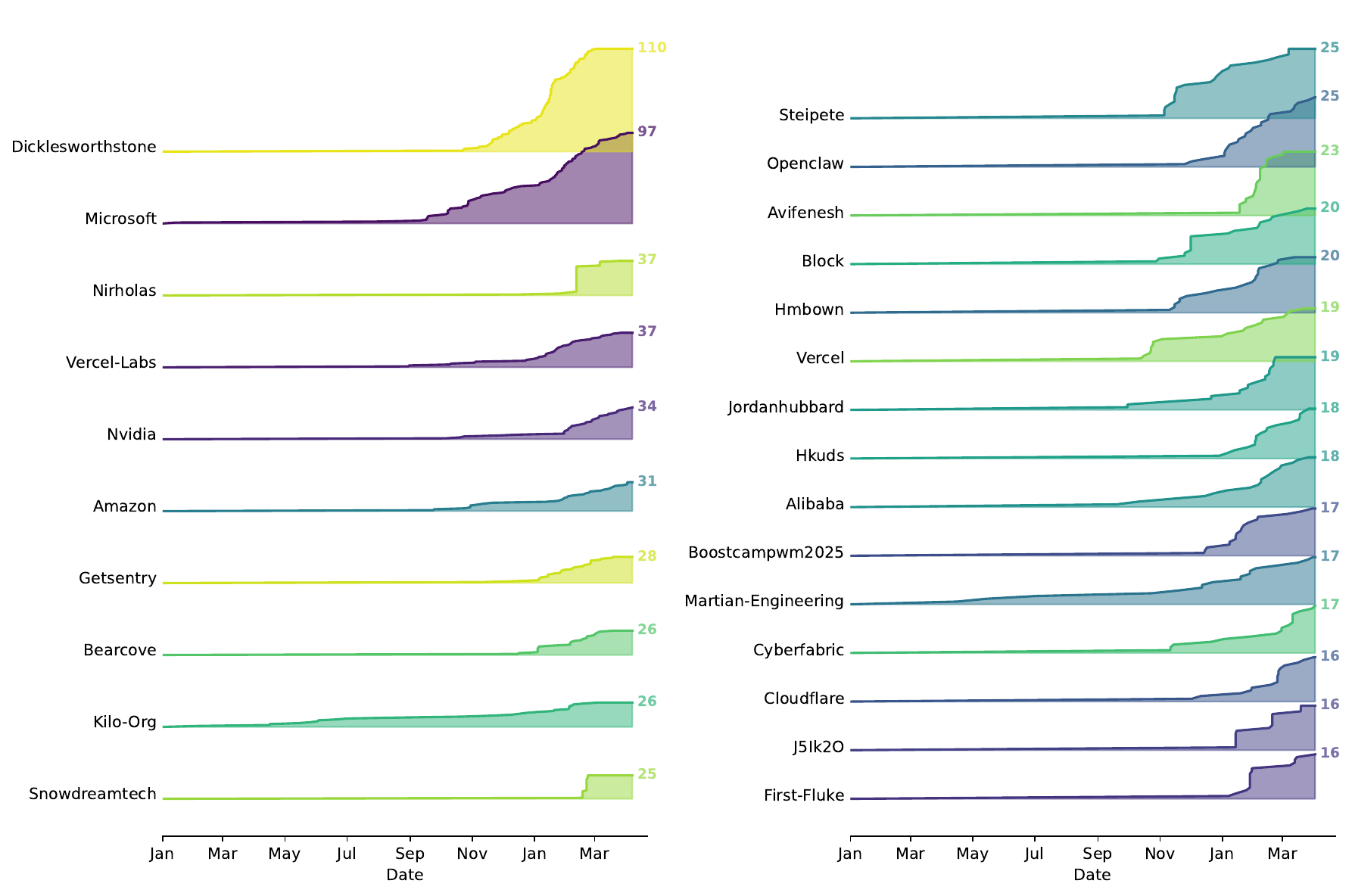}
    \caption{Adoption patterns across the top \VerydisplayedOrganizations\ organizations, with \VerytopOrganizationName\ leading at \VerytopOrganizationCount\ adoptions.}
    \label{fig:orgadoption-very}

    \vspace{1cm} 

    \includegraphics[width=0.8\textwidth]{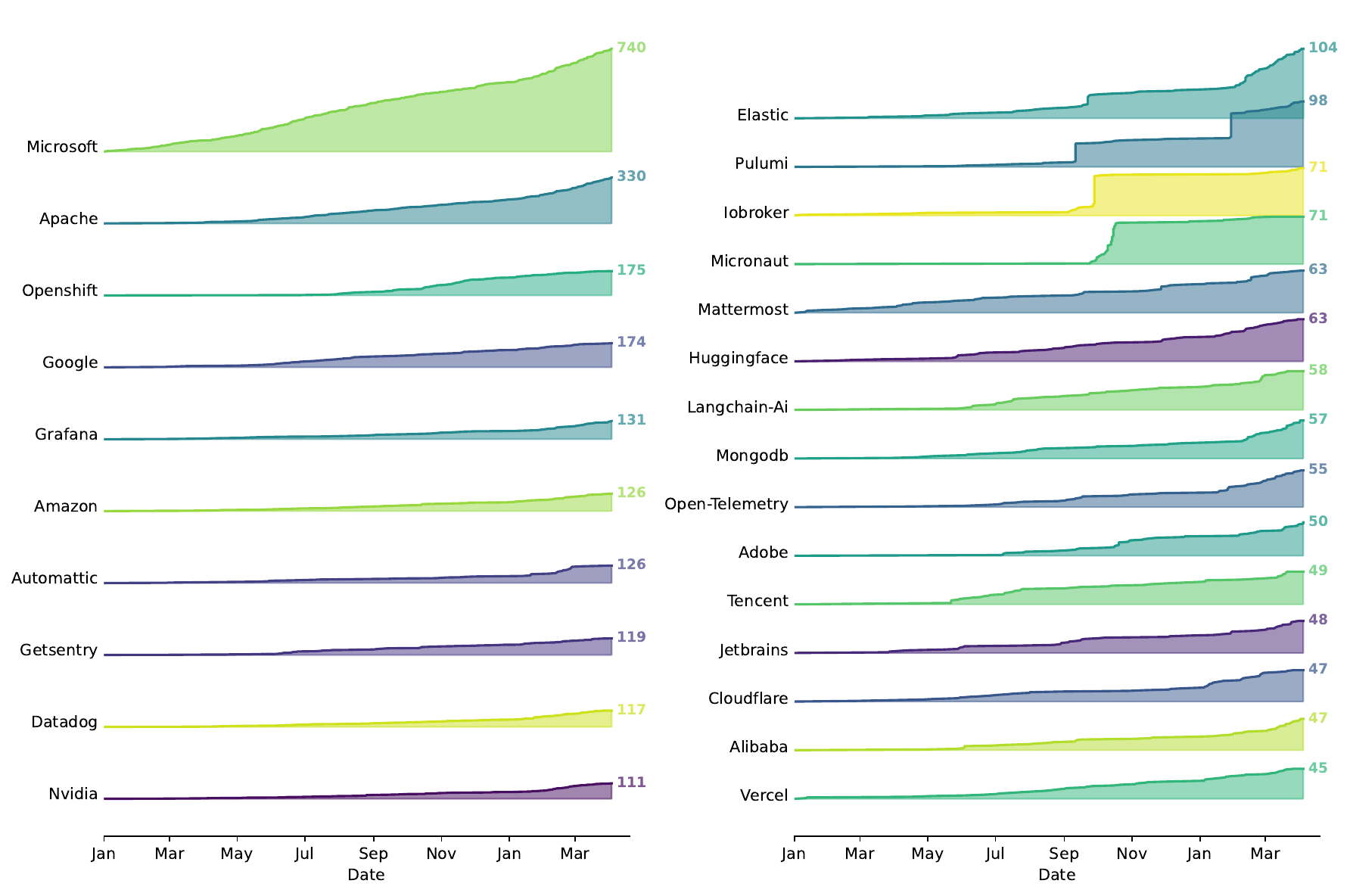}
    \caption{Adoption patterns across the top \AlldisplayedOrganizations\ organizations, with \AlltopOrganizationName\ leading at \AlltopOrganizationCount\ adoptions.}
    \label{fig:orgadoption-all}
\end{figure}

\Cref{fig:orgadoption-very} and \Cref{fig:orgadoption-all} contrast the adoption timeline for both newer and older projects, where the adoption is the number of repositories showing traces of adoption over time, for each entity. We see a rather surprising finding here: while for older projects, the entities that are plotted are established organizations (e.g. Microsoft, Apache, Google, Grafana, etc), for newer projects, a sizeable portion of the entities with the largest adoption (the largest number of adopting repositories) are \emph{individuals}. This is the case of the entity with the most repositories (``Dicklesworthstone''): during the study period, this single developer created more repositories with coding agent traces than Microsoft. Several other developers have the same behaviour, albeit to a lesser extent.

\section{RQ5: Size of AI-Assisted Contributions}

We analyze how tool adoption affects commit sizes across the studied repositories.


\newcommand{\VeryAddedLinesHumanMedian}{29}
\newcommand{\VeryAddedLinesHumanQOne}{6}
\newcommand{\VeryAddedLinesHumanQThree}{116}
\newcommand{\VeryAddedLinesHumanUpperWhisker}{281}

\newcommand{\VeryAddedLinesBotMedian}{6}
\newcommand{\VeryAddedLinesBotQOne}{1}
\newcommand{\VeryAddedLinesBotQThree}{20}
\newcommand{\VeryAddedLinesBotUpperWhisker}{48}

\newcommand{\VeryAddedLinesAIMedian}{42}
\newcommand{\VeryAddedLinesAIQOne}{9}
\newcommand{\VeryAddedLinesAIQThree}{156}
\newcommand{\VeryAddedLinesAIUpperWhisker}{376}

\newcommand{\VeryDeletedLinesHumanMedian}{9}
\newcommand{\VeryDeletedLinesHumanQOne}{3}
\newcommand{\VeryDeletedLinesHumanQThree}{30}
\newcommand{\VeryDeletedLinesHumanUpperWhisker}{70}

\newcommand{\VeryDeletedLinesBotMedian}{2}
\newcommand{\VeryDeletedLinesBotQOne}{1}
\newcommand{\VeryDeletedLinesBotQThree}{7}
\newcommand{\VeryDeletedLinesBotUpperWhisker}{16}

\newcommand{\VeryDeletedLinesAIMedian}{8}
\newcommand{\VeryDeletedLinesAIQOne}{3}
\newcommand{\VeryDeletedLinesAIQThree}{25}
\newcommand{\VeryDeletedLinesAIUpperWhisker}{58}

\newcommand{\VeryTotalFilesHumanMedian}{2}
\newcommand{\VeryTotalFilesHumanQOne}{1}
\newcommand{\VeryTotalFilesHumanQThree}{5}
\newcommand{\VeryTotalFilesHumanUpperWhisker}{11}

\newcommand{\VeryTotalFilesBotMedian}{1}
\newcommand{\VeryTotalFilesBotQOne}{1}
\newcommand{\VeryTotalFilesBotQThree}{2}
\newcommand{\VeryTotalFilesBotUpperWhisker}{4}

\newcommand{\VeryTotalFilesAIMedian}{2}
\newcommand{\VeryTotalFilesAIQOne}{1}
\newcommand{\VeryTotalFilesAIQThree}{4}
\newcommand{\VeryTotalFilesAIUpperWhisker}{8}


\newcommand{\VeryAddedLinesHumanBinOneToFive}{18.46}
\newcommand{\VeryAddedLinesHumanBinSixToTwoZero}{18.58}
\newcommand{\VeryAddedLinesHumanBinTwoOneToFiveZero}{14.21}
\newcommand{\VeryAddedLinesHumanBinFiveOneToOneZeroZero}{10.99}
\newcommand{\VeryAddedLinesHumanBinOneZeroOneToFiveZeroZero}{22.41}
\newcommand{\VeryAddedLinesHumanBinFiveZeroOneToOneZeroZeroZero}{6.58}
\newcommand{\VeryAddedLinesHumanBinOneZeroZeroOnePlus}{8.77}
\newcommand{\VeryAddedLinesBotBinOneToFive}{39.92}
\newcommand{\VeryAddedLinesBotBinSixToTwoZero}{22.88}
\newcommand{\VeryAddedLinesBotBinTwoOneToFiveZero}{21.07}
\newcommand{\VeryAddedLinesBotBinFiveOneToOneZeroZero}{4.66}
\newcommand{\VeryAddedLinesBotBinOneZeroOneToFiveZeroZero}{6.36}
\newcommand{\VeryAddedLinesBotBinFiveZeroOneToOneZeroZeroZero}{1.41}
\newcommand{\VeryAddedLinesBotBinOneZeroZeroOnePlus}{3.70}
\newcommand{\VeryAddedLinesAIBinOneToFive}{14.11}
\newcommand{\VeryAddedLinesAIBinSixToTwoZero}{17.36}
\newcommand{\VeryAddedLinesAIBinTwoOneToFiveZero}{14.50}
\newcommand{\VeryAddedLinesAIBinFiveOneToOneZeroZero}{11.62}
\newcommand{\VeryAddedLinesAIBinOneZeroOneToFiveZeroZero}{25.30}
\newcommand{\VeryAddedLinesAIBinFiveZeroOneToOneZeroZeroZero}{7.61}
\newcommand{\VeryAddedLinesAIBinOneZeroZeroOnePlus}{9.49}

\newcommand{\VeryDeletedLinesHumanBinOneToFive}{31.00}
\newcommand{\VeryDeletedLinesHumanBinSixToTwoZero}{25.61}
\newcommand{\VeryDeletedLinesHumanBinTwoOneToFiveZero}{14.88}
\newcommand{\VeryDeletedLinesHumanBinFiveOneToOneZeroZero}{8.93}
\newcommand{\VeryDeletedLinesHumanBinOneZeroOneToFiveZeroZero}{12.91}
\newcommand{\VeryDeletedLinesHumanBinFiveZeroOneToOneZeroZeroZero}{2.87}
\newcommand{\VeryDeletedLinesHumanBinOneZeroZeroOnePlus}{3.80}
\newcommand{\VeryDeletedLinesBotBinOneToFive}{54.99}
\newcommand{\VeryDeletedLinesBotBinSixToTwoZero}{22.86}
\newcommand{\VeryDeletedLinesBotBinTwoOneToFiveZero}{8.09}
\newcommand{\VeryDeletedLinesBotBinFiveOneToOneZeroZero}{4.07}
\newcommand{\VeryDeletedLinesBotBinOneZeroOneToFiveZeroZero}{5.75}
\newcommand{\VeryDeletedLinesBotBinFiveZeroOneToOneZeroZeroZero}{1.29}
\newcommand{\VeryDeletedLinesBotBinOneZeroZeroOnePlus}{2.95}
\newcommand{\VeryDeletedLinesAIBinOneToFive}{31.31}
\newcommand{\VeryDeletedLinesAIBinSixToTwoZero}{28.25}
\newcommand{\VeryDeletedLinesAIBinTwoOneToFiveZero}{15.52}
\newcommand{\VeryDeletedLinesAIBinFiveOneToOneZeroZero}{8.63}
\newcommand{\VeryDeletedLinesAIBinOneZeroOneToFiveZeroZero}{11.00}
\newcommand{\VeryDeletedLinesAIBinFiveZeroOneToOneZeroZeroZero}{2.19}
\newcommand{\VeryDeletedLinesAIBinOneZeroZeroOnePlus}{3.10}

\newcommand{\VeryTotalFilesHumanBinOne}{31.69}
\newcommand{\VeryTotalFilesHumanBinTwo}{18.21}
\newcommand{\VeryTotalFilesHumanBinThreeToFive}{15.84}
\newcommand{\VeryTotalFilesHumanBinSixToOneZero}{14.90}
\newcommand{\VeryTotalFilesHumanBinOneOneToTwoZero}{9.72}
\newcommand{\VeryTotalFilesHumanBinTwoOneToFiveZero}{6.40}
\newcommand{\VeryTotalFilesHumanBinFiveOnePlus}{3.24}
\newcommand{\VeryTotalFilesBotBinOne}{50.71}
\newcommand{\VeryTotalFilesBotBinTwo}{21.18}
\newcommand{\VeryTotalFilesBotBinThreeToFive}{12.18}
\newcommand{\VeryTotalFilesBotBinSixToOneZero}{8.23}
\newcommand{\VeryTotalFilesBotBinOneOneToTwoZero}{5.27}
\newcommand{\VeryTotalFilesBotBinTwoOneToFiveZero}{1.68}
\newcommand{\VeryTotalFilesBotBinFiveOnePlus}{0.75}
\newcommand{\VeryTotalFilesAIBinOne}{31.44}
\newcommand{\VeryTotalFilesAIBinTwo}{20.33}
\newcommand{\VeryTotalFilesAIBinThreeToFive}{18.22}
\newcommand{\VeryTotalFilesAIBinSixToOneZero}{15.52}
\newcommand{\VeryTotalFilesAIBinOneOneToTwoZero}{8.40}
\newcommand{\VeryTotalFilesAIBinTwoOneToFiveZero}{4.28}
\newcommand{\TotalFilesAIBinFiveOnePlus}{1.81}


\newcommand{\AllAddedLinesHumanMedian}{10}
\newcommand{\AllAddedLinesHumanQOne}{3}
\newcommand{\AllAddedLinesHumanQThree}{42}
\newcommand{\AllAddedLinesHumanUpperWhisker}{100}

\newcommand{\AllAddedLinesBotMedian}{4}
\newcommand{\AllAddedLinesBotQOne}{1}
\newcommand{\AllAddedLinesBotQThree}{10}
\newcommand{\AllAddedLinesBotUpperWhisker}{24}

\newcommand{\AllAddedLinesAIMedian}{30}
\newcommand{\AllAddedLinesAIQOne}{6}
\newcommand{\AllAddedLinesAIQThree}{110}
\newcommand{\AllAddedLinesAIUpperWhisker}{266}

\newcommand{\AllDeletedLinesHumanMedian}{5}
\newcommand{\AllDeletedLinesHumanQOne}{2}
\newcommand{\AllDeletedLinesHumanQThree}{17}
\newcommand{\AllDeletedLinesHumanUpperWhisker}{40}

\newcommand{\AllDeletedLinesBotMedian}{3}
\newcommand{\AllDeletedLinesBotQOne}{1}
\newcommand{\AllDeletedLinesBotQThree}{8}
\newcommand{\AllDeletedLinesBotUpperWhisker}{18}

\newcommand{\AllDeletedLinesAIMedian}{7}
\newcommand{\AllDeletedLinesAIQOne}{2}
\newcommand{\AllDeletedLinesAIQThree}{23}
\newcommand{\AllDeletedLinesAIUpperWhisker}{54}

\newcommand{\AllTotalFilesHumanMedian}{2}
\newcommand{\AllTotalFilesHumanQOne}{1}
\newcommand{\AllTotalFilesHumanQThree}{3}
\newcommand{\AllTotalFilesHumanUpperWhisker}{6}

\newcommand{\AllTotalFilesBotMedian}{1}
\newcommand{\AllTotalFilesBotQOne}{1}
\newcommand{\AllTotalFilesBotQThree}{2}
\newcommand{\AllTotalFilesBotUpperWhisker}{3}

\newcommand{\AllTotalFilesAIMedian}{2}
\newcommand{\AllTotalFilesAIQOne}{1}
\newcommand{\AllTotalFilesAIQThree}{4}
\newcommand{\AllTotalFilesAIUpperWhisker}{8}


\newcommand{\AllAddedLinesHumanBinOneToFive}{29.43}
\newcommand{\AllAddedLinesHumanBinSixToTwoZero}{23.05}
\newcommand{\AllAddedLinesHumanBinTwoOneToFiveZero}{14.33}
\newcommand{\AllAddedLinesHumanBinFiveOneToOneZeroZero}{9.86}
\newcommand{\AllAddedLinesHumanBinOneZeroOneToFiveZeroZero}{15.92}
\newcommand{\AllAddedLinesHumanBinFiveZeroOneToOneZeroZeroZero}{3.38}
\newcommand{\AllAddedLinesHumanBinOneZeroZeroOnePlus}{4.03}
\newcommand{\AllAddedLinesBotBinOneToFive}{43.41}
\newcommand{\AllAddedLinesBotBinSixToTwoZero}{28.03}
\newcommand{\AllAddedLinesBotBinTwoOneToFiveZero}{9.45}
\newcommand{\AllAddedLinesBotBinFiveOneToOneZeroZero}{6.14}
\newcommand{\AllAddedLinesBotBinOneZeroOneToFiveZeroZero}{8.04}
\newcommand{\AllAddedLinesBotBinFiveZeroOneToOneZeroZeroZero}{1.52}
\newcommand{\AllAddedLinesBotBinOneZeroZeroOnePlus}{3.42}
\newcommand{\AllAddedLinesAIBinOneToFive}{18.35}
\newcommand{\AllAddedLinesAIBinSixToTwoZero}{18.62}
\newcommand{\AllAddedLinesAIBinTwoOneToFiveZero}{14.70}
\newcommand{\AllAddedLinesAIBinFiveOneToOneZeroZero}{11.67}
\newcommand{\AllAddedLinesAIBinOneZeroOneToFiveZeroZero}{23.19}
\newcommand{\AllAddedLinesAIBinFiveZeroOneToOneZeroZeroZero}{6.09}
\newcommand{\AllAddedLinesAIBinOneZeroZeroOnePlus}{7.38}

\newcommand{\AllDeletedLinesHumanBinOneToFive}{40.34}
\newcommand{\AllDeletedLinesHumanBinSixToTwoZero}{25.28}
\newcommand{\AllDeletedLinesHumanBinTwoOneToFiveZero}{12.90}
\newcommand{\AllDeletedLinesHumanBinFiveOneToOneZeroZero}{7.50}
\newcommand{\AllDeletedLinesHumanBinOneZeroOneToFiveZeroZero}{9.78}
\newcommand{\AllDeletedLinesHumanBinFiveZeroOneToOneZeroZeroZero}{1.81}
\newcommand{\AllDeletedLinesHumanBinOneZeroZeroOnePlus}{2.38}
\newcommand{\AllDeletedLinesBotBinOneToFive}{48.22}
\newcommand{\AllDeletedLinesBotBinSixToTwoZero}{26.76}
\newcommand{\AllDeletedLinesBotBinTwoOneToFiveZero}{8.39}
\newcommand{\AllDeletedLinesBotBinFiveOneToOneZeroZero}{5.59}
\newcommand{\AllDeletedLinesBotBinOneZeroOneToFiveZeroZero}{6.99}
\newcommand{\AllDeletedLinesBotBinFiveZeroOneToOneZeroZeroZero}{1.18}
\newcommand{\AllDeletedLinesBotBinOneZeroZeroOnePlus}{2.87}
\newcommand{\AllDeletedLinesAIBinOneToFive}{34.89}
\newcommand{\AllDeletedLinesAIBinSixToTwoZero}{26.17}
\newcommand{\AllDeletedLinesAIBinTwoOneToFiveZero}{14.37}
\newcommand{\AllDeletedLinesAIBinFiveOneToOneZeroZero}{8.16}
\newcommand{\AllDeletedLinesAIBinOneZeroOneToFiveZeroZero}{10.90}
\newcommand{\AllDeletedLinesAIBinFiveZeroOneToOneZeroZeroZero}{2.28}
\newcommand{\AllDeletedLinesAIBinOneZeroZeroOnePlus}{3.22}

\newcommand{\AllTotalFilesHumanBinOne}{41.08}
\newcommand{\AllTotalFilesHumanBinTwo}{18.85}
\newcommand{\AllTotalFilesHumanBinThreeToFive}{15.52}
\newcommand{\AllTotalFilesHumanBinSixToOneZero}{12.51}
\newcommand{\AllTotalFilesHumanBinOneOneToTwoZero}{6.59}
\newcommand{\AllTotalFilesHumanBinTwoOneToFiveZero}{3.62}
\newcommand{\AllTotalFilesHumanBinFiveOnePlus}{1.84}
\newcommand{\AllTotalFilesBotBinOne}{41.78}
\newcommand{\AllTotalFilesBotBinTwo}{34.00}
\newcommand{\AllTotalFilesBotBinThreeToFive}{10.68}
\newcommand{\AllTotalFilesBotBinSixToOneZero}{5.99}
\newcommand{\AllTotalFilesBotBinOneOneToTwoZero}{3.88}
\newcommand{\AllTotalFilesBotBinTwoOneToFiveZero}{1.95}
\newcommand{\AllTotalFilesBotBinFiveOnePlus}{1.72}
\newcommand{\AllTotalFilesAIBinOne}{35.67}
\newcommand{\AllTotalFilesAIBinTwo}{18.00}
\newcommand{\AllTotalFilesAIBinThreeToFive}{17.14}
\newcommand{\AllTotalFilesAIBinSixToOneZero}{14.44}
\newcommand{\AllTotalFilesAIBinOneOneToTwoZero}{7.84}
\newcommand{\AllTotalFilesAIBinTwoOneToFiveZero}{4.53}
\newcommand{\AllTotalFilesAIBinFiveOnePlus}{2.38}

\begin{figure}[htbp]
    \centering
    \includegraphics[width=0.8\textwidth]{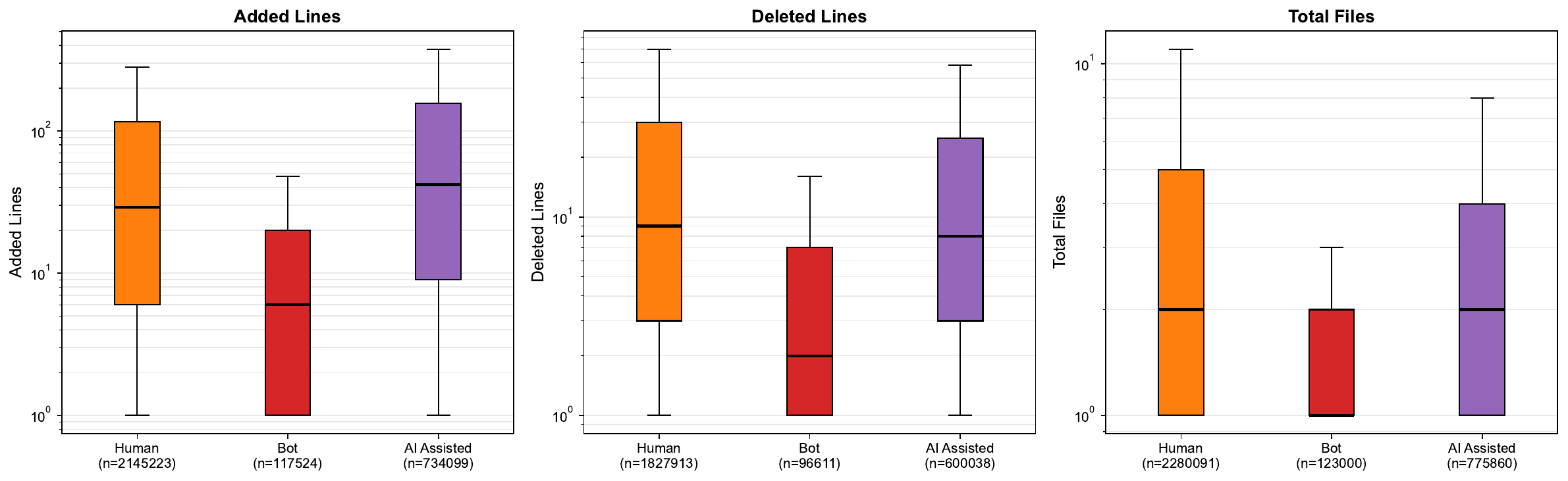}
    \caption{Boxplot comparison of added lines, deleted lines, and total files across human, bot, and AI-assisted commits for newer projects.}
    \label{fig:commit-sizes-very}
        \vspace{1cm} 
  \includegraphics[width=0.8\textwidth]{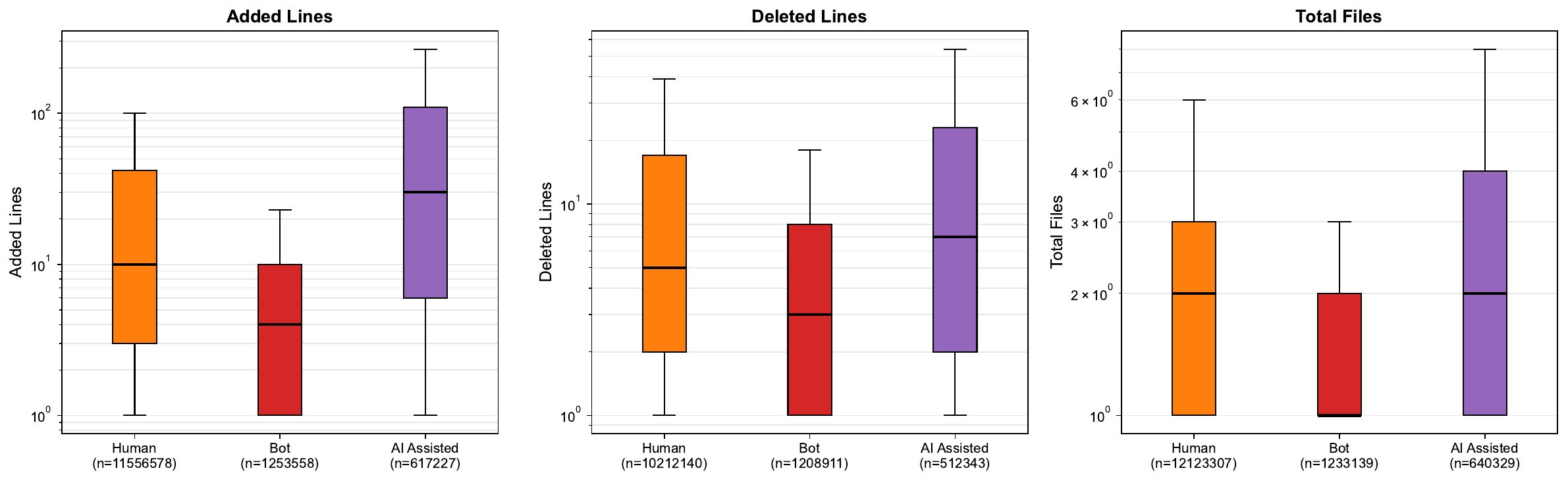}
    \caption{Boxplot comparison of added lines, deleted lines, and total files across human, bot, and AI-assisted commits for older projects.}
    \label{fig:commit-sizes-all}
\end{figure}

\Cref{fig:commit-sizes-very} and \Cref{fig:commit-sizes-all} contrast the sizes of contributions for both newer and older projects. We see a counter-intuitive behaviour: for newer projects, the size difference between AI-assisted and human-authored commits appears \emph{smaller} than for older projects. Looking into the details, the AI commits are actually \emph{larger} for newer projects, in terms of lines added: the median for the older projects is \AllAddedLinesAIMedian, versus \VeryAddedLinesAIMedian \ for newer projects, with a third quartile of \AllAddedLinesAIQThree \ for older projects, compared to \VeryAddedLinesAIQThree \ for newer projects.

However, the metrics for contributions classified as human-authored show a marked difference the median is only \AllAddedLinesHumanMedian (Q3: \AllAddedLinesHumanQThree) for the older projects, while the median rises to \VeryAddedLinesHumanMedian (Q3: \VeryAddedLinesHumanQThree). In a context where these projects have a very high level of AI adoption, we regard this drastic increase in the size of human-authored lines of code as implausible. We offer an alternative explanation: rather than reflecting an increase of human-authored productivty, this data reflects an increase of \emph{undetected agentic activity}.

Indeed, at the present we only track explicit signs of agent activity (refer to the original paper for methodological details). However, we know that agentic activity is only partially observable \cite{agentminingpaper}. In particular, it is very easy to switch off the commit-signing behaviour of an agent such as Claude Code. Likewise, agents such as Pi or OpenCode may simply not sign commits. Therefore, we think that a large amount of commits that we classify at the moment as human authored are in fact AI-assisted, and that this phenomenon is more prevalent for newer projects. Clearly, further study is needed to validate this hypothesis. Given this uncertainty, we refrain from analysing the commits further.


\section{Discussion}
\subsection{Limitations}

This study shares the same limitation as the original study \cite{robbes2026agentic}. We highlight two in particular:

The first is the prevalence of undetected agentic activity \cite{agentminingpaper}. While this activity is by definition very hard to detect, and thus to be certain of its prevalence, we think that \cref{fig:commit-sizes-very} and \cref{fig:commit-sizes-all} give us some insights on its prevalence. We think that the undetected agentic activity is much more prevalent in the dataset of newer projects. With that in mind, we think that our conclusions on the commit ratio, as presented here, are incomplete. Even more so than in the dataset of older projects, we think that the AI commit ratio is an undercount of the true AI commit ratio. While there is already a considerable difference in commit ratio between the older and the newer projects, it is likely to be even larger.

The second is possible sampling bias. We reused the same filtering criteria (more than 5,000 lines of code, more than 100 commits) as in the previous study for consistency. However, it is possible that these criteria, when applied to younger projects such as the ones here, tend to over-select for agentic projects. If agentic projects have an overall higher activity than non agentic projects, such a mechanism would tend to select them. We think that it is possible for it to be the case, particularly for the number of commits. We note that regular bots can also contribute to a higher commit count (e.g. dependabot), which mitigates this.

A similar study with different thresholds might thus arrive to different conclusion. Nevertheless, if smaller and less active projects have less agentic activity, it is still striking to observe so much agentic activity in the most active projects.

\subsection{Implications}

What can we say about this level of agentic activity?

This study shares the implication of the previous study, but the magnitude of the agentic activity in these newer projects makes them even more potent. If adoption is steadily increasing in older projects, coding agents are already ubiquitous in the newer projects that we selected. In this dataset, the transition towards coding agents is thus happening faster than we anticipated, with more intensity than we anticipated. Consider in particular  \Cref{fig:metrics-distribution-very}, compared to \Cref{fig:metrics-distribution-all}. There is already a very large difference in terms of commit ratio, but, if anything, we suspect that this difference is undercounted due to undetected agentic activity. As such, we renew our call for thorough studies of this transition towards agentic software development, as it is happening at an unprecedented speed.

Another implication concerns the other differences between projects. The choice of using agents is related to significant changes in the way the projects are built. This is already evident in the language distribution, where some languages (such as Rust) are far more common in the newer projects than others (such as Java). While modern coding agents are competent in most kinds of languages, some languages are clearly more favored than others. Rust in particular is a very good language for coding agents, as it has extensive static checks that give very rich feedback to coding agents. The broader implication is that, if such differences appear so quickly, other differences (in terms of tool usage, development styles, etc) might appear as well. These aspects should be extensively studied as well.

Yet another implication is the extreme usages observed in some cases. Some users rival, in terms of number of repositories created with agents, with entire organizations such as Microsoft. The fact that some users can be so active in so little time is an even more marked departure with earlier practice. While our previous study called for the study of outliers, we did not expect this kind of outliers, with this kind of intensity.

Finally, the last implication is an increased importance of designing approaches to detect coding agent activity. Before this study, we strongly suspected that a large portion of agentic activity is at the moment undetected. This study, and in particular the counter-intuitive result shown in \Cref{fig:commit-sizes-very} and \Cref{fig:commit-sizes-all}, comfort our opinion. As such, approaches to detect "secret" coding agent activity are extremely important to better study this phenomenon.

\section{Conclusion}

We presented a study of the adoption of coding agents for newly created projects. We found that, for our sample, coding agents are already ubiquitous. We found that a large proportion of projects are built almost entirely from coding agents, and that some users have extremely large levels of activity. This is in spite of the fact that a significant amount of this activity is as yet undetected, foreshadowing an even higher degree of true agentic activity. These findings have important implications, renewing and intensifying the calls for extensive study of this phenomenon, that is happening with breathtaking speed.

\bibliographystyle{alpha}
\bibliography{bib}

\end{document}